\UseRawInputEncoding

\documentclass[pra,groupedaddress,superscriptaddress,nofootinbib,floatfix,preprintnumbers,longbibliography,notitlepage,tightenlines,11pt]{revtex4-1}

\usepackage{amssymb,amsmath,graphicx,color,dsfont}
\usepackage{bm}
\usepackage[tight]{subfigure}
\usepackage[export]{adjustbox}
\usepackage{braket}
\usepackage[colorlinks=true,citecolor=blue,linkcolor=red]{hyperref}

\begin{document}

\dimen\footins=5\baselineskip\relax

\preprint{\vbox{
\hbox{UMD-PP-020-2}
}}

\title{Two-neutrino double-$\beta$ decay in pionless effective field theory
\\
from a Euclidean finite-volume correlation function
}
\author{Zohreh Davoudi{\footnote{\tt davoudi@umd.edu}}
}
\affiliation{Maryland Center for Fundamental Physics and Department of Physics, 
University of Maryland, College Park, MD 20742, USA}
\affiliation{
RIKEN Center for Accelerator-based Sciences,
Wako 351-0198, Japan}

\author{Saurabh V. Kadam{\footnote{\tt ksaurabh@umd.edu }}
}
\affiliation{Maryland Center for Fundamental Physics and Department of Physics, 
University of Maryland, College Park, MD 20742, USA}


\begin{abstract} 
	
Two-neutrino double-$\beta$ decay of certain nuclear isotopes is one of the rarest Standard Model processes observed in nature. Its neutrinoless counterpart is an exotic lepton-number nonconserving process that is widely searched for to determine if the neutrinos are Majorana fermions. In order to connect the rate of these processes to the Standard Model and beyond the Standard Model interactions, it is essential that the corresponding nuclear matrix elements are constrained reliably from theory. Lattice quantum chromodynamics (LQCD) and low-energy effective field theories (EFTs) are expected to play an essential role in constraining the matrix element of the two-nucleon subprocess, which could in turn provide the input into \emph{ab initio} nuclear-structure calculations in larger isotopes. Focusing on the two-neutrino process $nn \to pp \, (ee \bar{\nu}_e\bar{\nu}_e)$, the amplitude is constructed in this work in pionless EFT at next-to-leading order, demonstrating the emergence of a renormalization-scale independent amplitude and the absence of any new low-energy constant at this order beyond those present in the single-weak process. Most importantly, it is shown how a LQCD four-point correlation function in Euclidean and finite-volume spacetime can be used to constrain the Minkowski infinite-volume amplitude in the EFT. The same formalism is provided for the related single-weak process, which is an input to the double-$\beta$ decay formalism. The LQCD-EFT matching procedure outlined for the double-weak amplitude paves the road toward constraining the two-nucleon matrix element entering the neutrinoless double-$\beta$ decay amplitude with a light Majorana neutrino.

\end{abstract}
\maketitle

\section{INTRODUCTION  
\label{sec:Intro} 
}
\noindent
In certain nuclear isotopes with an even number of neutrons and protons, the rate of single-$\beta$ decay is suppressed compared with a double-$\beta$ decay. Consequently, two of the neutrons in a parent nucleus can convert to two protons while emitting two electrons and two antineutrinos: $(A,Z) \to (A,Z+2)+ee\bar{\nu}_e\bar{\nu}_e$. Here, $A$ and $Z$ are the atomic and proton numbers of the parent nucleus, respectively. Since the theoretical prediction of this process and the first estimation of its rate in the 1930s~\cite{PhysRev.48.512}, such a decay has been observed conclusively in a dozen isotopes ranging from $^{48}$Ca to $^{238}U$~\cite{Barabash:2010ie, Saakyan:2013yna}. This decay is one of the rarest Standard Model processes in nature, with experimental half-lives ranging from $\sim 10^{19}$ to $\sim 10^{24}$ years~\cite{Barabash:2010ie, Saakyan:2013yna}. The exotic beyond the Standard Model counterpart of this process, namely the neutrinoless mode $(A,Z) \to (A,Z+2)+ee$, remains at the center of vigorous theoretical studies and experimental searches~\cite{DellOro:2016tmg, Dolinski:2019nrj}, as its observation will unveil the nature of neutrinos and provides a mechanism for lepton-number violation, hence baryon-number violation, in the universe.

The two-neutrino mode ($2\nu\beta\beta$), nonetheless, continues to gain much attention, both theoretically and experimentally, for several reasons. First, this decay mode is a dominant background for the much less probable neutrinoless process which could occur in the same isotopes. Therefore, accurate constraints on its decay rate, and on the spectral shape of electron energies emitted in the decay, are crucial for better understanding of the background in neutrinoless double-$\beta$ ($0\nu\beta\beta$) decay searches~\cite{Saakyan:2013yna, Dolinski:2019nrj, Agostini:2017iyd}. Second, the measured decay rates can be converted to constraints on the corresponding nuclear matrix elements (MEs), under the assumption that only the Standard Model weak interactions are in play. Subsequently, these MEs can be compared against theoretical determinations to test the validity of the nuclear-structure models used~\cite{Moreno:2008dz, Simkovic:2018rdz, NEMO-3:2019gwo, Azzolini:2019yib}. This, along with theoretical calculations of the single-$\beta$ decay MEs, can refine these models, and inform the calculations of the exotic $0\nu\beta\beta$ decay process, for which the role of theory is crucial in constraining the new-physics mechanisms underlying the decay~\cite{DellOro:2016tmg, Dolinski:2019nrj}. Last but not least, the increased sensitivity of $0\nu\beta\beta$ decay searches in recent years has subsequently led to a number of high-statistics measurements of the two-neutrino mode~\cite{KamLAND-Zen:2019imh, Argyriades:2009ph, Arnold:2016qyg, Arnold:2016ezh, Arnold:2018tmo, NEMO-3:2019gwo}, opening a window for probing potential beyond the Standard Model scenarios through this decay mode as well (see e.g., Ref.~\cite{Deppisch:2020mxv} for constraints on right-handed currents from the energy distribution and angular correlation of outgoing electrons in the $2\nu\beta\beta$ decay). Such investigations require accurate nuclear MEs to be computed within the Standard Model, augmented with reliable uncertainties.

Computing nuclear MEs for the $2\nu\beta\beta$ process (and its neutrinoless counterpart) is a rather challenging task given the quantum many-body nature of the nuclear isotopes used, as well as uncertainties in nuclear interactions and currents entering the calculations~\cite{Barea:2013bz, Engel:2016xgb}. This article takes a bottom-up approach to this problem, laying out a possible framework for providing the Standard Model input for the ME involved at the microscopic level, namely that in the $nn \to pp\,(ee\bar{\nu}_e\bar{\nu}_e)$ process, using a combined LQCD and EFT approach. While such a process will not occur unless embedded in certain nuclear backgrounds, an apparent hierarchy of nuclear interactions and nuclear responses to external probes at low energies indicates that the most significant contribution to the nuclear decay may arise from single and correlated two-nucleon couplings, and corrections to this picture can be evaluated within an EFT power counting systematically. The contributions to the ME at each order in the EFT depend upon the low-energy constants (LECs) for interactions and currents, which can ideally be constrained by matching to the Standard Model determination of the same ME using LQCD methodology. Setting up this step is at the core of the present work. Upon successful completion of LQCD determinations of the double-weak two-nucleon ME~\cite{Shanahan:2017bgi,Tiburzi:2017iux}, the formalism presented can be employed to constrain EFTs that will be used in systematic \emph{ab initio} calculations of the decay rate in relevant isotopes. The compatibility of the EFT description in the few-nucleon sector with \emph{ab initio} nuclear many-body methods, and the validity of the adopted EFT power counting when applied to a large nuclear isotope, will require further study and are not the focus of the current work.\footnote{We refer the reader to a recent review in Ref.~\cite{Tews:2020hgp} on the status and prospect of EFT-based studies of nuclei using \emph{ab initio} nuclear many-body methods.}

Pionless EFT~\cite{Kaplan:1998tg, Kaplan:1998we, vanKolck:1998bw, Chen:1999tn}, a nuclear EFT relevant for processes with intrinsic momenta well below the pion mass, has shown notable success in studies of light nuclei and their electromagnetic and weak response, both at physical and unphysical quark masses (the latter accessible from LQCD), see Ref.~\cite{Hammer:2019poc} for a recent review. It is known, in particular, that within the Kaplan-Savage-Wise power counting of the pionless and pionfull EFT~\cite{Kaplan:1998tg, Kaplan:1998we}, a better convergence is observed as the EFT order is increased for observables in the two-nucleon (NN) spin-singlet channel~\cite{Fleming:1999ee, Beane:2001bc}. In a recent study, pionless EFT has further been applied to the problem of $0\nu\beta\beta$ decay in the two-nucleon system up to next-to-leading order (NLO), and is found to require a new short-distance coupling, or contact LEC, already at leading order (LO) to fully renormalize the two-nucleon $\Delta I =2$ transition amplitude, where $I$ denotes the total isospin~\cite{Cirigliano:2017tvr, Cirigliano:2018hja, Cirigliano:2019vdj}. This feature is not expected to occur for the two-neutrino mode, since the $1/q^2$-type neutrino potential induced by the exchange of a light neutrino among the nucleons is absent in the two-neutrino process. Here, $q$ denotes the momentum of the neutrino (i.e., antineutrino) exchanged. Nonetheless, there has been no explicit calculation of the $nn \to pp\,(ee\bar{\nu}_e\bar{\nu}_e)$ amplitude in the pionless EFT with nucleonic degrees of freedom (DOF) to date.

The first attempt at constructing and constraining the nuclear ME for this process was presented in Refs.~\cite{Shanahan:2017bgi,Tiburzi:2017iux}, in which a dibaryon formulation of the pionless EFT~\cite{Beane:2000fi, Phillips:1999hh} was considered. In this case, a contact LEC, called $h_{2,S}$ in this reference, was naturally introduced to account for the conversion of a spin-singlet dibaryon with $I_3=-1$ to that with $I_3=1$ at tree level, similar to the role of the so-called $l_{1,A}$ coupling of the dibaryon formalism that turns a spin-singlet dibaryon with $I_3=0$ to a spin-triplet dibaryon with $S_3=0$ at tree level. Here, $I_3$ ($S_3$) is the third component of the total isospin (spin) operator. Importantly, the first LQCD computation of the relevant ME at a pion mass of $\approx 800~$MeV was performed at the same time, and led to a constraint on this new short-distance coupling,\footnote{The effects of off-shell intermediate states had to be added to the tree-level coupling, leading to an effective short-distance coupling, called $\mathbb{H}_{2,S}$. It was this effective coupling that was constrained with LQCD input on the ME. See Ref.~\cite{Tiburzi:2017iux} for details.} demonstrating its roughly equal contribution to the ME when compared with the $l_{1,A}$ effects in the deuteron-pole contribution. It is valuable to obtain an EFT amplitude within the nucleonic formulation, which is a more suitable framework in connecting to nuclear-structure calculations of the ME in larger isotopes. Such an amplitude is calculated in the current work in Sec.~\ref{sec:doubleIV} for the $2\nu\beta\beta$ decay, and as a recap, in Sec.~\ref{sec:singleIV} for the single-$\beta$ decay process (for which the result had been known previously in the context of general single-weak processes in the two-nucleon systems, e.g., in Refs.~\cite{Kong:1999tw, Kong:2000px, Butler:2000zp}). It is shown that in both cases, the axial coupling of the nucleon and the correlated coupling of the two-nucleon system to an axial-vector current are sufficient to obtain a renormalization-scale independent amplitude at NLO. Such a conclusion for the power counting can be verified through simultaneous LQCD computations of both MEs.

Perhaps an even more significant component of this work is to provide the framework for obtaining the hadronic amplitude for the two-nucleon double-$\beta$ decay from four-point correlation functions of two nucleons obtained with LQCD in a finite Euclidean spacetime. It is well known that multi-hadron amplitudes cannot be directly accessed from LQCD Euclidean correlation functions~\cite{Maiani:1990ca}, and a mapping is required to constrain the amplitudes evaluated at eigenenergies of the system in a finite volume. These finite-volume energies, obtained from hadronic two-point functions, are sufficient for obtaining elastic and inelastic amplitudes with strong interactions in the two- and three-hadron sectors up to the next inelastic thresholds, as introduced in Refs.~\cite{Luscher:1986pf,Luscher:1990ux} and extended in Refs.~\cite{Rummukainen:1995vs, Kim:2005gf, He:2005ey, Davoudi:2011md, Leskovec:2012gb, Hansen:2012tf, Briceno:2012yi, Gockeler:2012yj, Briceno:2013lba, Feng:2004ua, Lee:2017igf, Bedaque:2004kc, Luu:2011ep, Briceno:2013hya, Briceno:2013bda, Briceno:2014oea, Briceno:2017max,Polejaeva:2012ut,Briceno:2012rv, Beane:2014qha, Hansen:2014eka, Hammer:2017uqm,Hammer:2017kms, Guo:2017ism,Mai:2017bge,Briceno:2017tce,Doring:2018xxx,Briceno:2018aml,Jackura:2019bmu,Hansen:2020zhy,Hansen:2019nir}. To constrain transition amplitudes involving external currents, finite-volume MEs, obtained from LQCD three-point functions, are additionally required. Such formalisms are developed in Refs.~\cite{Lellouch:2000pv, Detmold:2004qn, Meyer:2011um, Briceno:2012yi, Bernard:2012bi, Briceno:2014uqa, Feng:2014gba, Briceno:2015csa, Briceno:2015tza, Briceno:2015csa}. For nonlocal MEs derived from four-point functions, such as those relevant for rare kaon-decays and Compton amplitudes~\cite{Christ:2010gi, Christ:2014qaa, Christ:2015pwa,Christ:2012se, Bai:2014cva, Christ:2014qwa,Christ:2016mmq, Bai:2017fkh, Bai:2018hqu,Christ:2019dxu,Briceno:2019opb}, and hadronic double-$\beta$ decays~\cite{Shanahan:2017bgi,Tiburzi:2017iux,Feng:2018pdq,Tuo:2019bue,Detmold:2020jqv,Feng:2020nqj}, in addition to identifying the volume dependence of MEs, another crucial step is involved. Explicitly, due to the dependence of the time-ordered product of currents present in a four-point function on the spacetime signature, Euclidean and Minkowski MEs are fundamentally different when intermediate states with on-shell kinematics are present. These must be subtracted from the Euclidean correlation function, and be separately constructed from the knowledge of appropriate two- and three-point functions. These contributions must then be added back to the non-problematic contributions to obtain the complete Minkowski ME in a finite volume, which is used to construct the physical amplitude, see e.g., Ref.~\cite{Briceno:2019opb}.

Various components of the formalisms mentioned above, such as identifying volume corrections to external and intermediate two-hadron states, as well as reconstructing the Minkowski amplitude from the Euclidean four-point function, are relevant for the mapping of the nucleonic ME in the $2\nu\beta\beta$ transition as well. Nonetheless, none of the work developed so far can be applied directly to the problem at hand, mainly due to new single-body contributions to the two-body nonlocal ME.\footnote{In particular, a recent work~\cite{Feng:2020nqj} on finite-volume formalism for nonlocal $2 \to 2$ processes is missing the single-body contributions, making it inapplicable to nuclear double-$\beta$ decay transitions.} Furthermore, while the finite-volume matching for the single-weak processes, such as $pp$ fusion, has been presented in the past within the pionless-EFT framework~\cite{Briceno:2012yi}, such a matching for the EFT amplitude of the $2\nu\beta\beta$ transition with nucleonic DOF is developed for the first time in the current work, as outlined in Secs.~\ref{sec:doubleV}, \ref{sec:doubleVIV}, and \ref{sec:doubleEM} below. This parallels a related mapping with dibaryon formalism presented in Ref.~\cite{Tiburzi:2017iux} using a different approach. Finally, a new derivation is presented for the previously known result for the single-weak amplitude in the two-nucleon sector, combining the pionless-EFT result for the correlation function in a finite volume with the derivation of Ref.~\cite{Briceno:2015tza} for a general mapping for $2 \to 2$ single-weak processes, as outlined in Secs.~\ref{sec:singleV} and \ref{sec:singleVIV}. The conclusion and outlook of this work are presented in Sec.~\ref{sec:conclusion}.

\section{Setting up the formalism 
\label{sec:formalism} 
}
\noindent
In pionless EFT, the hadronic Lagrangian\footnote{All instances of Lagrangian in this paper are Lagrangian densities, which for brevity, will be called Lagrangian throughout.} is arranged according to the number of nucleons,
\begin{equation}
    \mathcal{L} = \mathcal{L}_{(1)} + \mathcal{L}_{(2)} +\cdots,
    \label{eq: EFT Lagrangian expansion}
\end{equation}
where $\mathcal{L}_{(n)}$ is composed of $n-$nucleon operators, and ellipsis denotes higher-nucleon operators. The single-nucleon Lagrangian
\begin{equation}
	{\cal L}_{(1)}=N^{\dagger }\bigg(i\partial_{t}+\frac{{\nabla}^2}{2M}\bigg)N+\cdots
	\label{eq: EFT 1 nucleon Lagrangian}
\end{equation}
is the nonrelativistic kinetic-energy operator of the nucleon at LO, and the ellipsis denotes relativistic corrections. Here, $\partial_t$ is the time derivative and ${\bf \nabla}$ is the spatial gradient operator. $N=(p,n)^T$ is an isospin doublet composed of the proton, $p$, and the neutron, $n$, fields, each with mass $M$. Isospin symmetry will be assumed throughout as the isospin-breaking effects are small and contribute at higher orders than considered. The Lagrangian with two-nucleon operators is given by
\begin{align}
	{\cal L}_{(2)}=
	&-C_{0}(N^{T}P_{i}N)^{\dagger }(N^{T}P_{i}N)
	-\widetilde{C}_{0}(N^{T}\widetilde{P}_{i}N)^{\dagger }(N^{T}\widetilde{P}_{i}N)
	\nonumber \\
	&+{\frac{C_{2}}{8}}\left[ (N^{T}P_{i}
	N)^{\dagger }[N^{T}(\overleftarrow{{\bf\nabla}}^{2}P_{i}-2
	\overleftarrow{{\bf\nabla}}\cdot P_{i}\overrightarrow{{\bf\nabla}}+
	P_{i}\overrightarrow{{\bf\nabla}}^{2})N]+{\rm h.c.}\right]
	\nonumber\\
	&+{\frac{\widetilde{C}_{2}}{8}}\left[ (N^{T}\widetilde{P}_{i}N)^{\dagger }(N^{T}[
	\overleftarrow{{\bf\nabla}}^{2}\widetilde{P}_{i}-2\overleftarrow{{\bf\nabla}}\cdot \widetilde{P}_{i}
	\overrightarrow{{\bf\nabla}}+\widetilde{P}_{i}\overrightarrow{{\bf\nabla}}^{2}]N)+{\rm h.c.}\right]+\cdots,
	\label{eq: EFT 2 nucleon Lagrangian}
	\end{align}
where the overhead arrow indicates which nucleon fields are being acted on by the derivative operator, and the ellipsis denotes higher-derivative operators. Here and everywhere, repeated indices are summed over where $i=1,2,3$. $P_{i}$ and $\widetilde{P}_{i}$ are the spin-isospin projection operators for spin-singlet $(^1S_0)$ and spin-triplet $(^3S_1)$ channels, respectively, with definition and normalization
\begin{eqnarray}
	P_{i} &\equiv &\frac{1}{\sqrt{8}}\sigma _{2}\tau _{2}\tau_{i},
	\quad \text{Tr}\left[P_{i}^{\dagger }P_{j}\right]=\frac{1}{2}\delta _{ij},
	\nonumber\\
	\widetilde{P}_{i} &\equiv &\frac{1}{\sqrt{8}}\sigma _{2}\sigma _{i}\tau _{2},\quad 
	\text{Tr}\left[\widetilde{P}_{i}^{\dagger }\widetilde{P}_{j}\right]=\frac{1}{2}\delta _{ij},
	\label{eq: spin-isospin projection operators}
\end{eqnarray}
where $\sigma_i$ $(\tau_i)$ are the Pauli matrices acting in spin (isospin) space. The term proportional to $C_0\;(\widetilde{C}_0)$ corresponds to the LO contact interaction in the $^1S_0$ ($^3S_1$) channel while the terms proportional to $C_2\;(\widetilde{C}_2)$ describe the NLO momentum-dependent interactions. The couplings in the EFT are generally renormalization-scale dependent, and their Renormalization Group flow is determined by requiring the physical observables such as amplitudes to be scale independent order by order in the EFT. The renormalization-scale dependence of the couplings in Eq.~(\ref{eq: EFT 2 nucleon Lagrangian}) is given in Eq.~\eqref{eq: C0, C2 scale dependence} below.

The effective Lagrangian for the charged-current (CC) weak interaction is given by
\begin{equation}
	{\cal L}_{\rm CC}\ =-G_{F} \; l_{-}^{\mu
	}J_{\mu }^{+}+{\rm h.c.} ,
	\label{eq: Charged current Lagrangian}
\end{equation}
where $G_{F}$ is the Fermi's constant. The leptonic current
\begin{equation}
    l_{-}^{\mu }=\overline{e}\gamma ^{\mu }(1-\gamma _{5})\nu
    \label{eq: Leptonic current}
\end{equation}
contains electron, $e$, and neutrino, $\nu$, fields, and the hadronic current can be written in terms of vector and axial contributions as
\begin{eqnarray}
	J_{\mu }^{+} =V_{\mu }^{+}-A_{\mu }^{+}
	=\frac{V_{\mu }^{1}+iV_{\mu }^{2}}{\sqrt{2}}-\frac{A_{\mu }^{1}+iA_{\mu }^{2}}{\sqrt{2}}.
	\label{eq: Full Hadronic current}
\end{eqnarray}
The superscript (subscript) in the hadronic (leptonic) current denotes isovector components while the subscript (superscript) denotes the spacetime vector components. The vector current mediates Fermi transitions, while the axial current governs Gamow-Teller transitions, which correspond to different isospin $(\Delta I)$ and spin $(\Delta S)$ selection rules.

The focus of this work is on isovector transitions that lead to (double) $\beta$ decays. Further, only low-energy processes are considered such that pionless EFT is a proper description. As a result, the only relevant currents are the spatial component of the axial-vector isovector currents.\footnote{The momentum-independent scalar-isovector current vanishes for $\beta$-decay processes. Furthermore, the leading momentum-dependent vector-isovector current does not contribute to the ME in the limit where the current carries zero spatial momenta. Finally, the leading momentum-dependent axial scalar-isovector current is proportional to $1/M$, and is therefore suppressed at the order of EFT considered here.} In fact, it is the nuclear ME of such currents, i.e., Gamow-Teller MEs, that are not constrained precisely phenomenologically and their determination will be the subject of a LQCD-EFT matching program.  At LO in pionless EFT, the nonrelativistic one-body operator with such quantum numbers is
\begin{equation}
	A_{k(1)}^{+} =\frac{g_{A}}{2} \, N^\dagger \tau _{+}\sigma_{k}N,
	\label{eq: axial current : one body}
\end{equation}
where $\tau_{+}=(\tau_{1}+i\tau_{2})/\sqrt{2}$ and $g_A$ is the nucleon axial charge. $k=1,2,3$ is the spatial Lorentz index. The momentum-independent two-body axial-vector/isovector current is
\begin{equation}
	A_{k(2)}^{+}=L_{1,A}\big( N^{T}\widetilde{P}_{k}N\big) ^{\dagger }\big( N^{T}P_{+}N\big),
	\label{eq: axial current : two body}
\end{equation}
where $P_{+} = (P_{1}+iP_{2})/{\sqrt{2}}$. This current contributes to the $\beta$-decay amplitude at NLO, as shown below, leading to a Renormalization-Group equation for $L_{1,A}$, obtained in terms of strong-interaction couplings in Eq.~\eqref{eq: EFT 2 nucleon Lagrangian} as well as the nucleon's axial charge, see Eqs.~\eqref{eq: L1A scale independent} and \eqref{eq: L1A tilde mu derivative}. All fields introduced in this subsection have spatial and temporal dependence.

\subsection{Infinite-volume amplitudes
\label{sec:formalismIV}
}
\begin{figure}[t!]
    \centering
    \includegraphics[scale=0.95]{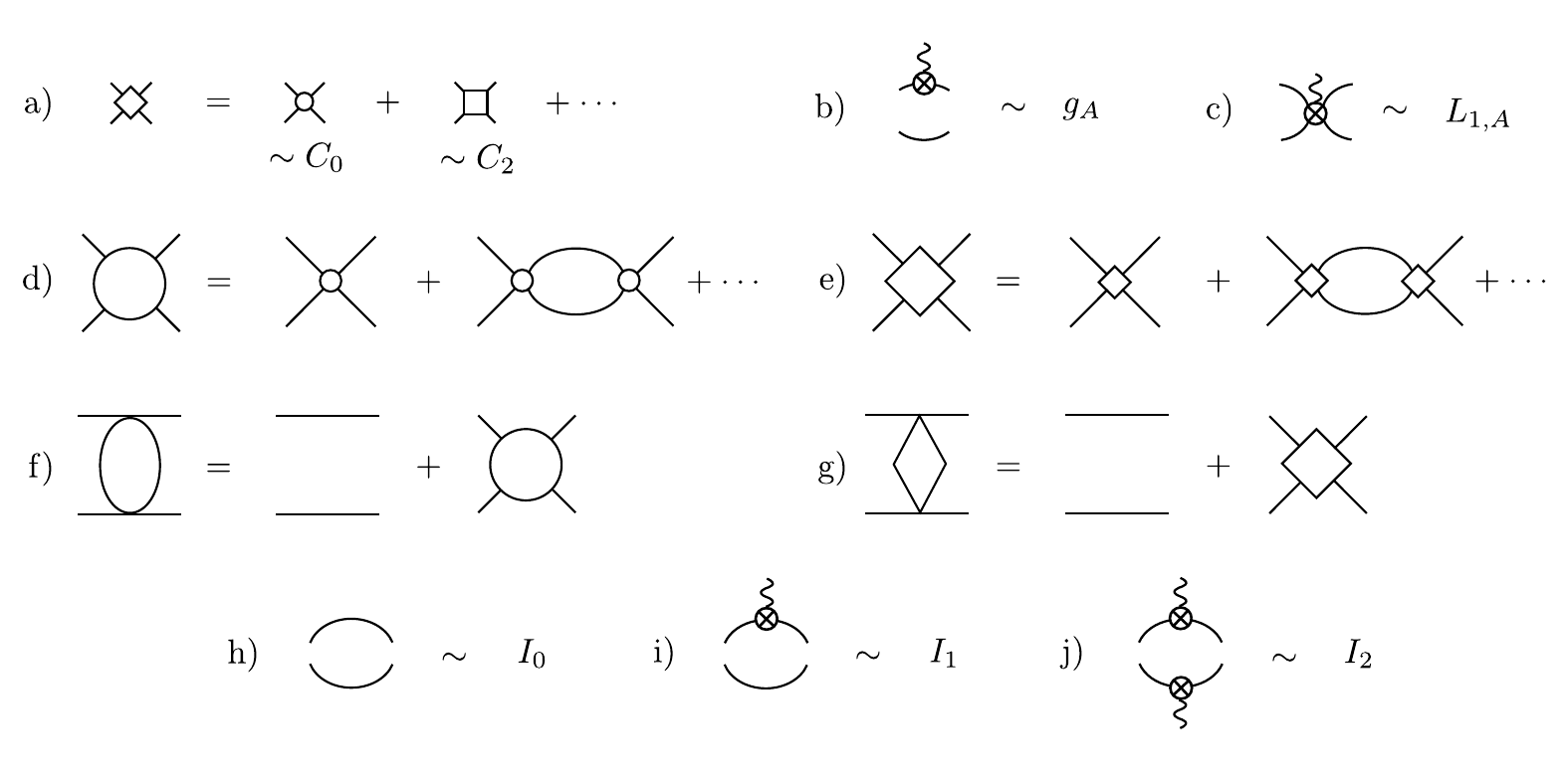}
    \caption{A summary of the building blocks used in the diagrammatic expansion of the amplitudes and correlation functions used in the formalism of this paper and presented in Figs.~\ref{fig: NN to NN full scattering amplitude}-\ref{fig: double weak finite V correlation}. a) represents the strong-interaction kernel in pionless EFT in the $^1S_0$ channel. The ellipsis denotes higher-order interactions in the EFT not appearing in this work. b) and c) are the single- and two-nucleon weak currents inducing the $^1S_0 \to {^3S_1}$ transition. The ellipses in d) and e) denote the s-channel diagrams included to all orders. h), i), and j) are the relevant loop functions. In all figures, the line represents the nucleon and the wavy line is a weak-current insertion. The counterparts of diagrams a), d), e), f), and g) for the $^3S_1$ channel are not shown for brevity. These are distinguished from the $^1S_0$ diagrams by the gray filled symbols in the upcoming figures.
    }
    \label{fig:notation}
\end{figure}
The nuclear weak-decay processes involve nonperturbative strong interactions between nucleons, governed by the terms in Eq.~\eqref{eq: EFT 2 nucleon Lagrangian}, and perturbative weak interactions between nucleons and leptons from $\mathcal{L}_{\rm CC}$ in Eq.~\eqref{eq: Charged current Lagrangian}. The Feynman amplitude of a weak process involving $n$-lepton currents is obtained by considering the $n$th order term in the perturbative expansion of the S-matrix with respect to the CC interaction. This amplitude can be decomposed into leptonic and hadronic parts as 
\begin{equation}
	i\mathcal{M}^{\rm full}_{X} = G_F^{n}\cdot i\mathcal{M}^{\rm lep.}_{X} \cdot i\mathcal{M}_{X},
\end{equation}
where $X$ denotes the transition considered, and all the amplitudes introduced have appropriate dependence on on-shell four-momenta of initial and final states that are suppressed for brevity. The leptonic amplitude, $\mathcal{M}^{\rm lep.}_X$, involves the kinematics of the electron(s) and neutrino(s) and the hadronic amplitude, $\mathcal{M}_X$, is the amplitude involving initial and final nucleonic states. The aim of this paper is to relate the MEs obtained in a finite Euclidean spacetime to this physical hadronic amplitude for the case of single- and double-$\beta$ decays. As a result, we turn our focus on the amplitude $\mathcal{M}_{X}$ and will not carry out the straightforward convolution\footnote{The term `convolution' is used in this paper to indicate the two functions are multiplied under an integral or a summation given the context, and the exact mathematical definition is not assumed.} of the hadronic contribution with the leptonic amplitude. This convolution is more involved for the case of $0\nu\beta\beta$ decay process, which includes a Majorana neutrino propagator, and will be studied elsewhere~\cite{Davoudi:2020}.

Given its nonperturbative nature, the hadronic amplitude $\mathcal{M}_{X}$ is obtained by solving the Schr\"{o}dinger's equation with the strong interaction potentials that act between nucleons in initial, final, and when relevant, intermediate nuclear states.\footnote{The potential language is appropriate as nucleons involved in low-energy processes are nonrelativistic to a good approximation. Relativistic corrections can be included systematically in the EFT} The $\beta$-decay processes considered in this work involve two-nucleon initial and final states and Gamow-Teller transitions, and pionless EFT will be used to construct the amplitudes up to and including NLO. Thus, as mentioned above, the only hadronic currents that contribute are those given in Eqs.~(\ref{eq: axial current : one body}) and (\ref{eq: axial current : two body}). The Hamiltonian for these processes is given by 
\begin{equation}
	\hat{H} \, = \, \hat{H}_{S} + \hat{H}_{\rm CC},\quad
	\hat{H}_{S} \, = \, \hat{H}_0 + \hat{V}_{S} ,
	\label{eq: Full and Strong Hamiltonian}
\end{equation}
where $\hat{H}_S$ is the strong Hamiltonian with $\hat{H}_0$ and $\hat{V}_S$ being the free two-nucleon Hamiltonian and the strong interaction potential, respectively. These are derived straightforwardly from the strong-interaction Lagrangian density in pionless EFT presented in Eq.~(\ref{eq: EFT Lagrangian expansion}).  The weak-Hamiltonian operator $\hat{H}_{\rm CC}$ can be understood as the product of a leptonic part and a hadronic part, and the latter is the only relevant part when acting on the state of two nucleons. In particular, assuming a final-state kinematic such that each hadronic current transfers zero spatial momenta (corresponding to pairs of electron and neutrino in a $n$-$\beta$ decay carrying away no spatial momenta), and at the EFT order considered in this work, the hadronic part, $\hat{A}_{\rm CC}$, can be written as
\begin{equation}
    \hat{A}_{\rm CC}=\hat{A}_{g_A}\bm{\otimes} \mathds{1} + \mathds{1} \bm{\otimes} \hat{A}_{g_A} + \hat{A}_{L_{1,A}} ,
    \label{eq: weak Hamiltonian HCC}
\end{equation}
where
\begin{equation}
    \hat{A}_{g_A}=\int \; d^3{\bm x} \; A_{k(1)}^{+}({x}), \quad
    \hat{A}_{L_{1,A}}=\int \; d^3 {\bm x}  \; A_{k(2)}^{+}({x}) ,
    \label{eq: CC hamiltonian, one-body and two-body}
\end{equation}
are constructed from the one-body and two-body current operators, respectively. The one-body operator is extended to two-nucleon Hilbert space by taking a tensor product $(\bm{\otimes})$\footnote{Not to be mixed with the $\otimes$-product notation introduced in later sections to indicate sum or integral convolution of functions.} with the identity operator, $\mathds{1}$, in the nonparticipating nucleon space. For a given $NN\to NN$ process, Lorentz index $k$ determines the change in spin along the spin quantization axis.

In order to treat strong interactions nonperturbatively, one needs to find the Feynman amplitude between two-nucleon eigenstates of $\hat{H}_S$, which are constructed from the eigenstates of $\hat{H}_0$~\cite{Kaplan:1996xu}. In the center-of-mass (CM) frame of the two-nucleon system, the $\hat{H}_0$ eigenstate in channel $NN$ ($^1S_0$ or $^3S_1$) with relative momentum $\bm{p}$ and energy $E$ is given by
\begin{equation}
    \hat{H}_0| \bm{p},NN \rangle  = \frac{\hat{\bm p}^2}{M}\, | \bm{p},NN \rangle=
    \frac{p^2}{M} | \bm{p},NN \rangle  = E\, | \bm{p},NN \rangle ,
    \label{eq: two-nucleon eigenstates of free Hamiltonian}
\end{equation}
where $\hat{\bm p}$ is the relative-momentum operator in the CM frame, and $p^2={\bm p} \cdot {\bm p}$. This state is normalized according to
\begin{equation}
    \langle \bm{p}',NN |\bm{p},NN \rangle = (2\pi)^3 \delta^3({\bm p}-{\bm p}'),
    \label{eq:free-state-norm}
\end{equation}
consistent with the normalization convention for nonrelativistic two-body states with zero total momentum. Using the free retarded Green's function
\begin{equation}
    G_0 (E^+) = \frac{1}{E-H_0 + i\epsilon},
    \label{eq: Green's function: Free}
\end{equation}
with $\epsilon > 0$, and the T-matrix for strong interaction
\begin{equation}
    T_S(E^+) = \hat{V}_{S} + \hat{V}_{S}\, G_0 (E^+)\,T_S(E^+),
    \label{eq: T-matrix for strong interaction}
\end{equation}
the $\hat{H}_S$ eigenstate in two-nucleon channel $NN$ with energy $E$ and relative momentum $\bm{p}$ is given by
\begin{equation}
    |\phi(E^+),NN \rangle = | {\bm p},NN \rangle + G_0 (E^+)T_S(E^+) \, | {\bm p},NN \rangle .
    \label{eq: Strong Hamiltonian Eigenstate}
\end{equation}

The hadronic Feynman amplitude, $\mathcal{M}_{X}$, between the initial and final two-nucleon states with CM energies $E_i$ and $E_f$, respectively, is then given by
\begin{equation}
	i\mathcal{M}_{X} \;=\; -i \langle \phi(E_f^-),NN| \;J_{X}|\phi(E_i^+),NN \rangle ,
	\label{eq: Hadronic amplitude in terms of T-matrix}
\end{equation}
where $E_f^-$ is defined through an advanced Green's function, i.e., $i\epsilon  \to -i\epsilon$ in Eq.~(\ref{eq: Green's function: Free}), and $J_X$ is the hadronic part of the T-matrix for the weak interaction. For the $n$th order weak process, $J_X$ is constructed using $n$ insertions of $\hat{A}_{\rm CC}$ and strong retarded Green's function, $G_S (E^+)$, defined as
\begin{equation}
	G_S (E^+) = \frac{1}{E-H_S + i\epsilon}
	= G_0 (E^+) + G_0 (E^+)\, T_S(E^+) \, G_0 (E^+).
	\label{eq: Green's function: Strong}
\end{equation}
The hadronic part of the T-matrix corresponding to the first- and second-order weak processes of this paper will be introduced in Eqs.~\eqref{eq: single beta decay T-matrix} and \eqref{eq: double beta T-matrix}, respectively.

Equations \eqref{eq: T-matrix for strong interaction}, \eqref{eq: Strong Hamiltonian Eigenstate}, and \eqref{eq: Hadronic amplitude in terms of T-matrix} imply that the hadronic Feynman amplitude can be expressed in terms of MEs of $\hat{V}_{S}$ and $\hat{A}_{\rm CC}$ between free eigenstates, defined in Eq.~(\ref{eq: two-nucleon eigenstates of free Hamiltonian}). For a given two-nucleon channel, these states can be constructed using the spin-isospin projection operators in Eq.~\eqref{eq: spin-isospin projection operators}. From the Lagrangian in Eq.~\eqref{eq: EFT 2 nucleon Lagrangian}, the MEs of $\hat{V}_S$ between such momentum eigenstates are given by
\begin{align}
    &\langle {\bm q}_1,{^1S_0}| \; \hat{V}_S^{\rm (LO)} \; | {\bm q}_2, {^1S_0} \rangle = C_0,\quad
        \label{eq: Strong potential MEs I}
    \\
    &\langle {\bm q}_1,{^1S_0}| \; \hat{V}_S^{\rm (NLO)} \; | {\bm q}_2, {^1S_0} \rangle = 
     \frac{C_{2}}{2}(q^2_1+q^2_2),
    \label{eq: Strong potential MEs II}
\end{align}
for LO and NLO interactions, respectively. Similar relations can be written for the spin-triplet channel with the corresponding couplings. One can similarly obtain the MEs of the one-body and two-body weak hadronic Hamiltonian using Eq.~\eqref{eq: axial current : one body},\eqref{eq: axial current : two body}, and \eqref{eq: CC hamiltonian, one-body and two-body}, which are only nonvanishing when taken between the spin-singlet and spin-triplet states:
\begin{align}
    &\langle {\bm q}_1,{^1S_0}| \; \hat{A}_{g_A}\bm{\otimes} \mathds{1} \; | {\bm q}_2, ^3S_1 \rangle = 
    \frac{g_A}{2} \, (2\pi)^3\, \delta^3({\bm q}_1-{\bm q}_2),
    \label{eq: Weak Hamiltonian MEs I}
    \\
   & \langle {\bm q}_1,{^1S_0}| \; \hat{A}_{L_{1,A}}\; | {\bm q}_2, ^3S_1 \rangle =  L_{1,A} .
    \label{eq: Weak Hamiltonian MEs II}
\end{align}

\begin{figure}
    \centering
    \includegraphics[scale=1.05]{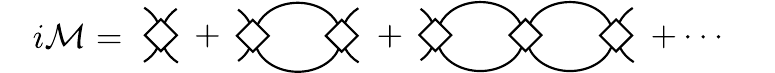}
    \caption{The full $NN\to NN$ scattering amplitude in the $^1S_0$ channel in infinite volume. The ellipsis denotes an expansion in arbitrary number of the Bethe-Salpeter kernel, defined in Fig.~\ref{fig:notation}-a. A similar expansion can be formed for the scattering in the $^3S_1$ channel with the corresponding Bethe-Salpeter kernel.}
    \label{fig: NN to NN full scattering amplitude}
\end{figure}

Using Eqs.~\eqref{eq: Strong potential MEs I} and (\ref{eq: Strong potential MEs II}) and inserting a complete set of free single- and multi-particle states in the iterative sum in Eq.~\eqref{eq: T-matrix for strong interaction}, one can obtain the $T_S$ MEs up to NLO:
\begin{align}
    &\langle{\bm q}_1,{{^1}S_0}| T_S^{\rm (LO)}(E^+) | {\bm q}_2,{{^1}S_0}\rangle  =
    \frac{C_0}{1-I_0(E^+)\,C_0},
    \label{eq: ME of strong T-matrix I}
    \\
    &\langle{\bm q}_1,{{^1}S_0}| T_S^{\rm (NLO)}(E^+) | {\bm q}_2,{{^1}S_0}\rangle =
    \frac{C_2}{2}\frac{(q^2_1+q^2_2-2p^2)}{1-I_0(E^+)\,C_0}+ \frac{C_2\,p^2}{[1-I_0(E^+)\,C_0]^2} .
    \label{eq: ME of strong T-matrix II}
\end{align}
Similar expressions can be formed for the spin-triplet channel upon replacements ${{^1}S_0} \to {{^3}S_1}$, $T_S \to \widetilde{T}_S$, $C_0 \to \widetilde{C}_0$, and $C_2 \to \widetilde{C}_2$. $I_0$ is an ultraviolet (UV) divergent integral that is regularized with the power-divergence subtraction (PDS) scheme introduced in Ref.~\cite{Kaplan:1998tg},
\begin{equation}
    I_0(E^+) = \int \frac{d^3{q}}{(2\pi)^3} \; \frac{1}{E-\frac{{\bm q}^2}{M}+i\epsilon} \xrightarrow{\text{PDS}} -\frac{M}{4\pi}\,(\mu+ip) ,
    \label{eq: I0 in infinite volume}
\end{equation}
where $p=\sqrt{ME}$, and $\mu$ is the renormalization scale. In a Feynman-diagram expansion, this divergence corresponds to the two-nucleon s-channel loop, see Fig.~\ref{fig: NN to NN full scattering amplitude}.

The on-shell $NN\to NN$ elastic scattering amplitude is defined as
\begin{align}
    i\mathcal{M}^{\rm (LO+NLO)}(E) &\equiv -i \; \langle {^1S_0},{\bm p}| \; T_S^{\rm (LO)}(E^+)+T_S^{\rm (NLO)}(E^+) \; | {\bm p},{^1S_0}\rangle ,
    \label{eq: LO+NLO 2to2 Amplitude }
\end{align}
with a similar expression for the amplitude in the spin-triplet channel upon replacements ${^1S_0} \to {^3S_1}$, $\mathcal{M} \to \widetilde{\mathcal{M}}$, and $T_S \to \widetilde{T}_S$. From the requirement of renormalization-scale invariance of this amplitude, the scale dependence of the strong interaction couplings can be deduced:
\begin{align}
    \mu\frac{dC_0}{d\mu} &= \frac{M \mu}{4 \pi}C_0^2,\quad
    \mu\frac{dC_2}{d\mu} = \frac{M \mu}{2 \pi}C_0 C_2,
    \label{eq: C0, C2 scale dependence}
\end{align}
with similar relations for the spin-triplet channel upon replacements $C_0 \to \widetilde{C}_0$, and $C_2 \to \widetilde{C}_2$.

In the remainder of this paper, two simplifying conventions will be applied to notation. First, the $NN$ channel index will be dropped from the states, and their assignment must be deduced from the ME under consideration. Instead, when helpful given the context, the kinematic dependence of the states will be displayed. Second, the energies are meant to be arising from a retarded propagator unless specified, hence the superscript `$+$' on energies will be dropped for brevity.

\subsection{Finite-volume methodology
\label{sec:formalismV}
}
In this section, the well-known formalism for relating two-body elastic scattering amplitudes will be reviewed, setting the stage for the derivation of the matching relation between finite-volume three- and four-point functions in the two-nucleon systems and the corresponding hadronic transition amplitudes. The derivation presented, drawn primarily from the approach of Refs.~\cite{Kim:2005gf,Briceno:2015csa,Briceno:2015tza,Briceno:2019opb}, provides a suitable ground for the finite-volume analysis of later sections. In all finite-volume quantities considered, the temporal extent of spacetime is taken to be infinite, while the spatial extents are taken to be finite. The spatial geometry is taken to be cubic with extents $L$, and periodic boundary conditions are assumed on the fields. As a result, while the energy remains a continuous variable, each Cartesian component of the spatial momentum of the non-interacting nucleons is discretized in units of $2\pi/L$.

In comparing assumptions of this work with the conditions on LQCD correlation functions, several differences can be highlighted. First, the spacetime is discretized in LQCD computations. In the following, therefore, it is assumed that the continuum limit of LQCD quantities are taken before matching to physical amplitudes through the matching relations provided in this work. Second, unless specified, the matching relations do not make any reference to the Euclidean spacetime and correlation functions/MEs are represented in Minkowski spacetime in both finite and infinite volume. Nonetheless, quantities computed with LQCD correspond to Euclidean spacetime. The distinction between Minkowski and Euclidean correlation functions becomes important only for the four-point functions, where the currents are separated in time, and the analytic structure of the MEs may differ significantly between different time signatures~\cite{Briceno:2019opb}. As a result, despite the two- and three-point functions, a straightforward analytic continuation does not transform quantities from one time signature to the other. Consequently, extra steps must be taken to use the matching relations of this work to extract physical amplitudes from LQCD four-point functions. These steps are outlined in Sec.~\ref{sec:doubleEM}, following and extending the protocol of Ref.~\cite{Briceno:2019opb} for nonlocal MEs of single-hadron states.
 
Consider the two-nucleon systems in the spin-singlet channel. Similar relations can be obtained identically for the spin-triplet, or any two-hadron channels for that matter, below the lowest three-hadron inelastic threshold. Nonrelativistic kinematic is assumed throughout although generalization to relativistic kinematic is straightforward and known~\cite{Kaplan:1998we}. The relations presented are general for all partial-wave amplitudes (that besides possible mixing of the physical amplitudes in the spin-singlet channel, get further mixed in a finite cubic volume). However, the s-wave limit will be made explicit when necessary, which is relevant for two-nucleon processes of this work that occur at low energies. The key idea of relating the finite-volume MEs to infinite-volume amplitudes is to equate two equivalent definitions of the $n$-point correlation function in a finite volume~\cite{Briceno:2015csa,Briceno:2015tza}, as will become evident shortly.
  
Consider the momentum-space finite-volume two-point correlation function of two nucleons projected to zero spatial momentum, 
\begin{align}
    C_L(P) &=
    \int_{L}d^3x\int dx_0\,e^{iP\cdot x}
    \left[\langle 0|\, T[B(x)B^\dagger(0)]\, |0\rangle \right]_L,
    \label{eq: two-point correlation (P)}
\end{align}
which can be expressed, alternatively, in the mixed momentum-time representation:
\begin{align}
    \label{eq: two point correlation(x-y) in C(P)}
    C_L(y_0-x_0,{\bm P}=0)
    &\equiv L^3 \int \frac{dE}{2\pi}\;e^{-iE(y_0-x_0)} \;C_L (P)
    \\
    \label{eq: two point correlation(x-y) in <BB>}
    &= \int_Ld^3x\;d^3y 
    \left[\langle 0|\, T[B(y)B^\dagger(x)]\, |0\rangle\right]_L
    \\
    &=L^6\sum_{E_n}\;e^{-iE_n(y_0-x_0)}\; \left[\langle 0|\,B(0) \vphantom{B^\dagger} \,| E_n;L\rangle\right]_L \left[\langle E_n;L|\,B^\dagger(0)\, |0\rangle\right]_L.
    \label{eq: two point correlation: dispersion}
\end{align}
In these equations, $B^{\dagger}$ and $B$ are two-nucleon source and sink interpolating operators for the spin-singlet channel, respectively. The operators in the second line have spacetime dependences and $x\equiv(x_0,{\bm x})$ and $y\equiv(y_0,{\bm y})$. $P$ denotes the total four-momentum of the two-nucleon system and $P=(E,\bm{P}=\bm{0})$. The Minkowski metric $g_{\mu\nu}={\rm diag}(1,-1,-1,-1)$ is assumed throughout. $T$ denotes the Minkowski time-ordering operator, and without loss of generality, it is assumed that $y_0>x_0$. The subscript $L$ on the spatial integral denotes the integral is performed over a finite cubic volume, and the $[\cdots]_L$ is used to denote the finite-volume nature of the ME enclosed. In going from the second to the third line, the relation $B(x) = e^{i\hat{P}_0x_0-i\hat{\bm P}\cdot  \bm{x}} B(0) e^{-i\hat{P}_0x_0+i\hat{\bm P}\cdot  \bm{x}}$ is used, where $\hat{P}_0$ and $\hat{\bm{P}}$ are, respectively, energy and momentum operators acting on the adjacent states. Furthermore, a complete set of discretized finite-volume two-nucleon states are inserted, with $n$ denoting the state index. These finite-volume states in the CM frame are characterized by the total energy of two nucleons, $E_n$, and an $L$ argument to remind the finite-volume nature of these states. They are chosen to satisfy the normalization condition
\begin{equation}
    \langle E_m;L|E_n;L\rangle  = \delta_{n,m},
    \label{eq: Finite V states normalization}
\end{equation}
to be compared with the normalization of the infinite-volume states in Eq.~(\ref{eq:free-state-norm}). Note that for on-shell states, $E_n=p_n^2/M$, where $p_n \equiv |\bm{p}_n|$ is the corresponding  interacting momentum of each nucleon in the CM frame.
\begin{figure}
    \centering
    \includegraphics[scale=1]{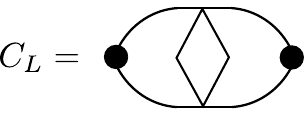}
    \caption{Diagrammatic representation of the finite-volume two-point correlation function corresponding to the expansion in Eq.~(\ref{eq: two point correlation analytic 1}). The black dots are the interpolating operators for the two-nucleon state in the spin-singlet channel. All other components are introduced in Fig.~\ref{fig:notation}. The s-channel loops are evaluated as a sum over discretized momenta, as discussed in the texts.
    \label{fig: two point correlation diagram}}
\end{figure}

The goal now is to express $C_L(P)$ in terms of a diagrammatic representation illustrated in Fig.~\ref{fig: two point correlation diagram}, with the building blocks provided in Fig.~\ref{fig:notation}. One can then perform the inverse Fourier integral in Eq.~\eqref{eq: two point correlation(x-y) in C(P)} to arrive at an equivalent form that can be compared against Eq.~(\ref{eq: two point correlation: dispersion}). The correlation function $C_L(P)$ contains an infinite coupled sum with s-channel two-nucleon loops, $iI_0^V$, as depicted in Fig.~\ref{fig:notation}-g, connected via the two-nucleon Bethe-Salpeter kernels, $\mathcal{K}$, as shown in Fig.~\ref{fig:notation}-a for pionless EFT. This expansion can be written in the compact notation of Refs.~\cite{Hansen:2019nir,Briceno:2019opb} as
\begin{equation}
    C_L(P) = \bar{B} \otimes iI^V_0 \sum_{n=0}^{\infty} \left[\otimes \; \mathcal{K} \otimes I^V_0 \right]^n \otimes \bar{B}^{\dagger}.
    \label{eq: two point correlation analytic 1}
\end{equation}
Here, all kinematic dependence of the functions are suppressed, but these will be made explicit shortly. $\bar{B}^\dagger$ and $\bar{B}$ are the zero-spatial-momentum projected source and sink interpolating operators for the two-nucleon system in the $^1S_0$ channel, i.e., $\bar{B}(P)=\int d^3x \, B(x)$ and $\bar{B}^\dagger(P)=\int d^3x \, B^\dagger(x)$. The symbol $\otimes$ should be interpreted as a sum/integral convolution of the adjacent functions. Explicitly, for two arbitrary functions $\chi$ and $\xi$ that are smooth in $q=(q^0,\bm{q})$:
\begin{eqnarray}
    \label{eq:I-convolution-1}
    \chi \otimes I^V_0 \otimes \xi &\equiv& 
    \frac{i}{L^3}\sum_{\bm{q}  \in \frac{2\pi}{L}\mathbb{Z}^3}\int\,\frac{dq_0}{2\pi}\; \chi(q) \frac{1}{q_0-\frac{{\bm q}^2}{2M}+i\epsilon}\;\frac{1}{E-q_0-\frac{{\bm q}^2}{2M}+i\epsilon}\xi(q) 
    \\
    \label{eq:I-convolution-2}
    &=& \frac{1}{L^3} \sum_{\bm q} \; \chi(E,\bm{q})\frac{1}{E-\frac{{\bm q}^2}{M}+i\epsilon}\xi(E,\bm{q}).
\end{eqnarray}
Since $\chi$ and $\xi$ are smooth functions, which is the case for interpolating fields and the kernels in Eq.~(\ref{eq: two point correlation analytic 1}) below inelastic thresholds, the only difference between the sum and the corresponding infinite-volume integral (beyond exponentially suppressed corrections) arise from singularities of the $1/(E-\frac{{\bm q}^2}{M})$ function in $\bm{q}$. Note that $i\epsilon$ is now redundant since the sum runs over discrete values of momenta. Subsequently, Eq.~(\ref{eq:I-convolution-2}) can be written as
\begin{eqnarray}
    \chi \otimes I^V_0 \otimes \xi &=&
   \chi \otimes I_0 \otimes \xi + \chi \; F_0 \; \xi,
    \label{eq:sum-int-form}
\end{eqnarray}
where the first term, $\chi \otimes I_0 \otimes \xi$, is defined in the same way as the right-hand side of Eq.~(\ref{eq:I-convolution-1}) except for the replacement $\frac{1}{L^3}\sum_{\bm q}\int\,\frac{dq_0}{2\pi} \to \int \frac{d^4q}{(2\pi)^4}$. On the other hand, the second term, $\chi \; F_0 \; \xi$, can be interpreted as the ordinary product of three functions, that are in general matrices in the angular momentum basis. In this basis, MEs of functions should be understood as projection of the original functions to a given partial-wave component with respect to the angular variable defined by $\bm{q}$. Furthermore, the $F_0$ function, defined as the difference between the sum and integral, projects the functions adjacent to it to on-shell values of momentum, corresponding to $|\bm{q}| \to \sqrt{ME} \equiv p$. For the s-wave projection that is a good approximation at low energies, the finite-volume function $F_0$ can be written in a simple form
\begin{align}
    F_0(E) = \frac{1}{L^3}\sum_{\bm{q}}\hspace{-.5cm}\int \; \frac{1}{E-\frac{{\bm q}^2}{M}+i\epsilon} 
    =\frac{M}{4\pi}\left[-4\pi \; c_{00}({\bm p}^2,L) +i p\right],
    \label{eq: F0 expression}
    \end{align}
where the sum-integral difference notation is defined as
\begin{align}
\frac{1}{L^3}\sum_{\bm{q}}\hspace{-.5cm}\int
    &\equiv \frac{1}{L^3}\sum_{\bm{q}  \in \frac{2\pi}{L}\mathbb{Z}^3}-\int\frac{d^3\bm{q}}{(2\pi)^3}.
    \label{eq:sumintegral}
\end{align}
Furthermore, $c_{lm}$ is a function closely related to L\"uscher's $\mathcal{Z}$ function:
\begin{align}
c_{00}({\bm p}^2,L)=\frac{\sqrt{4\pi}}{L^3}\left(\frac{2\pi}{L}\right)^{l-2} \mathcal{Z}_{00}\left[1;(pL/2\pi)^2\right],~\text{with}~~\mathcal{Z}_{00}[s;x^2]=\sum_{\bm n \in \mathbb{Z}^3}\frac{Y_{00}(\widehat{\bm n})}{\left(\bm{n}^2-x^2\right)^s},
\end{align}
where $\bm{n}$ is a Cartesian vector with integer components, and $\widehat{\bm{n}}$ denotes the direction of $\bm{n}$. In the remainder of this paper, the product of functions can be interpreted either as an ordinary product of scalar quantities in the s-wave approximation, or a matrix product of quantities expressed in the angular-momentum basis, in which case the generalized form of $F_0$ should be used, see Refs.~\cite{Luscher:1986pf,Luscher:1990ux,Kim:2005gf}. For the spin-triplet channel, the spin quantum number must be taken into account, giving rise to the finite-volume function derived in Ref.~\cite{Briceno:2013lba}. The s-wave approximation of this function, nonetheless, is the same as that presented in Eq.~(\ref{eq: F0 expression}). Expressions for $F_0$ generalized to moving frames, hadrons with unequal masses, systems with relativistic kinematic, volumes with elongated sides, and general twisted boundary conditions exist, see Refs.~\cite{Briceno:2017max} for a review.\footnote{Despite having used the s-wave approximation of finite-volume relations, the original ordering of the functions is preserved in this paper so that generalization to higher partial waves could be achieved straightforwardly, i.e., by promoting the functions to their matrix form in the angular-momentum basis.}

Now using Eq.~(\ref{eq:sum-int-form}) to expand and rearrange Eq.~(\ref{eq: two point correlation analytic 1}), it is straightforward to see that $C_L(P)$ can be written as
\begin{equation}
    C_L (P) = C_{\infty}(P) + \mathcal{B} \; iF_0 \sum_{n=0}^{\infty} \left[ -\mathcal{M} F_0\right]^n\mathcal{B}^{\dagger}.
    \label{eq: two point correlation analytic 2}
\end{equation}
The first term is obtained by collecting all terms with $I_0$ in Eq.~(\ref{eq: two point correlation analytic 1}) to produces the infinite-volume two-point correlation function. All the functions in the second term are evaluated at the on-shell values of energy, i.e., they are solely a function of only one kinematic variable, $E$. $\mathcal{B}$ and $\mathcal{B}^{\dagger}$ are the end cap functions which are built out of infinite-volume quantities. Explicitly,
\begin{equation}
	\mathcal{B} \,=\, \bar{B} \sum_{n=0}^{\infty} \left[\otimes  \; I_0 \otimes \mathcal{K}\right]^n,
	~~ \mathcal{B}^{\dagger} \,=\, \sum_{n=0}^{\infty} \left[ \mathcal{K} \otimes I_0\; \otimes\right]^n\bar{B}^{\dagger}.
	\label{eq: in-and-out states in inf V}
\end{equation}
$i\mathcal{M}$ is the full on-shell $NN\to NN$ scattering amplitude in the spin-singlet channel, as shown in Fig.~\ref{fig: NN to NN full scattering amplitude}, and is given by
\begin{equation}
    i\mathcal{M} = -i\mathcal{K} \sum_{n=0}^{\infty} \left[\otimes \; I_0 \otimes \mathcal{K}\right]^n.
    \label{eq: NN to NN full amplitude}
\end{equation}
Given that the product of the functions in the second term of Eq.~\eqref{eq: two point correlation analytic 2} is now an ordinary (matrix) product, these terms can be summed to all orders in $F_0$:
\begin{align}
	C_L(P)&= C_{\infty}(P) \,+\, \mathcal{B}(E) \, i\mathcal{F}(E)\, \mathcal{B}^\dagger(E),
	\label{eq: two point correlation analytic 3}
\end{align}
where a new finite-volume function $\mathcal{F}$ is defined for convenience in later discussions:
\begin{align}
	\mathcal{F} & \equiv \frac{1}{F^{-1}_0+\mathcal{M}}.
	\label{eq: definition of mathcal F}
\end{align}
Equation~\eqref{eq: two point correlation analytic 3} has only singularities at interacting energies of the two-nucleon system in the finite volume~\cite{Briceno:2015csa}, $E_n$. These arise from L\"uscher's `quantization condition': 
\begin{equation}
    F^{-1}_0(E)+\mathcal{M}(E)=0,~~\text{for}~~E=E_n.
    \label{eq: Luscher condition}
\end{equation}
For a generic case with no s-wave approximation, the condition reads ${\rm det}\left[F^{-1}_0+\mathcal{M}\right]=0$, where the determinant is taken in the angular-momentum space. In any case, a cutoff is required on the partial waves included, as otherwise the relation is not of practical use. In summary, L\"uscher's quantization condition provides a constraint on the physical elastic amplitude of two nucleons at the finite-volume eigenenergies of the two-nucleon system, which are accessible from LQCD computations of Euclidean two-point correlation functions in a finite volume. The condition is valid up to exponential corrections that go as $e^{-L/R}$, where $R \sim M_\pi^{-1}$ is the range of strong interactions and $M_\pi$ denotes the mass of the pion.

Having arrived at the form in Eq.~\eqref{eq: two point correlation analytic 3} for $C_L(P)$, one can now perform integration over $E$ in Eq.~\eqref{eq: two point correlation(x-y) in C(P)} to obtain
\begin{equation}
    C_L(y_0-x_0,{\bm P}=0)=L^3\sum_{E_n}\;e^{-iE_n(y_0-x_0)}\;\mathcal{B}(E_n) \, \mathcal{R}(E_n)\, \mathcal{B}^\dagger(E_n),
    \label{eq: two point function integrated in p0}
\end{equation}
where the generalized Lellouch-L\"uscher residue matrix is given by:
\begin{equation}
    \mathcal{R}(E_n) \,=\, \lim_{E \to E_n} (E-E_n)\;\mathcal{F}(E).
    \label{eq: definition of residue}
\end{equation}
Note that in order to arrive at this result in Minkowski space, the $i\epsilon$ prescription in the correlation functions must be recovered, i.e., $E \to E+i \epsilon$. Now comparing Eq.~(\ref{eq: two point function integrated in p0}) with Eq.~\eqref{eq: two point correlation: dispersion} for each value of $E_n$, one obtains the following matching condition:\footnote{Note that $\mathcal{B}^\dagger$ ($\mathcal{B}$) represents the matrix element of the interpolating operator between vacuum and an ``in'' (``out'') asymptotic state, hence one should essentially distinguish $\mathcal{B}$ and $\mathcal{B}^\dagger$ in notation. While this distinction is not carried out in the following for presentational brevity, it should be implicit that $\mathcal{B} \neq (\mathcal{B}^\dagger)^\dagger$. In particular, given this distinction, it can be shown that the matching relation in Eq.~(\ref{eq: matching condition for NN to NN}) is real: the Watson phases in $\mathcal{B}$ and $\mathcal{B}^\dagger$ associated with two-nucleon ``in'' and ``out'' states fully cancel the phase of $\mathcal{R}$, leaving the right-hand side real, see Ref.~\cite{Briceno:2015tza} for explicit examples. The left-hand side is manifestly real since such a distinction does not hold for $B$ and $B^\dagger$. In particular, $B = (B^\dagger)^\dagger$.}
\begin{equation}
	L^3\left[\langle 0|\,B(0) \vphantom{B^\dagger} \,| E_n;L\rangle\right]_L \left[\langle E_n;L|\,B^\dagger(0)\, |0\rangle\right]_L = \mathcal{B}(E_n) \, \mathcal{R}(E_n) \,\mathcal{B}^\dagger(E_n).
	\label{eq: matching condition for NN to NN}
\end{equation}
This relation, and its spin-singlet counterpart, will be used in later sections to simplify the matching condition for three- and four-point functions obtained from the same source and sink two-nucleon interpolating operators.

\section{Single-$\beta$ decay process 
\label{sec:single} 
}
\noindent
Low-energy single-weak processes in the two-nucleon sector, including $pp$ fusion: $pp \to de^+\nu_e$, neutrino(antineutrino)-induced disintegration of the deuteron: $\nu(\bar{\nu}) d\to np\nu(\bar{\nu})$ and $\bar{\nu}d \to e^+nn$, and muon capture on the deuteron: $\mu d \to nn \nu_{\mu}$ have been studied in the past within the framework of pionless EFT~\cite{Kong:1999tw, Kong:2000px, Butler:2000zp, Chen:2002pv, Chen:2005ak, Acharya:2019fij}. The Gamow-Teller ME in these processes is dominated by the single-nucleon contribution. At LO, the ME is characterized by the nucleon's axial charge, $g_A$. The two-nucleon contribution, characterized by the $L_{1,A}$ LEC in Eq.~(\ref{eq: axial current : two body}), is only a few percent of the full ME. It, nonetheless, constitutes the dominant source of theoretical uncertainty in the determinations of relevant cross sections~\cite{Adelberger:2010qa}. A precise LQCD determination of the $L_{1,A}$ LEC at the physical values of the quark masses will be a critical goal of the next-generation nuclear LQCD studies in the coming years. The two-nucleon contribution to the Gamow-Teller ME can also be constrained from the half-life of the tritium~\cite{De-Leon:2016wyu, Baroni:2016xll}. With the finite-volume LQCD technology developed in this work, simultaneous fits to single- and double-$\beta$ decay processes in the two-nucleon sector will be possible, which could provide better constraints on this unknown LEC, see also Refs.~\cite{Savage:2016kon,Shanahan:2017bgi,Tiburzi:2017iux}. Alternatively, the constraint on $L_{1,A}$ from single-weak processes could be used to evaluate the significance of the higher-order two-weak two-nucleon contribution in the double-$\beta$ decay, hence testing the validity of the EFT power counting of this work. Although the single-weak process $nn \to np({^3}S_1)$ does not occur in free space, its Gamow-Teller ME in the isospin-symmetric limit is the same as that in all the single-weak processes mentioned above. Since this process constitutes a subprocess of the double-$\beta$ decay of the two-neutron system, the subscript $nn \to np$ is adopted for the amplitudes and correlation functions below, but the formalism is general for any $^3S_1 \to {^1}S_0$ or time-reversed transitions in two-nucleon systems. Since these will be ingredients to the double-$\beta$ decay formalism, single-weak amplitudes and their matching to finite-volume correlation functions will be studied in detail in this section. The derivation of the matching relation is new and follows that of Ref.~\cite{Briceno:2015tza} for general $2(J) \to 2$ processes. The result obtained is in agreement with the earlier results on this problem presented in Refs.~\cite{Detmold:2004qn, Briceno:2012yi}.
 
\subsection{Physical single-$\beta$ decay amplitude
\label{sec:singleIV}}
\begin{figure}
    \centering
    \includegraphics[scale=0.95]{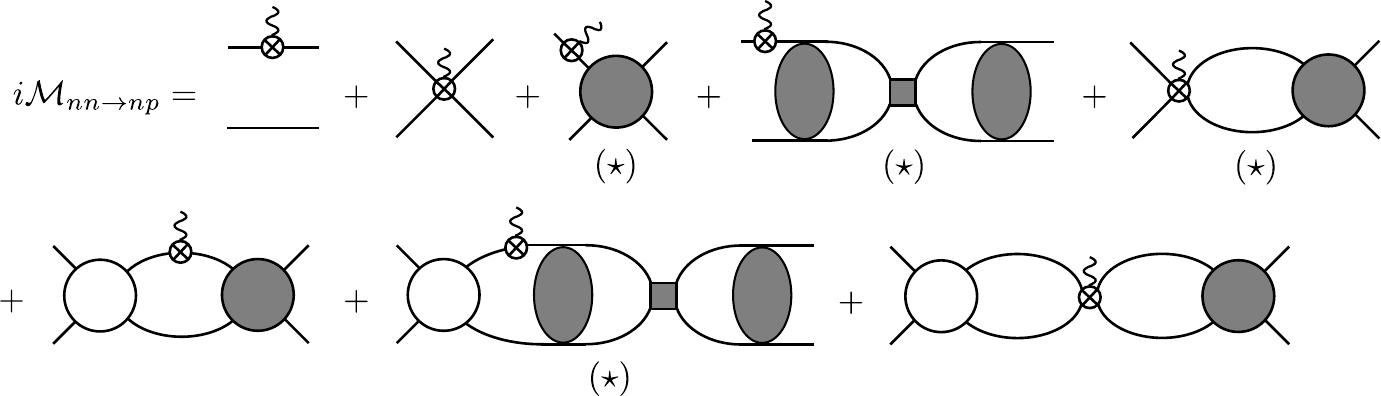}
    \caption{Diagrams contributing to the single-$\beta$ decay hadronic amplitude in Eq.~(\ref{eq: single beta decay full amplitude in inf V}) at NLO in pionless EFT. All building blocks of this amplitude are introduced in Fig.~\ref{fig:notation}. The ($\star$) symbol under a given diagram indicates that a counterpart of the diagram must also be considered in which the diagram is reversed and the replacement ${^1}S_0 \leftrightarrow {^3}S_1$ is applied to all components except the initial and final two-nucleon states. For an on-shell amplitude, the external legs must be evaluated at on-shell kinematics.}
    \label{fig: single beta decay amplitude diagrams}
\end{figure}
Following the formalism presented in Sec.~\ref{sec:formalismIV}, the single-$\beta$ decay amplitude will be constructed in this section at NLO in pionless EFT. The hadronic transition considered in Eq.~(\ref{eq: Hadronic amplitude in terms of T-matrix}) is $X \equiv nn\to np$. For simplicity, we consider the initial two-nucleon state to be at rest with energy $E_i$ and further set the four-momentum carried away by the weak current to $(E,{\bf 0})$, so the two-nucleon system in the final state remains unboosted with energy $E_f$, where $E_f=E_i-E$. The initial neutron-neutron state is a spin-singlet state while the final neutron-proton state is a spin-triplet state with $m=0,\pm1$, where $m$ is the eigenvalue of $\sigma_3$. This state arise from the $(k=)m$th component of the currents in Eqs.~(\ref{eq: axial current : one body}) and (\ref{eq: axial current : two body}). Given the rotational symmetry, transition amplitudes with different $m$ values are equal. Thus, we fix $m=0$ for the final state and omit the $m$ index from the notation.

The hadronic part of the T-matrix for the $nn\to np$ transition, a first-order process in the weak Hamiltonian, is given by
\begin{equation}
	J_{nn\rightarrow np} \equiv \int d^3x \, \mathcal{J}_{nn \to pp}(x) \;=\; \hat{A}_{\rm CC},
	\label{eq: single beta decay T-matrix}
\end{equation}
where $\hat{A}_{\rm CC}$ is given in Eq.~\eqref{eq: weak Hamiltonian HCC}. Substituting this in Eq.~\eqref{eq: Hadronic amplitude in terms of T-matrix} and using Eq.~\eqref{eq: Strong Hamiltonian Eigenstate}, one can decompose the amplitude into different terms depending on the number of $T_S$ insertions. Diagrammatically, the contributions to the hadronic amplitude are shown in Fig.~\ref{fig: single beta decay amplitude diagrams}. This amplitude can be computed up to NLO in strong interactions by inserting a complete set of hadronic states between the operators and using Eqs.~\eqref{eq: Green's function: Free}, \eqref{eq: Weak Hamiltonian MEs I}, \eqref{eq: Weak Hamiltonian MEs II}, \eqref{eq: ME of strong T-matrix I}, and \eqref{eq: ME of strong T-matrix II}. A straightforward calculation gives
\begin{align}
    i\mathcal{M}_{nn\to np} (E_i,E_f) &= -ig_A (2\pi)^3 \delta^3(\bm{p}_i-\bm{p}_f)+ i\mathcal{M}^{\rm DF}_{nn\to np} (E_i,E_f)
    \nonumber\\
    &+ \frac{g_A}{E_i-E_f}\; \left[ i\mathcal{M}^{\rm (LO+NLO)}(E_i)-i\widetilde{\mathcal{M}}^{\rm (LO+NLO)}(E_f)\right],
    \label{eq: single beta decay full amplitude in inf V}
\end{align}
where $\bm{p}_{i(f)} = \sqrt{ME_{i(f)}}\;\widehat{\bm{p}}_{i(f)}$, with the wide hat used to denote the directionality of the three-vector. The amplitude is conveniently split into three terms, the free-neutron decay contribution, a divergence-free (DF) term and a divergent term, where divergence refers to the behavior of the terms in the $E_i\to E_f$ limit. Only the second term, the divergence-free amplitude, is relevant for the finite-volume calculation of this process, as shown in Sec.~\ref{sec:singleV}. It is given by
\begin{align}
    i\mathcal{M}^{\rm DF}_{nn\to np} &= i\widetilde{\mathcal{M}}^{\rm (LO+NLO)}(E_f) \; [\,i\,g_A\,I_1(E_f,E_i)\,] \; i\mathcal{M}^{\rm (LO+NLO)}(E_i)
    \nonumber\\
    & + i\widetilde{\mathcal{M}}^{\rm LO}(E_f)\; [i\widetilde{L}_{1,A}]\; i\mathcal{M}^{\rm LO}(E_i).
    \label{eq: single beta decay DF amplitude}
\end{align}
Here, $I_1$ is a loop introduced in Fig.~\ref{fig:notation}-i with three propagators, and is given by
\begin{equation}
    I_1(E_1,E_2) = \int \frac{d^3{q}}{(2\pi)^3} \frac{1}{E_1-\frac{{\bm q}^2}{M}+i\epsilon}\frac{1}{E_2-\frac{{\bm q}^2}{M}+i\epsilon}=\frac{1}{E_1-E_2}\,\left[I_0(E_2)-I_0(E_1)\right].
    \label{eq: I1 definition}
\end{equation}
Furthermore, in arriving at Eq.~(\ref{eq: single beta decay DF amplitude}), the identity
\begin{align}
     \int \frac{d^3{q}}{(2\pi)^3} \frac{q^2}{(E_1-\frac{{\bm q}^2}{M}+i\epsilon)(E_2-\frac{{\bm q}^2}{M}+i\epsilon)}
     &=M\left[-I_0(E_1)+E_2I_1(E_1,E_2)\right]
     \nonumber\\
     &=M\left[-I_0(E_2)+E_1I_1(E_1,E_2)\right]
    \label{eq: I1 identity}
\end{align}
is used. The $\widetilde{L}_{1,A}$ is a renormalization-scale independent combination of strong and weak couplings,\footnote{$\widetilde{L}_{1,A}$ is the only exception to the rule in this paper that all quantities with a tilde over them correspond to the spin-triplet channel. The convention for this quantity is maintained to be compatible with the literature.}
\begin{equation}
    \widetilde{L}_{1,A} =
    \frac{L_{1,A}}{C_0\,\widetilde{C}_0}-\frac{g_A\,M}{2}\frac{(C_2+\widetilde{C}_2)}{C_0\,\widetilde{C}_0}.
    \label{eq: L1A scale independent}
\end{equation}
This is because, first of all, from Eqs.~\eqref{eq: I1 definition} and \eqref{eq: I0 in infinite volume}, $I_1(E_i,E_f)$ can be seen to be $\mu$ independent. Second, divergent terms in Eq.~\eqref{eq: single beta decay full amplitude in inf V} are also renormalization-scale independent, as the $g_A$ coupling and the $NN \to NN$ elastic scattering amplitudes are scale independent. Thus, for the physical amplitude in Eq.~\eqref{eq: single beta decay full amplitude in inf V} to be independent of the renormalization scale, $\widetilde{L}_{1,A}$ should satisfy
\begin{equation}
    \mu\frac{d}{d\mu} \widetilde{L}_{1,A} =0.
    \label{eq: L1A tilde mu derivative}
\end{equation}
This, along with Eq.~\eqref{eq: C0, C2 scale dependence} and its spin-triplet counterpart, determines the $\mu$ dependence of $L_{1,A}$. This result was first obtained in Refs.~\cite{Kong:2000px, Butler:2000zp}.
\subsection{Finite-volume correlation function
\label{sec:singleV}}
\begin{figure}
    \centering
    \includegraphics[scale=1]{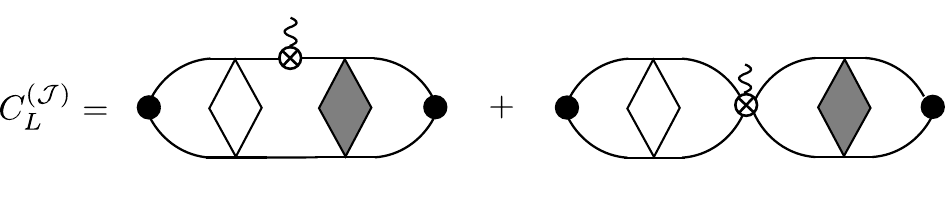}
    \caption{Diagrammatic representation of the finite-volume correlation function with a single insertion of the weak current corresponding to the expansion in Eq.~(\ref{eq:CLJ}). The black dots are the interpolating operators for the two-nucleon channels in spin-singlet or triplet states. All other components are introduced in Fig.~\ref{fig:notation}. The loops are evaluated as a sum over discretized momenta, as discussed in the text.}
    \label{fig: singleV finite diagrams}
\end{figure}

The first step in establishing a matching relation for the infinite-volume single-$\beta$ decay amplitude calculated in Sec.~\ref{sec:singleIV} is to construct the momentum-space three-point function in a finite volume, $C_L^{(\mathcal{J})}(P_i,P_f)$. $P_{i(f)}$ denotes the four-momentum of the two-nucleon state in the spin-singlet (triplet) channel at rest with $P_{i(f)}=(E_{i(f)},\bm{P}_{i(f)}=\bm{0})$. This correlation function can be expanded diagrammatically as shown in Fig.~\ref{fig: singleV finite diagrams}. The result can be expressed as
\begin{equation}
    C_L^{(\mathcal{J})}(P_i,P_f) = \widetilde{\bar{B}}(P_f) \sum_{n=0}^{\infty} \big[\otimes \; I^V_0 \otimes \widetilde{\mathcal{K}} \big]^n \otimes \left[ig_A I^V_1+iL_{1,A}I^V_0\otimes I^V_0\right] \otimes \sum_{n=0}^{\infty} \left[ \mathcal{K} \otimes I^V_0 \otimes \right]^n \bar{B}^{\dagger}(P_i).
    \label{eq:CLJ}
\end{equation}
Here, $\bar{B}^{\dagger}(\widetilde{\bar{B}}^{\dagger})$ and $\bar{B}(\widetilde{\bar{B}})$ are two-nucleon source and sink interpolating operators for the spin-singlet (triplet) channel, respectively, all projected to zero spatial momentum. $I_0^V$, $\mathcal{K}\;,\widetilde{\mathcal{K}}$, as well as $\otimes$ product notation, are defined in Sec.~\ref{sec:formalismV}. $I_1^V$ is the finite-volume counterpart of the $I_1$ loop defined in Sec.~\ref{sec:singleIV}, i.e., the replacement $\frac{d^3q}{(2\pi)^3} \to\frac{1}{L^3}\sum_{\bm q} $ must be applied. It is important to expand the correlation function in Eq.~(\ref{eq:CLJ}) in fixed orders in the EFT to ensure the renormalization-scale independence of physical observables extracted from the correlation function, including the energy eigenvalues and the finite-volume MEs. For convenience, this simple form will be carried out for now but a pionless EFT expansion at NLO will be performed shortly.

The goal is to separate the infinite-volume correlation function and to re-arrange purely finite-volume contributions in a manner similar to that outlined in Sec.~\ref{sec:formalismV}. First, as was shown in Eq.~(\ref{eq:sum-int-form}), the sum convolution of functions adjacent to $I_0^V$ can be separated into two contributions, the infinite-volume piece and a remnant finite-volume piece in which $\otimes$ products turn into ordinary (matrix) products of functions, and where the $F_0$ function puts any adjacent functions on shell. The sum convolution of functions adjacent to $I_1^V$ for two generic left and right functions $\chi$ and $\xi$, defined as
\begin{equation}
    \chi \otimes I^V_1 \otimes \xi \equiv \frac{1}{L^3}\sum_{\bm{q}}
    \chi(\bm q)\frac{1}{E_i-\frac{{\bm q}^2}{M}}
    \frac{1}{E_f-\frac{{\bm q}^2}{M}}\xi(\bm{q}),
    \label{eq: I1 V LR definition}
\end{equation}
can proceed similarly, except now the left and right sides of $I_1^V$ have, in general, different on-shell kinematics. The general case of the convolution of functions with arbitrary momentum dependence adjacent to a (relativistic) $I_1^V$ is worked out in Ref.~\cite{Briceno:2015tza}. Here, the situation is simpler, as at NLO in the pionless-EFT expansion of the three-point function, there are only two relevant scenarios to consider. First, one encounters two contact (momentum-independent) kernels or current on both sides of the $I_1^V$ loop, in which case the convolution becomes trivially the ordinary product of (on-shell) functions. The function $I_1^V$ in such ordinary products is that defined in Eq.~(\ref{eq: I1 definition}) upon replacements $\int \frac{d^3{q}}{(2\pi)^3} \to \frac{1}{L^3} \sum_{\bm q}$, $I_1 \to I_1^V$, and $I_0 \to I_0^V$,\footnote{Note that we have overloaded the $I_1^V$ symbol, used both as a sum with an arbitrary summand (made of left and right functions) times the corresponding propagators, and as the sum evaluated with just the propagators. When $I_1^V$ is not convoluted with adjacent functions with the $\otimes$ product sign, it is meant to be a function on its own as defined in this paragraph.} with $I_0^V=I_0+F_0$, where $I_0$ and $F_0$ are defined in Eqs.~(\ref{eq: I0 in infinite volume}) and (\ref{eq: F0 expression}), respectively. It is evident that $ I^V_1$ is UV convergent and is hence scale independent. Note also that this function is regular in the limit $E_i \to E_f$. The other possibility for the convolution of $I_1^V$ with two adjacent functions at NLO is when there is one momentum-dependent kernel $C_2 q^2$ (or $\widetilde{C}_2 q^2$) on either side of $I_1^V$ (with $\bm{q}$ being the summed momentum), in which case the application of the finite-volume counterpart of the identity in Eq.~(\ref{eq: I1 identity}) leads to ordinary product of on-shell functions.

With these considerations, the expansion in Eq.~(\ref{eq:CLJ}) can be organized as follows. First, collecting all terms with only infinite-volume loop contributions isolates the infinite-volume three-point function. Second, one can collect all terms that contain at least one $F_0$ in the $\sum_{n=0}^{\infty} \left[\otimes \; \widetilde{\mathcal{K}} \otimes I^V_0 \right]^n$ and $\sum_{n=0}^{\infty} \left[ \mathcal{K} \otimes I^V_0 \otimes \right]^n$ factors. These can then be expanded in a geometric sum to give $\widetilde{\mathcal{F}}(E_f)$ and $\mathcal{F}(E_i)$, respectively, as outlined in Sec.~\ref{sec:formalismV}. Now expanding only up to NLO in the pionless EFT as in Sec.~\ref{sec:singleIV}, Eq.~(\ref{eq:CLJ}) can be shown to reduce to
\begin{equation}
   C_L^{(\mathcal{J})}(P_i,P_f) = C_{\infty}^{(\mathcal{J})}(P_i,P_f) + 
   \widetilde{\mathcal{B}}(E_f)\,i\widetilde{\mathcal{F}}(E_f)
   \left[i\mathcal{M}^{\rm{DF},V}_{nn\to np} (E_i,E_f)\right]
   i\mathcal{F}(E_i)\,\mathcal{B}^\dagger(E_i)+\cdots.
   \label{eq: three-point correlation final}
\end{equation}
Here, the ellipsis denotes all other terms in Eq.~(\ref{eq:CLJ}) that contribute neither to $C_{\infty}^{(\mathcal{J})}$ nor to the finite-volume terms in which factors of $\widetilde{\mathcal{F}}$ and $\mathcal{F}$ are both present. The reason for such a rearrangement of the terms in the finite-volume correlation function will become evident shortly. $\mathcal{M}^{\rm{DF},V}_{nn\to np} $ is the finite-volume counterpart of the single-$\beta$ decay divergence-free amplitude $\mathcal{M}^{\rm DF}_{nn\to np} $ defined in Eq.~(\ref{eq: single beta decay DF amplitude}), in which $I_1$ is replaced with $I_1^V$. Explicitly,
\begin{eqnarray}
   i \mathcal{M}^{\rm{DF},V}_{nn\to np} (E_i,E_f)=
   i\mathcal{M}^{\rm DF}_{nn\to np} (E_i,E_f)+
   i\widetilde{\mathcal{M}}^{\rm (LO+NLO)}(E_f) \; \left[ig_A\,F_1(E_f,E_i)\right] \; i\mathcal{M}^{\rm (LO+NLO)}(E_i).
   \nonumber\\
   \label{eq: DF amplitude: finite V}
\end{eqnarray}
Here, $F_1=I_1^V-I_1$, or in terms of the sum-integral notation defined in Eq.~(\ref{eq:sumintegral}),
\begin{equation}
    F_1(E_f,E_i) = \frac{1}{L^3}\sum_{\bm{q}}\hspace{-.5cm}\int
    \frac{1}{E_i-\frac{{\bm q}^2}{M}+i\epsilon}
    \frac{1}{E_f-\frac{{\bm q}^2}{M}+i\epsilon}
    =\frac{1}{E_i-E_f}\,\left[F_0(E_f)-F_0(E_i)\right],
    \label{eq:F1def}
\end{equation}
where $F_0$ is defined in Eq.~(\ref{eq: F0 expression}). Finally, it should be noted that all terms in Eq.~(\ref{eq: three-point correlation final}) are evaluated at the on-shell kinematics, given their proximity to $F_0$ functions, and given the convolution rules with the $I_1^V$ loop explained above. Equation~(\ref{eq: three-point correlation final}) will be used in the next subsection to derive the matching relation between the finite-volume ME and the physical hadronic amplitude for the single-$\beta$ decay process.

\subsection{The matching relation
\label{sec:singleVIV}}
To obtain the infinite-volume hadronic transition amplitude for the $nn\to np$ process from the corresponding ME in a finite volume, once again one could inspect the finite-volume correlation function. In momentum space, the correlation function is given by:
\begin{equation}
    C_L^{(\mathcal{J})} (P_i,P_f) =
    \int_Ld^3x\,d^3y \int dx_0\,dy_0
    \;e^{-iP_i\cdot x}\;e^{iP_f\cdot y}
    \left[\langle 0|\, T[\widetilde{B}(y)\,\mathcal{J}_{nn \to np}(0)\, B^\dagger(x)]\,|0\rangle \right]_L,
    \label{eq: three-point correlation (Pi,Pf)}
\end{equation}
while the momentum-time representation of this correlation function can be written in different ways:
\begin{eqnarray}
    C_L^{(\mathcal{J})}(y_0-z_0,z_0-x_0)
    &\equiv& L^3 \int\, \frac{dE_i}{2\pi}\, \frac{dE_f}{2\pi} \; e^{-iE_i(z_0-x_0)} \; e^{-iE_f(y_0-z_0)}\;C_L
   ^{(\mathcal{J})}(P_i,P_f)
    \label{eq: three point correlation in diagrammatic definition}
    \\
    & = & \int_L d^3x\; d^3y\; d^3z\;
    \left[\langle 0|\, T[\widetilde{B}(y)\,\mathcal{J}_{nn\rightarrow np}(z)\,B^\dagger(x)]\, |0\rangle\right]_L
    \\
    &= &L^9\,\sum_{E_{n_i},E_{n_f}} e^{-iE_{n_i}(z_0-x_0)}\,e^{-iE_{n_f}(y_0-z_0)} \left[\langle 0|\,\widetilde{B}(0)\,| E_{n_f},L\rangle\right]_L
    \nonumber\\
    && \hspace{1.85cm} \left[\vphantom{B^\dagger} \langle E_{n_f},L|\,\mathcal{J}_{nn\rightarrow np}(0)\,|E_{n_i},L\rangle\right]_L \left[\langle E_{n_i},L|\,B^\dagger(0)\, |0\rangle\right]_L.
    \label{eq: three-point correlation dispersive expression}
\end{eqnarray}
In the last equality $y_0>z_0>x_0$ is assumed, complete sets of finite-volume states are inserted between the operators, and the operators in the Heisenberg picture are expressed in the Schr\"odinger picture. Now Eq.~(\ref{eq: three-point correlation final}) can be used in Eq.~(\ref{eq: three point correlation in diagrammatic definition}) to perform the energy integrations. This gives
\begin{align}
    &C_L^{(\mathcal{J})}(y_0-z_0,z_0-x_0)
    =L^3\,\sum_{E_{n_i},E_{n_f}}
    e^{-iE_{n_i}(z_0-x_0)}\,e^{-iE_{n_f}(y_0-z_0)}
    \nonumber\\
    &\hspace{6cm} \times 
    \widetilde{\mathcal{B}}(E_{n_f})\widetilde{\mathcal{R}}(E_{n_f})
    \left[i\mathcal{M}^{\rm{DF},V}_{nn\to np} (E_{n_i},E_{n_f})\right]
    \mathcal{R}(E_{n_i}) \mathcal{B}(E_{n_i}).
    \label{eq: three point correlation integrated in pi pf}
\end{align}
Here, the only contributions to the energy Fourier integrals arise from the poles of the $\widetilde{\mathcal{F}}$ and $\mathcal{F}$ functions, $E_{n_f}$ and $E_{n_i}$, which are the finite-volume energy eigenvalues of the spin-triplet and spin-singlet two-nucleon systems at rest, respectively, see Eq.~(\ref{eq: Luscher condition}). $\mathcal{R}$  ($\widetilde{\mathcal{R}}$) is the residue of  $\mathcal{F}$ ($\widetilde{\mathcal{F}}$) at the corresponding finite-volume energies, as defined in Eq.~(\ref{eq: definition of residue}). This step makes it clear why only the contributions with both factors of $\widetilde{\mathcal{F}}$ and $\mathcal{F}$ were collected in the correlation function explicitly, as the inverse Fourier integrals of the remaining contributions vanish. For a comprehensive proof of the absence of any poles beside the finite-volume two-nucleon energies in the finite-volume correlation function, see Ref.~\cite{Briceno:2015tza}.

Having arrived at Eq.~\eqref{eq: three point correlation integrated in pi pf}, it can be compared with its equivalent form in Eq.~\eqref{eq: three-point correlation dispersive expression}. For each $E_{n_i}$ and $E_{n_f}$, one obtains
\begin{align}
    &L^6\left[\langle 0|\,\widetilde{B}(0)\,| E_{n_f},L\rangle\right]_L \left[\vphantom{B^\dagger}\langle E_{n_f},L|\,\mathcal{J}_{nn\rightarrow np}(0)\,|E_{n_i},L\rangle\right]_L \left[\langle E_{n_i},L|\,B^\dagger(0)\, |0\rangle\right]_L
    \nonumber\\[10pt]
    &\hspace{4.5 cm} =\widetilde{\mathcal{B}}(E_{n_f})\widetilde{\mathcal{R}}(E_{n_f})
    \left[i\mathcal{M}^{\rm{DF},V}_{nn\to np} (E_{n_i},E_{n_f})\right]
    \mathcal{R}(E_{n_i}) \,\mathcal{B}^\dagger(E_{n_i}).
    \label{eq: matching relation for single 1}
\end{align}
Multiplying this equation by its complex conjugate and using Eq.~\eqref{eq: matching condition for NN to NN}, one finally arrives at the relation that connects the hadronic  ME in a finite volume with the divergence-free part of the physical amplitude:
\begin{align}
    L^6&\left|\left[\vphantom{B^\dagger}\langle E_{n_i},L|\,\mathcal{J}_{nn\rightarrow np}({0})\,|E_{n_f},L\rangle \right]_L\right|^2 = \left|\widetilde{\mathcal{R}}(E_{n_f})\right| \left|\vphantom{\widetilde{\mathcal{R}}(E_{n_f})}\mathcal{M}^{\rm{DF},V}_{nn\to np} (E_{n_i},E_{n_f})\right|^2  \left|\vphantom{\widetilde{\mathcal{R}}(E_{n_f})}\mathcal{R}(E_{n_i})\right|.
    \label{eq: matching relation for single 2}
\end{align}
This equation is the main result of this section. Note that the divergence-free part of the physical hadronic amplitude is embedded in $\mathcal{M}^{\rm{DF},V}_{nn\to np}$, see Eq.~(\ref{eq: DF amplitude: finite V}). To construct the full hadronic transition amplitude, the divergent parts must be added to $\mathcal{M}^{\rm{DF}}_{nn\to np}$. The divergent amplitudes, nonetheless, are composed of contributions that can be determined from the nucleon single-$\beta$ decay amplitude ($g_A$ coupling), as well as the strong-interaction two-nucleon scattering amplitudes. This relation, therefore, provides a means to constrain the $L_{1,A}$ LEC from a LQCD calculation of the finite-volume ME of the axial-vector current in the two-nucleon system, see Ref.~\cite{Savage:2016kon} for a first calculation along this direction.

\section{Double-$\beta$ decay process 
\label{sec:double} 
}
\noindent
The analysis of the single-$\beta$ decay in infinite and finite volumes can now be extended to the case of the double-$\beta$ decay. While the overall strategy is the same as that outlined in Sec.~\ref{sec:single}, the  double-$\beta$ decay problem presents new features originating from the presence of two spacetime-displaced axial-vector currents in the MEs. This, first of all, makes distinct contributions arising from different time ordering of the currents for general kinematics. Second, and more significantly, it necessitates further steps to be taken to connect the Minkowski finite-volume ME to that calculated in Euclidean spacetime, which is the case with LQCD computations. The first two subsections follow the procedure of Sec.~\ref{sec:single}, while the last subsection addresses this latter point. It is found that no new contact two-nucleon two-current LEC is needed to renormalize the physical hadronic amplitude at NLO, therefore the only two-nucleon axial-current LEC that is to be constrained from the matching relation is $L_{1,A}$. LQCD computations of this ME will be able to test the validity of this power counting.

\subsection{Physical double-$\beta$ decay amplitude
\label{sec:doubleIV}}
\begin{figure}
    \centering
    \includegraphics[scale=0.74]{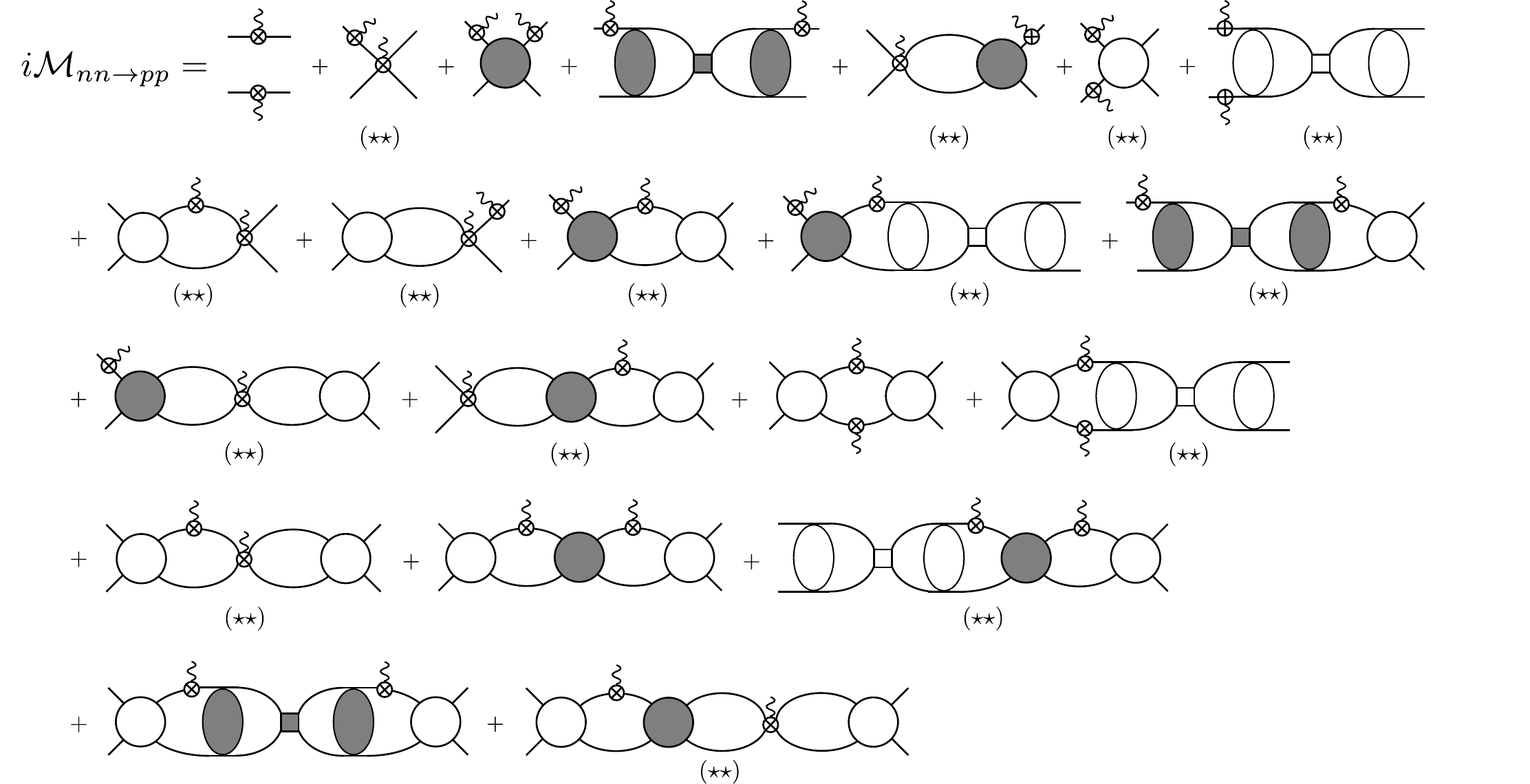}
    \caption{Diagrammatic representation of the hadronic transition amplitude contributing to the $nn \to pp$ process. All components of the diagrams are introduced in Fig.~\ref{fig:notation}, where the external legs are evaluated at the on-shell kinematics. The ($\star\star$) symbol under a given diagram indicates that the reversed diagrams must be included as well (without changing the initial and final states). An exact set of diagrams must be computed but with the first and the second currents reversed in order, which is equivalent to setting $E_1 \leftrightarrow E_2$ in these diagrams, where $E_1$ and $E_2$ are the energies carried out by each of the two currents, see the discussions in the text.}
    \label{fig: double weak infinite V amplitude diagrams}
\end{figure}
In this section, the hadronic amplitude for the double-$\beta$ decay will be calculated. This is the amplitude given in Eq.~(\ref{eq: Hadronic amplitude in terms of T-matrix}) for  $X=nn\to pp$. The two currents are the same for this transition, and in the spin-isospin symmetric limit, all three spin components of the intermediate spin-triplet state contribute equally. In the following, only the amplitude for transitions with the $k=3$ component of the currents in Eq.~(\ref{eq: CC hamiltonian, one-body and two-body}) is considered, and in the full amplitude this contribution must be multiplied by a factor of 3. Once again, the currents are assumed to carry out no spatial momenta and their kinematics is given by $Q_1=(E_1,{\bm 0 })$ and $Q_2=(E_2,{\bm 0 })$. The initial and final two-nucleon states are at rest, which constrains the intermediate states between the current to have zero spatial momentum. The hadronic contribution to the T-matrix is given by:
\begin{align}
	J_{nn\rightarrow pp} \equiv \int d^3x \, \mathcal{J}_
	{nn  \to pp}(x) = \hat{A}_{\rm CC}\,G_S\,\hat{A}_{\rm CC}.
	\label{eq: double beta T-matrix}
\end{align}
Here, $\hat{A}_{\rm CC}$ is the hadronic contribution to the weak Hamiltonian defined in Eq.~(\ref{eq: weak Hamiltonian HCC}), and $G_S$ is the strong retarded Green's function defined in Eq.~\eqref{eq: Green's function: Strong}. There will be two possibilities for the intermediate states, one with total energy $E_{*1}=E_i-E_1$ and one with total energy $E_{*2}=E_i-E_2$. These should be considered separately in the time-independent Lippmann-Schwinger formalism that is adopted for computing the amplitudes. The two contributions correspond to different time orderings of the currents in a time-dependent quantum field theory approach that is used in the definition of the finite-volume correlation functions.

$J_{nn \to pp}$ can now be used in Eq.~\eqref{eq: Hadronic amplitude in terms of T-matrix} to obtain the hadronic amplitude up to NLO in pionless EFT. The diagrammatic expansion of this amplitude is shown in Fig.~\ref{fig: double weak infinite V amplitude diagrams}. Following the method described in Sec.~\ref{sec:formalismIV}, one obtains
\begin{align}
    i\mathcal{M} _{nn\to pp} (E_i,E_1,E_f)&= i\mathcal{M}^{\rm DF}_{nn\to pp}(E_i,E_1,E_f) +
    i\,\frac{g_A^2}{4} (2\pi)^3 \delta^3(\bm{p}_i-\bm{p}_f)\left[ \frac{1}{E_1}+\frac{1}{E_2} \right]+\frac{g_A^2}{2(E_1+E_2)}
    \nonumber\\[10 pt]
    &\hspace{-1.35 cm} \left[\frac{i\mathcal{M}^{\rm (LO+NLO)}(E_i)-i\widetilde{\mathcal{M}}^{\rm (LO+NLO)}(E_{*1})}{E_2}-\frac{i\widetilde{\mathcal{M}}^{\rm (LO+NLO)}(E_{*1})-i\mathcal{M}^{\rm (LO+NLO)}(E_f)}{E_1}\right] 
    \nonumber\\
    & \qquad~~ + \frac{g_A}{2} \left[\frac{i\mathcal{M}^{\rm DF}_{nn\to np} (E_i,E_{*1})}{E_2}- \frac{i\mathcal{M}^{\rm DF}_{np\to pp}(E_{*1},E_f)}{E_1} \right]+ (E_1 \leftrightarrow E_2).
    \label{eq: double amplitude full IV}
\end{align}
$(E_1 \leftrightarrow E_2)$ indicates that the counterpart of every term where $E_1$ is exchanged with $E_2$ must be included as well. Similar to Eq.~\eqref{eq: single beta decay full amplitude in inf V}, terms in Eq.~\eqref{eq: double amplitude full IV} have been grouped into a divergence-free $nn\to pp$ amplitude, as well as divergent amplitudes which are singular in the limit $E_1,E_2\to0$. The divergent part contains the divergence-free single-weak transition amplitude, which is given by Eq.~\eqref{eq: single beta decay DF amplitude},\footnote{Note that $\mathcal{M}^{\rm DF}_{nn\to np}$ and $\mathcal{M}^{\rm DF}_{np\to pp}$ are the same in the isospin-symmetric limit.} as well as $NN\to NN$ elastic scattering amplitudes for both spin-singlet and spin-triplet channels. The divergence-free part of the $nn\to pp$ amplitude, $\mathcal{M}^{\rm DF}_{nn\to pp}$, is given by
\begin{align}
    i\mathcal{M}^{\rm DF}_{nn\to pp}  (E_i,E_1,E_f) &=
    \frac{g_A}{2}\,i\mathcal{M}^{\rm (LO+NLO)}(E_f)
    \bigg[g_A\,i\,I_1(E_f,E_{*1})\,i\widetilde{\mathcal{M}}^{\rm (LO+NLO)}(E_{*1})
    \,i\,I_1(E_i,E_{*1})\,+
    \nonumber\\
    &\hspace{-3 cm} i\,\widetilde{L}_{1,A}\,i\widetilde{\mathcal{M}}^{\rm (LO+NLO)}(E_{*1}) \left[i\,I_1(E_i,E_{*1})+i\,I_1(E_{*1},E_f)\right]
   +i\,g_A\,I_2(E_i,E_{*1},E_f) \bigg] i\mathcal{M}^{\rm (LO+NLO)}(E_i),
    \label{eq: double amplitude DF IV}
\end{align}
where $I_1$ is given by Eq.~\eqref{eq: I1 definition} and the new type of loop arising from four propagators and two weak currents, as shown in Fig.~\ref{fig:notation}-j, is given by
\begin{align}
    I_2(E_i,E_{*1},E_f) &= \int \frac{d^3{q}}{(2\pi)^3}
    \frac{1}{E_i-\frac{{\bm q}^2}{M}+i\epsilon}
    \frac{1}{E_{*1}-\frac{{\bm q}^2}{M}+i\epsilon}
    \frac{1}{E_f-\frac{{\bm q}^2}{M}+i\epsilon}
    \nonumber\\
    &=\frac{1}{E_i-E_f}\,[I_1(E_{*1},E_f)-I_1(E_i,E_{*1})],
    \label{eq: I2 definition}
\end{align}
and a useful identity is used to simplify the expressions:
\begin{align}
    \int \frac{d^3{q}}{(2\pi)^3}
    \frac{1}{E_i-\frac{{\bm q}^2}{M}+i\epsilon}
    \frac{\bm{q}^2}{E_{*1}-\frac{{\bm q}^2}{M}+i\epsilon}
    \frac{1}{E_f-\frac{{\bm q}^2}{M}+i\epsilon} 
    =-MI_1(E_{*1},E_f)+ME_iI_2(E_i,E_{*1},E_f).
    \label{eq: I2q2identity}
\end{align}
As the only new ingredient of the double-$\beta$ decay amplitude compared with the single-$\beta$ decay is $\mathcal{M}^{\rm DF}_{nn\to pp}$, it is this quantity that is aimed to be constrained from a finite-volume matching relation in the next subsections. At the order considered in the EFT, the physical hadronic amplitude in Eq.~(\ref{eq: double amplitude full IV}) is evidently renormalization-scale independent (note that the $I_2$ loop is UV convergent), and as a result, no new LEC beyond those present in the single-$\beta$ decay amplitude is needed to renormalize the amplitude, as stressed before. A contact two-weak two-nucleon operator, in particular, should appear at next-to-NLO using a naive-dimensional analysis, and no new UV divergence at the previous order is observed to require its promotion to a lower order.

\subsection{Finite-volume correlation function
\label{sec:doubleV}}
\begin{figure}
    \centering
    \includegraphics[scale=0.85]{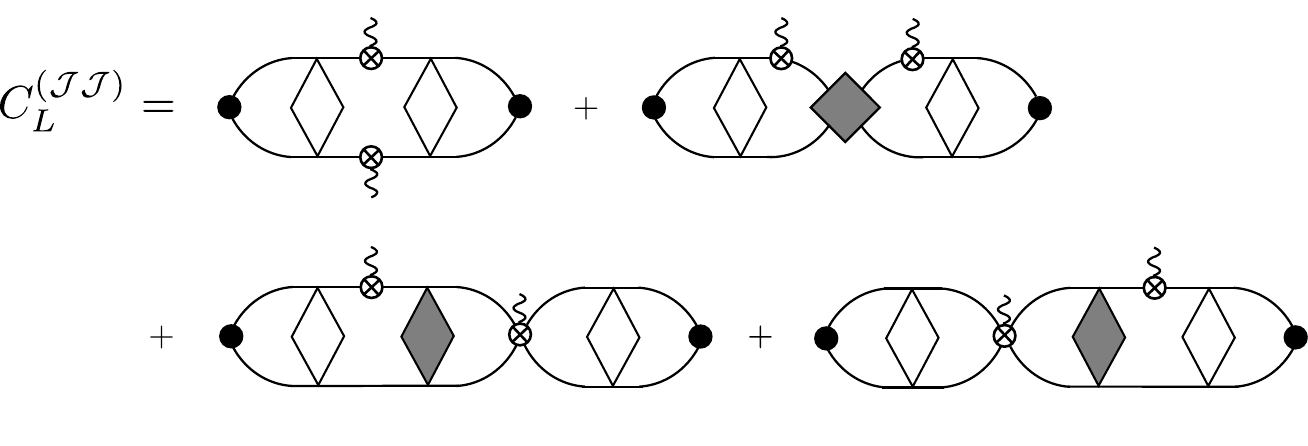}
    \caption{Diagrammatic representation of the finite-volume correlation function with two insertions of the weak current corresponding to the expansion in Eq.~(\ref{eq:CLJJ}). The black dots are the interpolating operators for the spin-singlet two-nucleon states. All other components are introduced in Fig.~\ref{fig:notation}. The loops are evaluated as a sum over discretized momenta, as discussed in the texts. An exact set of diagrams must be computed but with the first and the second currents reversed in order, which is equivalent to setting $E_1 \leftrightarrow E_2$ in these diagrams, where $E_1$ and $E_2$ are the energies carried out by each of the two currents, see the discussions in the text.}
    \label{fig: double weak finite V correlation}
\end{figure}
The relevant finite-volume correlation function for the double-$\beta$ decay process in momentum space can be represented by diagrams shown in Fig.~\ref{fig: double weak finite V correlation}, where an expansion at NLO in the pionless EFT can be performed later. These diagrams correspond to the following expansion
\begin{align}
    C_L^{(\mathcal{J}\mathcal{J})}(P_i,P_f) &= \bar{B}_{pp}(P_f) \sum_{n=0}^{\infty} \left[\otimes \, I^V_0 \otimes \mathcal{K} \right]^n \otimes
    \bigg\{\frac{g_A^2}{2}\,iI_2^V+i\frac{g_A^2}{2}\,I^V_1 \otimes \widetilde{\mathcal{K}} \sum_{n=0}^{\infty} \big[\otimes \, I^V_0 \otimes \widetilde{\mathcal{K}} \big]^n \otimes I_1^V
    \nonumber\\
    &+i\frac{g_A}{2}\,L_{1,A} I^V_1 \sum_{n=0}^{\infty} \big[\otimes \, \widetilde{\mathcal{K}} \otimes I^V_0 \big]^n \otimes I_0^V
    +i\frac{g_A}{2}\,L_{1,A} \,I_0^V \otimes \sum_{n=0}^{\infty} \big[ I^V_0 \otimes \widetilde{\mathcal{K}} \otimes \big]^n  I^V_1\bigg\}
    \nonumber\\
    &\hspace{7.25 cm}\otimes \sum_{n=0}^{\infty} \left[ \mathcal{K} \otimes I^V_0 \otimes \right]^n \bar{B}^{\dagger}_{nn}(P_i)+\cdots.
    \label{eq:CLJJ}
\end{align}
Here, the kinematic dependence of the functions is suppressed, except for the dependence of the interpolating functions on total initial and final momenta. The ellipsis indicates that a similar expansion with $E_1 \leftrightarrow E_2$ must be included. All ingredients in Eq.~(\ref{eq:CLJJ}) are introduced in the previous sections, except for $\otimes I_2^V \otimes$, where $I_2^V$ is the finite-volume counterpart of the loop function with two insertions of the single-body weak current. For generic left and right functions $\chi$ and $\xi$, the $\otimes$-product sign is defined as
\begin{equation}
    \chi \otimes I^V_2 \otimes \xi \equiv  \frac{1}{L^3}\sum_{\bm{q}}
    \chi(\bm q)\frac{1}{E_i-\frac{{\bm q}^2}{M}}
    \frac{1}{E_{*1}-\frac{{\bm q}^2}{M}}
    \frac{1}{E_f-\frac{{\bm q}^2}{M}}\xi(\bm{q}),
    \label{eq: I2 V definition II}
\end{equation}
and similarly for $E_{*2}$. For the analysis of this work at NLO in the EFT, the left and right convoluting functions are either momentum independent kernels/current, in which case the $\otimes$ sign trivially becomes an ordinary product, or one of the convoluting functions is proportional to $\bm{q}^2$ and one is constant, in which case the finite-volume counterpart of the identity in Eq.~(\ref{eq: I2q2identity}) (in which $\int \frac{d^3{q}}{(2\pi)^3} \to \frac{1}{L^3} \sum_{\bm q}$, $I_1 \to I_1^V$, and $I_2 \to I_2^V$) can be used to return to the ordinary product of functions. When appearing in an ordinary product, $I_2^V$ is simply given by Eq.~(\ref{eq: I2 definition}) upon replacements just mentioned. Note that this function is regular in the limit $E_i \to E_f$. 

The expansion in Eq.~(\ref{eq:CLJJ}) can now be turned into a useful form, in a similar fashion to Sec.~\ref{sec:singleV}. The idea is to first isolate the infinite-volume contribution, which arises from separating the loop functions to an infinite-volume integral and a sum-integral difference, and collecting all terms with exclusively infinite-volume loops. Next, given the identity in Eq.~(\ref{eq:sum-int-form}), factors of $\mathcal{F}(E_f)$ and $\mathcal{F}(E_i)$ can be formed out of $\sum_{n=0}^{\infty} \left[\otimes \, \mathcal{K} \otimes I^V_0 \right]^n$ adjacent to $B$ and $B^\dagger$, up to additional contributions that do not concern us. The $\mathcal{F}$ functions enforce on-shell kinematics on the adjacent function, which combined with identities on the $\otimes I_1^V \otimes$ and $\otimes I_2^V \otimes$ as described before, ensure that all functions in such a term are evaluated on-shell. These features, along with straightforward algebra, lead to 
\begin{align}
   &C_L^{(\mathcal{J}\mathcal{J})}(P_i,Q_1,P_f) = C_{\infty}^{(\mathcal{J}\mathcal{J})}(P_i,Q_1,P_f) +
   \mathcal{B}_
   {pp}(E_f)\,i\mathcal{F}(E_f)\bigg[
   i\mathcal{M}^{\rm{DF},V}_{nn\to pp} (E_i,E_1,E_f)
   \nonumber\\
   &\hspace{0.6 cm}+\frac{1}{2}\, i\mathcal{M}^{\rm{DF},V}_{np\to pp}(E_{*1},E_f)\,
   i\widetilde{\mathcal{F}}(E_{*1})\,
   i\mathcal{M}^{\rm{DF},V}_{nn\to np}(E_i,E_{*1}) + (E_1 \leftrightarrow E_2)\bigg]
   i\mathcal{F}(E_i)\,\mathcal{B}^\dagger_{nn}(E_i)+\cdots,
   \label{eq: four point correlation final}
\end{align}
where now the ellipsis denotes any other terms left out while expanding Eq.~(\ref{eq:CLJJ}) in terms of infinite- and finite-volume contributions. The reason behind this peculiar rearrangement of the terms, as discussed in Sec.~\ref{sec:singleVIV}, is to keep only contributions with poles in $E_i$ and $E_f$ (which are the finite-volume energy eigenvalues of two nucleons in the spin-singlet channel). This fact will be used in the matching relation that will be derived shortly. $\mathcal{M}^{\rm{DF},V}_{nn\to np}$ in Eq.~(\ref{eq: four point correlation final}) is related to the physical divergence-free transition amplitude, $\mathcal{M}^{\rm{DF}}_{nn\to np}$, which are defined in Eqs.~(\ref{eq: DF amplitude: finite V}) and (\ref{eq: single beta decay DF amplitude}), respectively. The only new quantity here that needs to be defined is $\mathcal{M}^{\rm{DF},V}_{nn \to pp}$, which is the finite-volume counterpart of $\mathcal{M}^{\rm{DF}}_{nn \to pp}$ in Eq.~(\ref{eq: double amplitude DF IV}), with replacements $I_1 \to I_1^V$ and $I_2 \to I_2^V$ . Explicitly,
\begin{align}
   &i\mathcal{M}^{\rm{DF},V}_{nn\to pp}
   (E_i,E_1,E_f)
  =i\mathcal{M}^{\rm DF}_{nn\to pp} (E_i,E_1,E_f)
  \nonumber\\
   &\hspace{0.4cm} + \frac{g_A}{2}\,i\mathcal{M}^{\rm (LO+NLO)}(E_f)\bigg[i\,g_A\,F_2(E_1,E_{*1},E_f)
   -g_A\,F_1(E_f,E_{*1})\,i\widetilde{\mathcal{M}}^{\rm (LO+NLO)}(E_{*1})\,I_1(E_i,E_{*1})
   \nonumber\\
   &\hspace{0.1cm} - g_A\,I_1(E_f,E_{*1})\,i\widetilde{\mathcal{M}}^{\rm (LO+NLO)}(E_{*1})\,F_1(E_i,E_{*1})
   -g_A \,F_1(E_f,E_{*1}) \, i\widetilde{\mathcal{M}}^{\rm (LO+NLO)}(E_{*1}) \,F_1(E_i,E_{*1})
   \nonumber\\
& \hspace{3.0cm} -\widetilde{L}_{1,A}\,i\widetilde{\mathcal{M}}^{\rm (LO+NLO)}(E_{*1})
   \left[F_1(E_i,E_{*1})+ F_1(E_{*1},E_f)\right]\bigg] i\mathcal{M}^{\rm (LO+NLO)}(E_i).
   \label{eq: double DF in V}
\end{align}
Here, $F_2=I_2^V-I_2$, or in terms of the sum-integral notation defined in Eq.~(\ref{eq:sumintegral}),
\begin{eqnarray}
    F_2(E_i,E_{*1},E_f) &=&  \frac{1}{L^3}\sum_{\bm{q}}\hspace{-.5cm}\int
    \frac{1}{E_i-\frac{{\bm q}^2}{M}+i\epsilon}
    \frac{1}{E_{*1}-\frac{{\bm q}^2}{M}+i\epsilon}
    \frac{1}{E_f-\frac{{\bm q}^2}{M}+i\epsilon}
    \nonumber\\
    &=&\frac{1}{E_i-E_f}\,[F_1(E_{*1},E_f)-F_1(E_i,E_{*1})],
\end{eqnarray}
where $F_1$ is defined in Eq.~(\ref{eq:F1def}). Equation~(\ref{eq: double DF in V}) provides a direct relation between the finite-volume quantities and the physical divergence-free double-$\beta$ decay amplitude.

\subsection{The matching relation
\label{sec:doubleVIV}}
\noindent
The momentum-space four-point correlation function in the finite volume that was constructed in the previous subsection can be written as
\begin{align}
    C_L^{(\mathcal{J}\mathcal{J})}(P_i,Q_1,P_f) &=
    \int_Ld^3x \, d^3y \, d^3z
    \int dx_0 \, dy_0 \, dz_0
    \;e^{-iP_i\cdot x}\;e^{iP_f\cdot y}\;e^{iQ_1\cdot z}
    \nonumber\\
    &\hspace{2cm}\times
    \bigg[\langle 0|\, T[B_{pp}(y)\,\mathcal{J}_{pp \to np}(z)\,\mathcal{J}_{pp \to np}(0)\, B_{nn}^\dagger(x)]\,|0\rangle \bigg]_L.
    \label{eq: four-point correlation (Pi,Pf,Q)}
\end{align}
The time-momentum representation of the same correlation function is 
\begin{align}
     C_L^{(\mathcal{J}\mathcal{J})}&(x_0,y_0,Q_1)
     \equiv \int \frac{dE_i}{2\pi}\,\frac{dE_f}{2\pi} \, e^{iE_i\cdot x_0} \, e^{-iE_f y_0}\,
    C_L^{(\mathcal{J}\mathcal{J})}(P_i,Q_1,P_f)
    \label{eq: four point correlation diagrammatic definition}
    \\
    & = \int_L d^3x\,d^3y\,
    \int  dz_0\int_L d^3z\, e^{iQ_1\cdot z}
    \bigg[\langle 0|\, T[B_{pp}(y)\,\mathcal{J}_{pp \to np}(z)\,\mathcal{J}_{pp \to np}(0)\,B^\dagger_{nn}(x)]\, |0\rangle\bigg]_L
    \label{eq: four point correlation dispersive definition 1}
    \\
    &=L^6\sum_{E_{n_i},E_{n_f}}
    e^{iE_{n_i}x_0}\,e^{-iE_{n_f}y_0}\,\int dz_0\int_L d^3z\, e^{iQ_1\cdot z}\left[\vphantom{B^\dagger}\langle 0|\,B_{pp}(0)\,| E_{n_f},L\rangle\right]_L
    \nonumber\\
    &\hspace{2.0cm}\times \, \left[\vphantom{B^\dagger}\langle E_{n_f},L|\,T[\mathcal{J}_{pp \to np}(z)\,\mathcal{J}_{pp \to np}(0)]\,|E_{n_i},L\rangle \right]_L \left[\langle E_{n_i},L|\,B^\dagger_{nn}(0)\, |0\rangle\right]_L,
    \label{eq: four point correlation dispersive expression 2}
\end{align}
written in different forms following similar steps as in the previous sections. In the last equality, it is it assumed that $y_0>z_0>0>x_0$ or $y_0>0>z_0>x_0$.

The result in Eq.~(\ref{eq: four point correlation final}) for $C_L^{(\mathcal{J}\mathcal{J})}$ can now be input in Eq.~(\ref{eq: four point correlation diagrammatic definition}). Integration over $E_i$ and $E_f$ leads to nonvanishing contributions from the only poles of the function, which are the finite-volume eigenenergies in the spin-singlet channel, $E_{n_i}$ and $E_{n_f}$, respectively. Note that these are solutions to the quantization condition in Eq.~(\ref{eq: Luscher condition}). This gives rise to
\begin{align}
    &C_L^{(\mathcal{J}\mathcal{J})}(x_0,y_0,Q_1)
    =\sum_{E_{n_i},E_{n_f}}e^{iE_{n_i}x_0} \, e^{-iE_{n_f}y_0}\,
    \mathcal{B}_{pp}(E_{n_f}) \mathcal{R}(E_{n_f})\bigg[i\mathcal{M}^{\rm{DF},V}_{nn\to pp} (E_{n_i},E_1,E_{n_f})
    \nonumber\\
    &\hspace{1.0cm} +\frac{1}{2}\,i\mathcal{M}^{\rm{DF},V}_{nn\to np} (E_{n_i},E_{*1})\, i\widetilde{\mathcal{F}}(E_{*1})\, i\mathcal{M}^{\rm{DF},V}_{np\to pp} (E_{*1},E_{n_f}) + (E_1\leftrightarrow E_2)\bigg]
    \mathcal{R}(E_{n_i}) \,\mathcal{B}^\dagger_{nn}(E_{n_i}).
    \label{eq: four point correlation integrated}
\end{align}
Here, $\mathcal{R}$ is the residue of the finite-volume function $\mathcal{F}$ evaluated at the corresponding finite-volume energy, as defined in Eq.~(\ref{eq: definition of residue}). Equating this for each $E_{n_i}$ and $E_{n_f}$ with Eq.~\eqref{eq: four point correlation dispersive expression 2} and using Eq.~\eqref{eq: matching condition for NN to NN}, one finally obtains the desired matching relation for the double-$\beta$ decay amplitude:
\begin{align}
    &L^{6}\left|\int dz_0\int_L d^3z\, e^{iQ_1\cdot z}
    \left[\vphantom{B^\dagger}\langle E_{n_f},L|\, T[\mathcal{J}_{pp \to np}(z)\,\mathcal{J}_{pp \to np}(0)]\, |E_{n_i},L\rangle\right]_L \right|^2 =\left|\mathcal{R}(E_{n_f})\right|\,\times
    \nonumber\\
    & \left|i\mathcal{M}^{\rm{DF},V}_{nn\to pp} (E_{n_i},E_1,E_{n_f})
    +\frac{1}{2}i\mathcal{M}^{\rm{DF},V}_{nn\to np} (E_{n_i},E_{*1}) i\widetilde{\mathcal{F}}_{(np)}(E_{*1}) i\mathcal{M}^{\rm{DF},V}_{np\to pp} (E_{*1},E_{n_f}) + (E_1\leftrightarrow E_2) \right|^2
    \nonumber
   \\
   &\hspace{13.2 cm} \times \, \left|\mathcal{R}(E_{n_i})\right|,
    \label{eq: matching relation double}
\end{align}
with the finite-volume divergence-free amplitudes for single- and double-$\beta$ decays defined in Eqs.~(\ref{eq: DF amplitude: finite V}) and (\ref{eq: double DF in V}), respectively. This relation is a main result of this work as it connects the (Minkowski) finite-volume ME of two time-ordered currents to the physical (Minkowski) amplitudes in infinite volume. The only subtlety is that LQCD calculations do not have direct access to the former, i.e., the left-hand side of this equation. Instead, they provide the Euclidean counterpart of this ME. It is, therefore, necessary to describe how to construct the Minkowski finite-volume ME from its Euclidean counterpart. Once this construction is carried out, Eq.~(\ref{eq: matching relation double}) enables constraining the physical double-$\beta$ decay amplitude.

\subsection{Constructing the Minkowski correlation function from its Euclidean counterpart 
\label{sec:doubleEM}}
The quantity derived in Eq.~(\ref{eq: matching relation double}) in connection to the physical hadronic amplitude for double-$\beta$ decay is defined in Minkowski spacetime. In particular, the ME on the left-hand side of Eq.~(\ref{eq: matching relation double}) has the following spectral representation:
\begin{align}
    &\mathcal{T}_L^{(\rm M)} \equiv \int dz_0\int_L d^3z\, e^{iQ_1\cdot z}
    \left[\vphantom{B^\dagger}\langle E_f,L|\, T[\mathcal{J}_{pp \to np}(z)\,\mathcal{J}_{pp \to np}(0)]\, |E_i,L\rangle\right]_L =L^3\sum_{m=0}^{\infty}\int dz_0 \, \times
    \nonumber\\
   & \hspace{1.0 cm} \bigg[ e^{i(E_f+E_1-E_{*m}+i\epsilon) z_0}
   \left[\vphantom{B^\dagger}\langle E_f,L|\, \mathcal{J}_{pp \to np}(0)\,|E_{*m},L\rangle\right]_L\left[\vphantom{B^\dagger}\langle E_{*m},L |\, \mathcal{J}_{pp \to np}(0)\,|E_i,L\rangle\right]_L \theta(z_0)\,+
    \nonumber\\
    & \hspace{1.25 cm} e^{i(E_1-E_i+E_{*m}-i\epsilon) z_0}
   \left[\vphantom{B^\dagger}\langle E_f,L|\, \mathcal{J}_{pp \to np}(0)\,|E_{*m},L\rangle\right]_L\left[\vphantom{B^\dagger}\langle E_{*m},L |\, \mathcal{J}_{pp \to np}(0)\,|E_i,L\rangle\right]_L \theta(-z_0)\bigg]
    \nonumber\\
    & \hspace{0.75 cm} = iL^3\sum_{m=0}^{\infty} \left[ \frac{\left[\vphantom{B^\dagger}\langle E_f,L|\, \mathcal{J}_{pp \to np}(0)\,|E_{*m},L\rangle\right]_L\left[\vphantom{B^\dagger}\langle E_{*m},L |\, \mathcal{J}_{pp \to np}(0)\,|E_i,L\rangle\right]_L}{E_f+E_1-E_{*m}+i\epsilon} \right . 
    \nonumber\\
    &  \hspace{4.35 cm}  \left . - \frac{\left[\vphantom{B^\dagger}\langle E_f,L|\, \mathcal{J}_{pp \to np}(0)\,|E_{*m},L\rangle\right]_L\left[\vphantom{B^\dagger}\langle E_{*m},L |\, \mathcal{J}_{pp \to np}(0)\,|E_i,L\rangle\right]_L}{E_1-E_i+E_{*m}-i\epsilon} \right],
  \label{eq: }
\end{align}
where the kinematic dependence of $\mathcal{T}_L^{(\rm M)}$ on the left-hand side is left implicit. $m$ is an index corresponding to the tower of finite-volume energy eigenstates with quantum numbers of $\mathcal{J}_{nn \to np}|E_i,L\rangle$ (or identically $\langle E_i,L|\mathcal{J}_{nn \to np}$). In contrary, such a spectral decomposition is ill-defined in Euclidean spacetime as integration over the Euclidean time $\tau \equiv z_0^{(E)} = iz_0$ may diverge depending on the energy of intermediate states. More explicitly, if there are intermediate states that could go on-shell given $E_i$, $E_n$, and $E_1$, the Minkowski and Euclidean correlation functions have fundamentally different analyticity properties. Since a LQCD computation obtains the Euclidean finite-volume four-point correlation function, it is essential that the connection to this Minkowski relation is established.

Extending the procedure of Ref.~\cite{Briceno:2019opb} to nonlocal two-nucleon MEs of this work, one can start by defining the time-momentum representation of the Euclidean correlation function:
\begin{align}
   G_L(\tau)=\int_L d^3z \left[\langle E_f,L|\, T^{(\rm{E})}[\mathcal{J}^{(\rm E)}_{pp \to np}(\tau,\bm{z})\,\mathcal{J}^{(\rm E)}_{pp \to np}(0)]\, |E_i,L\rangle\right]_L,
  \label{eq:GL}
\end{align}
which can be directly accessed via a LQCD computation. The initial and final states are at zero spatial momentum and hence the $\bm{p}_i$ and $\bm{p}_f$ dependence of the function is dropped. $T^{(\rm E)}$ means time ordering must be implemented with respect to the Euclidean-time variable. Moreover, the Heisenberg-picture operator in Euclidean spacetime satisfies $\mathcal{J}^{(\rm E)}_{pp \to np}(\tau,\bm{z})=e^{\hat{P}_0\tau -i \hat{\bm{P}}\cdot \bm{z}} \, \mathcal{J}^{(\rm E)}_{pp \to np}(0) \,$ $ e^{-\hat{P}_0\tau+i \hat{\bm{P}}\cdot \bm{z}}$, where $\hat{P}_0$ and $\hat{\bm{P}}$ are energy (Hamiltonian) and momentum operators, respectively. In the following, without loss of generality we assume that $\mathcal{J}^{(\rm E)}_{pp \to np}(0)$ is the same as its Minkowski counterpart, i.e., it has no phase relative to the Minkowski current at origin. We thus drop the Euclidean superscript on the Schr\"odinger-picture currents.

Now assuming that there are $N$ lowest-lying intermediate states (indexed as $m= 0,\cdots,N-1$) with energy $E_{*m} \leq E_f+E_1$, and $N'$ lowest-lying intermediate states (indexed as $m= 0,\cdots,N'-1$) with energy $E_{*m} \leq E_i-E_1$, the contribution from all states except the lowest $N$ or $N'$ states in the theory can be directly analytically continued to Minkowski space. With this observation, one can break the Minkowski  ME $\mathcal{T}_L^{(\rm M)}$ to two parts as following:
\begin{align}
    \mathcal{T}_L^{(\rm M)} =i\mathcal{T}_L^{(\rm{E})\,<}+i\mathcal{T}_L^{(\rm{E})\,{\geq}},
  \label{eq:TLMtwoparts}
\end{align}
where the first term, defined as
\begin{align}
    &\mathcal{T}_L^{(\rm{E})\,<} \equiv L^3 \left[\sum_{m=0}^{N-1} \frac{\left[\vphantom{B^\dagger}\langle E_f,L|\, \mathcal{J}_{pp \to np}(0)\,|E_{*m},L\rangle\right]_L\left[\vphantom{B^\dagger}\langle E_{*m},L |\, \mathcal{J}_{pp \to np}(0)\,|E_i,L\rangle\right]_L}{E_f+E_1-E_{*m}} - \right .
    \nonumber\\
    &  \hspace{2.175 cm} \left . \sum_{m=0}^{N'-1}  \frac{\left[\vphantom{B^\dagger}\langle E_f,L|\, \mathcal{J}_{pp \to np}(0)\,|E_{*m},L\rangle\right]_L\left[\vphantom{B^\dagger}\langle E_{*m},L |\, \mathcal{J}_{pp \to np}(0)\,|E_i,L\rangle\right]_L}{E_1-E_i+E_{*m}}\right],
  \label{eq: }
\end{align}
cannot be directly accessed via the Euclidean four-point correlation function, and can only be reconstructed from the knowledge of $N$ or $N'$ (whichever larger) finite-volume two-nucleon eigenenergies and the MEs of the weak current between initial/final two-nucleon states and the $N$ or $N'$ allowed intermediate states. These must be evaluated using LQCD computations of two- and three-point correlation functions in a finite volume. The second term in Eq.~(\ref{eq:TLMtwoparts}) needs the values of the four-point function evaluated with LQCD, as well as the two- and three-point functions that need to be evaluated in obtaining the first term. Explicitly,
\begin{align}
    \mathcal{T}_L^{(\rm{E})\,\geq} \equiv \int d\tau \,e^{E_1\tau} \left[ G_L(\tau)-G_L^{<}(\tau) \right],
  \label{eq: }
\end{align}
where $G_L$ is the Euclidean bi-local MEs defined in Eq.~(\ref{eq:GL}) and $G_L^{<}$ is:
\begin{align}
   G_L^{<}(\tau)=\sum_{m=0}^{N} c_m \, \theta(\tau) \, e^{-(E_{*m}-E_f)|\tau|}+\sum_{m=0}^{N'} c_m \, \theta(-\tau) \, e^{-(E_{*m}-E_i)|\tau|},
  \label{eq:GLsmaller}
\end{align}
with the $c_m$ function defined in terms of finite-volume single-$\beta$ decay MEs  satisfying either $E_{*m} \leq E_f+E_1$ or $E_{*m} \leq E_i-E_1$, corresponding to the first and second sums in Eq.~(\ref{eq:GLsmaller}), respectively:
\begin{align}
   c_{m} \equiv L^3 \left[\vphantom{B^\dagger}\langle E_f,L|\, \mathcal{J}_{pp \to np}(0)\,|E_{*m},L\rangle\right]_L\left[\vphantom{B^\dagger}\langle E_{*m},L |\, \mathcal{J}_{pp \to np}(0)\,|E_i,L\rangle\right]_L.
  \label{eq: }
\end{align}

Given all these ingredients, the physical (infinite Minkowski spacetime) double-$\beta$ decay amplitude can be extracted from a (finite) set of relevant two-, three-, and four-point correlation functions in a Euclidean finite spacetime, quantities that are all accessible from LQCD. In fact, as pointed out in Ref.~\cite{Briceno:2019opb}, it will be essential that the matching relations, including Minkowski spacetime construction and the finite-volume corrections to the ME, are all fit to LQCD data on the required $n$-point functions simultaneously to ensure the exact cancellations among various contributions that lead to the correct analytic structure of the infinite-volume amplitude.

To close, it should be noted that a closely-related procedure was implemented in Refs.~\cite{Shanahan:2017bgi,Tiburzi:2017iux} to isolate the deuteron-pole contribution from the Euclidean four-point correlation function, resulting in the desired double-$\beta$ decay ME. Nonetheless, the initial and final states in that study were bound nuclear states given larger-than-physical values of the quark masses used in the study, with only one bound on-shell intermediate state (the deuteron). This, therefore, required a simpler formalism to extract the physical ME with no need for a finite-volume matching. The complete framework of this work, therefore, will be necessary in future LQCD calculations of the same process at lighter values of the quark masses, where two-nucleon states will no longer be bound. The limitations of the applicability of  the formalism presented are its reliance on a finite-order EFT expansion, which can be straightforwardly improved by including higher-order contributions in the analysis, and its validity only below three-hadron thresholds. This latter limitation will be more challenging to relax, but progress is plausible given the finite-volume technology developed in recent years for the three-hadron problem from LQCD, see Ref.~\cite{Hansen:2019nir} for a review.

\section{Conclusion and outlook
\label{sec:conclusion} 
}
\noindent
Lattice quantum chromodynamics coupled with applicable effective field theories is becoming a critical component of modern and systematic approaches to understanding and predicting certain nuclear phenomena from first principles by supplementing quantum many-body methods in nuclear physics~\cite{Beane:2010em,Beane:2014oea,Briceno:2014tqa,Detmold:2019ghl,Cirigliano:2019jig,Kronfeld:2019nfb,Cirigliano:2020yhp,Drischler:2019xuo}. This paper provides the elementary building blocks for implementing such a perspective in the case of single- and double-weak processes in nuclei, with a focus on the less-developed process of two-neutrino double-$\beta$ decay. This work consists of two components:
\begin{itemize}
\item[$\triangleright$]{Within the framework of pionless EFT with nucleonic degrees of freedom, which is a good description of the $nn \to pp \, (ee \bar{\nu}_e\bar{\nu}_e)$ process at low energies, the physical amplitude is constructed for the first time in this work. While this process is not expected to occur in nature given the short lifetime of the neutron compared with the inverse rate of the double-weak process, such a transition is still the primary process occurring in isotopes undergoing a double-$\beta$ decay of ordinary type. As a result, any constraint on the relevant nuclear matrix element at the two-nucleon level serves as the microscopic input to quantum many-body calculations of the rate. Investigating various contributions to the amplitude, including those with electron-neutrino emission on the external nucleons, are shown to confirm the working principles of the pionless EFT as a renormalized approach to nuclear observables when applied to a double-weak amplitude.

No new low-energy constants beyond the nucleon ($g_A$) and the two-nucleon ($L_{1,A}$) axial-vector couplings are shown to be needed to guarantee a renormalization-scale independent amplitude at next-to-leading order. This is in contrary to the case of pionless EFT with dibaryon DOF in Refs.~\cite{Shanahan:2017bgi,Tiburzi:2017iux}, where a new short-distance two-nucleon two-weak current coupling, $h_{2,S}$, was introduced, and its effect on the amplitude at unphysically large values of the quark masses (determined with LQCD) was shown to be comparable to the contributions from the $l_{1,A}$ coupling (of the dibaryon formalism). By examining the consistency of $L_{1,A}$ values when constrained from single- and double-weak processes independently, future analysis of LQCD results will determine the validity of the pionless EFT power counting with nucleonic DOF as derived in this paper, and may have implications for `quenching' of $g_A$ in double-$\beta$ decays.
}
\item[$\triangleright$]{Despite the lack of experimental constraints on the process, LQCD can still give access to the nonlocal correlation functions of two-nucleon systems with two insertions of the weak current, hence determining the relevant MEs contributing to the physical amplitude. There is, however, a nontrivial procedure involved to perform this matching, as developed for the first time in this paper for this problem. The procedure builds upon the previously developed finite-volume formalisms for the local MEs of two nucleons and nonlocal MEs of a single hadron. It requires the determination of the finite-volume two-nucleon spectrum in the spin-singlet and spin-triplet channels, and the three-point and four-point correlation functions corresponding to the single-$\beta$ and double-$\beta$ decay processes from LQCD for a range of kinematics. The reason is that the Euclidean and Minkowski correlation functions can not be mapped to each other straightforwardly when on-shell multi-hadron intermediate states are present, whose effects must be isolated and treated separately~\cite{Briceno:2019opb}. Furthermore, by assuming that only two-nucleon intermediate states can go on shell, power-law volume corrections to the amplitude are derived within the pionless-EFT approach at NLO, completing the matching relation.

This formalism brings us one step closer to the ultimate goal of constraining the physical amplitude for the neutrinoless process $nn \to pp\,(ee)$ with the light Majorana exchange, or in turn the new short-distance LEC in pionless EFT that is present at LO~\cite{Cirigliano:2018hja}. Obtaining this constraint from a Euclidean four-point correlation function involves the convolution of the nuclear ME with the neutrino propagator, further complicating the relation to the Minkowski correlation function. Nonetheless, upon slight modifications, identification of the finite-volume corrections to the reconstructed Minkowski amplitude follows the same procedure as outlined in this work. These steps will be presented in future work~\cite{Davoudi:2020}.} 
\end{itemize}

Finally, an interesting observation is regarding the utility of Eq.~(\ref{eq: matching relation double}) even when the amplitude is computed using a quantum simulator or a quantum computer. In a quantum simulation, the most natural approach is to construct the states and evolve and measure them using analog or digital simulation protocols. As a result, no Monte Carlo importance sampling of quantum field configurations is required to demand a Wick rotation to Euclidean spacetime, as is the case in LQCD computations. Nonetheless, a quantum processor does not provide infinite computing resources, requiring truncations on the physical Hilbert space of the fields. Among the unavoidable truncations is that of the spatial extent of the system, bringing in the need to determine the infinite-volume limit of observables from finite-volume quantities. When it comes to multi-hadron observables, approaching such a limit can be involved. One approach is to take the ordered double limit $L \to \infty$ and $\epsilon \to 0$ in the finite-volume Minkowski amplitude, which as shown in Ref.~\cite{Briceno:2020rar} in given examples, may require large spatial volumes to allow convergence to the infinite-volume value. Another approach employs the finite-volume formalisms such as that developed in this work, e.g., Eq.~(\ref{eq: matching relation double}), to provide the mapping between the Minkowski but finite-volume ME and the physical (Minkowski and infinite-volume) amplitude at a given volume. The latter circumvents the need for the involved procedure described in Sec.~\ref{sec:doubleEM}, but is limited to low-energy kinematics below three-hadron inelastic thresholds, which, however, can be generalized. This is an example of the applicability of a range of theoretical developments in LQCD for hadronic and nuclear physics to the quantum simulation of these strongly-interacting systems in nature.

\noindent
\subsection*{Acknowledgments}
ZD thanks members of the NPLQCD collaboration for numerous discussions regarding the topic of double-$\beta$ decays from LQCD. She particularly acknowledges valuable discussions with William Detmold and Martin Savage regarding formal aspects of the LQCD-EFT matching program for this problem. ZD and SVK are supported by the Alfred P. Sloan fellowship, and by the Maryland Center for Fundamental Physics at the University of Maryland, College Park. ZD is further supported by the U.S. Department of Energy's Office of Science Early Career Award, under award no. DE-SC0020271. This work was initiated at Massachusetts Institute of Technology and was supported by the U.S. Department of Energy Early Career Research Award DE-SC0010495 and grant number DE-SC0011090

\bibliography{bibi.bib}

\begin{thebibliography}{111}%
\makeatletter
\providecommand \@ifxundefined [1]{%
 \@ifx{#1\undefined}
}%
\providecommand \@ifnum [1]{%
 \ifnum #1\expandafter \@firstoftwo
 \else \expandafter \@secondoftwo
 \fi
}%
\providecommand \@ifx [1]{%
 \ifx #1\expandafter \@firstoftwo
 \else \expandafter \@secondoftwo
 \fi
}%
\providecommand \natexlab [1]{#1}%
\providecommand \enquote  [1]{``#1''}%
\providecommand \bibnamefont  [1]{#1}%
\providecommand \bibfnamefont [1]{#1}%
\providecommand \citenamefont [1]{#1}%
\providecommand \href@noop [0]{\@secondoftwo}%
\providecommand \href [0]{\begingroup \@sanitize@url \@href}%
\providecommand \@href[1]{\@@startlink{#1}\@@href}%
\providecommand \@@href[1]{\endgroup#1\@@endlink}%
\providecommand \@sanitize@url [0]{\catcode `\\12\catcode `\$12\catcode
  `\&12\catcode `\#12\catcode `\^12\catcode `\_12\catcode `\%12\relax}%
\providecommand \@@startlink[1]{}%
\providecommand \@@endlink[0]{}%
\providecommand \url  [0]{\begingroup\@sanitize@url \@url }%
\providecommand \@url [1]{\endgroup\@href {#1}{\urlprefix }}%
\providecommand \urlprefix  [0]{URL }%
\providecommand \Eprint [0]{\href }%
\providecommand \doibase [0]{http://dx.doi.org/}%
\providecommand \selectlanguage [0]{\@gobble}%
\providecommand \bibinfo  [0]{\@secondoftwo}%
\providecommand \bibfield  [0]{\@secondoftwo}%
\providecommand \translation [1]{[#1]}%
\providecommand \BibitemOpen [0]{}%
\providecommand \bibitemStop [0]{}%
\providecommand \bibitemNoStop [0]{.\EOS\space}%
\providecommand \EOS [0]{\spacefactor3000\relax}%
\providecommand \BibitemShut  [1]{\csname bibitem#1\endcsname}%
\let\auto@bib@innerbib\@empty
\bibitem [{\citenamefont {Goeppert-Mayer}(1935)}]{PhysRev.48.512}%
  \BibitemOpen
  \bibfield  {author} {\bibinfo {author} {\bibfnamefont {M.}~\bibnamefont
  {Goeppert-Mayer}},\ }\bibfield  {title} {\enquote {\bibinfo {title} {Double
  beta-disintegration},}\ }\href {\doibase 10.1103/PhysRev.48.512} {\bibfield
  {journal} {\bibinfo  {journal} {Phys. Rev.}\ }\textbf {\bibinfo {volume}
  {48}},\ \bibinfo {pages} {512--516} (\bibinfo {year} {1935})}\BibitemShut
  {NoStop}%
\bibitem [{\citenamefont {Barabash}(2010)}]{Barabash:2010ie}%
  \BibitemOpen
  \bibfield  {author} {\bibinfo {author} {\bibfnamefont {A.S.}\ \bibnamefont
  {Barabash}},\ }\bibfield  {title} {\enquote {\bibinfo {title} {{Precise
  half-life values for two neutrino double beta decay}},}\ }\href {\doibase
  10.1103/PhysRevC.81.035501} {\bibfield  {journal} {\bibinfo  {journal} {Phys.
  Rev. C}\ }\textbf {\bibinfo {volume} {81}},\ \bibinfo {pages} {035501}
  (\bibinfo {year} {2010})},\ \Eprint {http://arxiv.org/abs/1003.1005}
  {arXiv:1003.1005 [nucl-ex]} \BibitemShut {NoStop}%
\bibitem [{\citenamefont {Saakyan}(2013)}]{Saakyan:2013yna}%
  \BibitemOpen
  \bibfield  {author} {\bibinfo {author} {\bibfnamefont {Ruben}\ \bibnamefont
  {Saakyan}},\ }\bibfield  {title} {\enquote {\bibinfo {title} {{Two-Neutrino
  Double-Beta Decay}},}\ }\href {\doibase 10.1146/annurev-nucl-102711-094904}
  {\bibfield  {journal} {\bibinfo  {journal} {Ann. Rev. Nucl. Part. Sci.}\
  }\textbf {\bibinfo {volume} {63}},\ \bibinfo {pages} {503--529} (\bibinfo
  {year} {2013})}\BibitemShut {NoStop}%
\bibitem [{\citenamefont {Dell'Oro}\ \emph {et~al.}(2016)\citenamefont
  {Dell'Oro}, \citenamefont {Marcocci}, \citenamefont {Viel},\ and\
  \citenamefont {Vissani}}]{DellOro:2016tmg}%
  \BibitemOpen
  \bibfield  {author} {\bibinfo {author} {\bibfnamefont {Stefano}\ \bibnamefont
  {Dell'Oro}}, \bibinfo {author} {\bibfnamefont {Simone}\ \bibnamefont
  {Marcocci}}, \bibinfo {author} {\bibfnamefont {Matteo}\ \bibnamefont {Viel}},
  \ and\ \bibinfo {author} {\bibfnamefont {Francesco}\ \bibnamefont
  {Vissani}},\ }\bibfield  {title} {\enquote {\bibinfo {title} {{Neutrinoless
  double beta decay: 2015 review}},}\ }\href {\doibase 10.1155/2016/2162659}
  {\bibfield  {journal} {\bibinfo  {journal} {Adv. High Energy Phys.}\ }\textbf
  {\bibinfo {volume} {2016}},\ \bibinfo {pages} {2162659} (\bibinfo {year}
  {2016})},\ \Eprint {http://arxiv.org/abs/1601.07512} {arXiv:1601.07512
  [hep-ph]} \BibitemShut {NoStop}%
\bibitem [{\citenamefont {Dolinski}\ \emph {et~al.}(2019)\citenamefont
  {Dolinski}, \citenamefont {Poon},\ and\ \citenamefont
  {Rodejohann}}]{Dolinski:2019nrj}%
  \BibitemOpen
  \bibfield  {author} {\bibinfo {author} {\bibfnamefont {Michelle~J.}\
  \bibnamefont {Dolinski}}, \bibinfo {author} {\bibfnamefont {Alan~W.P.}\
  \bibnamefont {Poon}}, \ and\ \bibinfo {author} {\bibfnamefont {Werner}\
  \bibnamefont {Rodejohann}},\ }\bibfield  {title} {\enquote {\bibinfo {title}
  {{Neutrinoless Double-Beta Decay: Status and Prospects}},}\ }\href {\doibase
  10.1146/annurev-nucl-101918-023407} {\bibfield  {journal} {\bibinfo
  {journal} {Ann. Rev. Nucl. Part. Sci.}\ }\textbf {\bibinfo {volume} {69}},\
  \bibinfo {pages} {219--251} (\bibinfo {year} {2019})},\ \Eprint
  {http://arxiv.org/abs/1902.04097} {arXiv:1902.04097 [nucl-ex]} \BibitemShut
  {NoStop}%
\bibitem [{\citenamefont {Agostini}\ \emph {et~al.}(2017)\citenamefont
  {Agostini} \emph {et~al.}}]{Agostini:2017iyd}%
  \BibitemOpen
  \bibfield  {author} {\bibinfo {author} {\bibfnamefont {M.}~\bibnamefont
  {Agostini}} \emph {et~al.},\ }\bibfield  {title} {\enquote {\bibinfo {title}
  {{Background-free search for neutrinoless double-$\beta$ decay of $^{76}$Ge
  with GERDA}},}\ }\href {\doibase 10.1038/nature21717} {\bibfield  {journal}
  {\bibinfo  {journal} {Nature}\ }\textbf {\bibinfo {volume} {544}},\ \bibinfo
  {pages} {47} (\bibinfo {year} {2017})},\ \Eprint
  {http://arxiv.org/abs/1703.00570} {arXiv:1703.00570 [nucl-ex]} \BibitemShut
  {NoStop}%
\bibitem [{\citenamefont {Moreno}\ \emph {et~al.}(2009)\citenamefont {Moreno},
  \citenamefont {Alvarez-Rodriguez}, \citenamefont {Sarriguren}, \citenamefont
  {Moya~de Guerra}, \citenamefont {Simkovic},\ and\ \citenamefont
  {Faessler}}]{Moreno:2008dz}%
  \BibitemOpen
  \bibfield  {author} {\bibinfo {author} {\bibfnamefont {O.}~\bibnamefont
  {Moreno}}, \bibinfo {author} {\bibfnamefont {R.}~\bibnamefont
  {Alvarez-Rodriguez}}, \bibinfo {author} {\bibfnamefont {P.}~\bibnamefont
  {Sarriguren}}, \bibinfo {author} {\bibfnamefont {E.}~\bibnamefont {Moya~de
  Guerra}}, \bibinfo {author} {\bibfnamefont {F.}~\bibnamefont {Simkovic}}, \
  and\ \bibinfo {author} {\bibfnamefont {A.}~\bibnamefont {Faessler}},\
  }\bibfield  {title} {\enquote {\bibinfo {title} {{Single and low-lying states
  dominance in two-neutrino double-beta decay}},}\ }\href {\doibase
  10.1088/0954-3899/36/1/015106} {\bibfield  {journal} {\bibinfo  {journal} {J.
  Phys. G}\ }\textbf {\bibinfo {volume} {36}},\ \bibinfo {pages} {015106}
  (\bibinfo {year} {2009})},\ \Eprint {http://arxiv.org/abs/0811.0319}
  {arXiv:0811.0319 [nucl-th]} \BibitemShut {NoStop}%
\bibitem [{\citenamefont {\v~Simkovic}\ \emph {et~al.}(2018)\citenamefont
  {\v~Simkovic}, \citenamefont {Dvornický}, \citenamefont {Stefánik},\ and\
  \citenamefont {Faessler}}]{Simkovic:2018rdz}%
  \BibitemOpen
  \bibfield  {author} {\bibinfo {author} {\bibfnamefont {Fedor}\ \bibnamefont
  {\v~Simkovic}}, \bibinfo {author} {\bibfnamefont {Rastislav}\ \bibnamefont
  {Dvornický}}, \bibinfo {author} {\bibfnamefont {Du\v sa\v~n}\ \bibnamefont
  {Stefánik}}, \ and\ \bibinfo {author} {\bibfnamefont {Amand}\ \bibnamefont
  {Faessler}},\ }\bibfield  {title} {\enquote {\bibinfo {title} {{Improved
  description of the $2\nu\beta\beta$ -decay and a possibility to determine the
  effective axial-vector coupling constant}},}\ }\href {\doibase
  10.1103/PhysRevC.97.034315} {\bibfield  {journal} {\bibinfo  {journal} {Phys.
  Rev. C}\ }\textbf {\bibinfo {volume} {97}},\ \bibinfo {pages} {034315}
  (\bibinfo {year} {2018})},\ \Eprint {http://arxiv.org/abs/1804.04227}
  {arXiv:1804.04227 [nucl-th]} \BibitemShut {NoStop}%
\bibitem [{\citenamefont {Arnold}\ \emph {et~al.}(2019)\citenamefont {Arnold}
  \emph {et~al.}}]{NEMO-3:2019gwo}%
  \BibitemOpen
  \bibfield  {author} {\bibinfo {author} {\bibfnamefont {R.}~\bibnamefont
  {Arnold}} \emph {et~al.} (\bibinfo {collaboration} {NEMO-3}),\ }\bibfield
  {title} {\enquote {\bibinfo {title} {{Detailed studies of $^{100}$Mo
  two-neutrino double beta decay in NEMO-3}},}\ }\href {\doibase
  10.1140/epjc/s10052-019-6948-4} {\bibfield  {journal} {\bibinfo  {journal}
  {Eur. Phys. J. C}\ }\textbf {\bibinfo {volume} {79}},\ \bibinfo {pages} {440}
  (\bibinfo {year} {2019})},\ \Eprint {http://arxiv.org/abs/1903.08084}
  {arXiv:1903.08084 [nucl-ex]} \BibitemShut {NoStop}%
\bibitem [{\citenamefont {Azzolini}\ \emph {et~al.}(2019)\citenamefont
  {Azzolini} \emph {et~al.}}]{Azzolini:2019yib}%
  \BibitemOpen
  \bibfield  {author} {\bibinfo {author} {\bibfnamefont {O.}~\bibnamefont
  {Azzolini}} \emph {et~al.},\ }\bibfield  {title} {\enquote {\bibinfo {title}
  {{Evidence of Single State Dominance in the Two-Neutrino Double-$\beta$ Decay
  of $^{82}$Se with CUPID-0}},}\ }\href {\doibase
  10.1103/PhysRevLett.123.262501} {\bibfield  {journal} {\bibinfo  {journal}
  {Phys. Rev. Lett.}\ }\textbf {\bibinfo {volume} {123}},\ \bibinfo {pages}
  {262501} (\bibinfo {year} {2019})},\ \Eprint
  {http://arxiv.org/abs/1909.03397} {arXiv:1909.03397 [nucl-ex]} \BibitemShut
  {NoStop}%
\bibitem [{\citenamefont {Gando}\ \emph {et~al.}(2019)\citenamefont {Gando}
  \emph {et~al.}}]{KamLAND-Zen:2019imh}%
  \BibitemOpen
  \bibfield  {author} {\bibinfo {author} {\bibfnamefont {A.}~\bibnamefont
  {Gando}} \emph {et~al.} (\bibinfo {collaboration} {KamLAND-Zen}),\ }\bibfield
   {title} {\enquote {\bibinfo {title} {{Precision measurement of the
  $^{136}$Xe two-neutrino $\beta\beta$ spectrum in KamLAND-Zen and its impact
  on the quenching of nuclear matrix elements}},}\ }\href {\doibase
  10.1103/PhysRevLett.122.192501} {\bibfield  {journal} {\bibinfo  {journal}
  {Phys. Rev. Lett.}\ }\textbf {\bibinfo {volume} {122}},\ \bibinfo {pages}
  {192501} (\bibinfo {year} {2019})},\ \Eprint
  {http://arxiv.org/abs/1901.03871} {arXiv:1901.03871 [hep-ex]} \BibitemShut
  {NoStop}%
\bibitem [{\citenamefont {Argyriades}\ \emph {et~al.}(2010)\citenamefont
  {Argyriades} \emph {et~al.}}]{Argyriades:2009ph}%
  \BibitemOpen
  \bibfield  {author} {\bibinfo {author} {\bibfnamefont {J.}~\bibnamefont
  {Argyriades}} \emph {et~al.} (\bibinfo {collaboration} {NEMO-3}),\ }\bibfield
   {title} {\enquote {\bibinfo {title} {{Measurement of the two neutrino double
  beta decay half-life of Zr-96 with the NEMO-3 detector}},}\ }\href {\doibase
  10.1016/j.nuclphysa.2010.07.009} {\bibfield  {journal} {\bibinfo  {journal}
  {Nucl. Phys. A}\ }\textbf {\bibinfo {volume} {847}},\ \bibinfo {pages}
  {168--179} (\bibinfo {year} {2010})},\ \Eprint
  {http://arxiv.org/abs/0906.2694} {arXiv:0906.2694 [nucl-ex]} \BibitemShut
  {NoStop}%
\bibitem [{\citenamefont {Arnold}\ \emph
  {et~al.}(2016{\natexlab{a}})\citenamefont {Arnold} \emph
  {et~al.}}]{Arnold:2016qyg}%
  \BibitemOpen
  \bibfield  {author} {\bibinfo {author} {\bibfnamefont {R.}~\bibnamefont
  {Arnold}} \emph {et~al.} (\bibinfo {collaboration} {NEMO-3}),\ }\bibfield
  {title} {\enquote {\bibinfo {title} {{Measurement of the 2$\nu\beta\beta$
  decay half-life of $^{150}$Nd and a search for 0$\nu\beta\beta$ decay
  processes with the full exposure from the NEMO-3 detector}},}\ }\href
  {\doibase 10.1103/PhysRevD.94.072003} {\bibfield  {journal} {\bibinfo
  {journal} {Phys. Rev. D}\ }\textbf {\bibinfo {volume} {94}},\ \bibinfo
  {pages} {072003} (\bibinfo {year} {2016}{\natexlab{a}})},\ \Eprint
  {http://arxiv.org/abs/1606.08494} {arXiv:1606.08494 [hep-ex]} \BibitemShut
  {NoStop}%
\bibitem [{\citenamefont {Arnold}\ \emph
  {et~al.}(2016{\natexlab{b}})\citenamefont {Arnold} \emph
  {et~al.}}]{Arnold:2016ezh}%
  \BibitemOpen
  \bibfield  {author} {\bibinfo {author} {\bibfnamefont {R.}~\bibnamefont
  {Arnold}} \emph {et~al.} (\bibinfo {collaboration} {NEMO-3}),\ }\bibfield
  {title} {\enquote {\bibinfo {title} {{Measurement of the double-beta decay
  half-life and search for the neutrinoless double-beta decay of $^{48}$Ca with
  the NEMO-3 detector}},}\ }\href {\doibase 10.1103/PhysRevD.93.112008}
  {\bibfield  {journal} {\bibinfo  {journal} {Phys. Rev. D}\ }\textbf {\bibinfo
  {volume} {93}},\ \bibinfo {pages} {112008} (\bibinfo {year}
  {2016}{\natexlab{b}})},\ \Eprint {http://arxiv.org/abs/1604.01710}
  {arXiv:1604.01710 [hep-ex]} \BibitemShut {NoStop}%
\bibitem [{\citenamefont {Arnold}\ \emph {et~al.}(2018)\citenamefont {Arnold}
  \emph {et~al.}}]{Arnold:2018tmo}%
  \BibitemOpen
  \bibfield  {author} {\bibinfo {author} {\bibfnamefont {R.}~\bibnamefont
  {Arnold}} \emph {et~al.},\ }\bibfield  {title} {\enquote {\bibinfo {title}
  {{Final results on $^{82}{Se}$ double beta decay to the ground state of
  $^{82}{Kr}$ from the NEMO-3 experiment}},}\ }\href {\doibase
  10.1140/epjc/s10052-018-6295-x} {\bibfield  {journal} {\bibinfo  {journal}
  {Eur. Phys. J. C}\ }\textbf {\bibinfo {volume} {78}},\ \bibinfo {pages} {821}
  (\bibinfo {year} {2018})},\ \Eprint {http://arxiv.org/abs/1806.05553}
  {arXiv:1806.05553 [hep-ex]} \BibitemShut {NoStop}%
\bibitem [{\citenamefont {Deppisch}\ \emph {et~al.}(2020)\citenamefont
  {Deppisch}, \citenamefont {Graf},\ and\ \citenamefont
  {\v~Simkovic}}]{Deppisch:2020mxv}%
  \BibitemOpen
  \bibfield  {author} {\bibinfo {author} {\bibfnamefont {Frank~F.}\
  \bibnamefont {Deppisch}}, \bibinfo {author} {\bibfnamefont {Lukas}\
  \bibnamefont {Graf}}, \ and\ \bibinfo {author} {\bibfnamefont {Fedor}\
  \bibnamefont {\v~Simkovic}},\ }\bibfield  {title} {\enquote {\bibinfo {title}
  {{Searching for New Physics in Two-Neutrino Double Beta Decay}},}\
  }\href@noop {} {\  (\bibinfo {year} {2020})},\ \Eprint
  {http://arxiv.org/abs/2003.11836} {arXiv:2003.11836 [hep-ph]} \BibitemShut
  {NoStop}%
\bibitem [{\citenamefont {Barea}\ \emph {et~al.}(2013)\citenamefont {Barea},
  \citenamefont {Kotila},\ and\ \citenamefont {Iachello}}]{Barea:2013bz}%
  \BibitemOpen
  \bibfield  {author} {\bibinfo {author} {\bibfnamefont {J.}~\bibnamefont
  {Barea}}, \bibinfo {author} {\bibfnamefont {J.}~\bibnamefont {Kotila}}, \
  and\ \bibinfo {author} {\bibfnamefont {F.}~\bibnamefont {Iachello}},\
  }\bibfield  {title} {\enquote {\bibinfo {title} {{Nuclear matrix elements for
  double-$\beta$ decay}},}\ }\href {\doibase 10.1103/PhysRevC.87.014315}
  {\bibfield  {journal} {\bibinfo  {journal} {Phys. Rev. C}\ }\textbf {\bibinfo
  {volume} {87}},\ \bibinfo {pages} {014315} (\bibinfo {year} {2013})},\
  \Eprint {http://arxiv.org/abs/1301.4203} {arXiv:1301.4203 [nucl-th]}
  \BibitemShut {NoStop}%
\bibitem [{\citenamefont {Engel}\ and\ \citenamefont
  {Menéndez}(2017)}]{Engel:2016xgb}%
  \BibitemOpen
  \bibfield  {author} {\bibinfo {author} {\bibfnamefont {Jonathan}\
  \bibnamefont {Engel}}\ and\ \bibinfo {author} {\bibfnamefont {Javier}\
  \bibnamefont {Menéndez}},\ }\bibfield  {title} {\enquote {\bibinfo {title}
  {{Status and Future of Nuclear Matrix Elements for Neutrinoless Double-Beta
  Decay: A Review}},}\ }\href {\doibase 10.1088/1361-6633/aa5bc5} {\bibfield
  {journal} {\bibinfo  {journal} {Rept. Prog. Phys.}\ }\textbf {\bibinfo
  {volume} {80}},\ \bibinfo {pages} {046301} (\bibinfo {year} {2017})},\
  \Eprint {http://arxiv.org/abs/1610.06548} {arXiv:1610.06548 [nucl-th]}
  \BibitemShut {NoStop}%
\bibitem [{\citenamefont {Shanahan}\ \emph {et~al.}(2017)\citenamefont
  {Shanahan}, \citenamefont {Tiburzi}, \citenamefont {Wagman}, \citenamefont
  {Winter}, \citenamefont {Chang}, \citenamefont {Davoudi}, \citenamefont
  {Detmold}, \citenamefont {Orginos},\ and\ \citenamefont
  {Savage}}]{Shanahan:2017bgi}%
  \BibitemOpen
  \bibfield  {author} {\bibinfo {author} {\bibfnamefont {Phiala~E.}\
  \bibnamefont {Shanahan}}, \bibinfo {author} {\bibfnamefont {Brian~C.}\
  \bibnamefont {Tiburzi}}, \bibinfo {author} {\bibfnamefont {Michael~L.}\
  \bibnamefont {Wagman}}, \bibinfo {author} {\bibfnamefont {Frank}\
  \bibnamefont {Winter}}, \bibinfo {author} {\bibfnamefont {Emmanuel}\
  \bibnamefont {Chang}}, \bibinfo {author} {\bibfnamefont {Zohreh}\
  \bibnamefont {Davoudi}}, \bibinfo {author} {\bibfnamefont {William}\
  \bibnamefont {Detmold}}, \bibinfo {author} {\bibfnamefont {Kostas}\
  \bibnamefont {Orginos}}, \ and\ \bibinfo {author} {\bibfnamefont {Martin~J.}\
  \bibnamefont {Savage}},\ }\bibfield  {title} {\enquote {\bibinfo {title}
  {{Isotensor Axial Polarizability and Lattice QCD Input for Nuclear
  Double-$\beta$ Decay Phenomenology}},}\ }\href {\doibase
  10.1103/PhysRevLett.119.062003} {\bibfield  {journal} {\bibinfo  {journal}
  {Phys. Rev. Lett.}\ }\textbf {\bibinfo {volume} {119}},\ \bibinfo {pages}
  {062003} (\bibinfo {year} {2017})},\ \Eprint
  {http://arxiv.org/abs/1701.03456} {arXiv:1701.03456 [hep-lat]} \BibitemShut
  {NoStop}%
\bibitem [{\citenamefont {Tiburzi}\ \emph {et~al.}(2017)\citenamefont
  {Tiburzi}, \citenamefont {Wagman}, \citenamefont {Winter}, \citenamefont
  {Chang}, \citenamefont {Davoudi}, \citenamefont {Detmold}, \citenamefont
  {Orginos}, \citenamefont {Savage},\ and\ \citenamefont
  {Shanahan}}]{Tiburzi:2017iux}%
  \BibitemOpen
  \bibfield  {author} {\bibinfo {author} {\bibfnamefont {Brian~C.}\
  \bibnamefont {Tiburzi}}, \bibinfo {author} {\bibfnamefont {Michael~L.}\
  \bibnamefont {Wagman}}, \bibinfo {author} {\bibfnamefont {Frank}\
  \bibnamefont {Winter}}, \bibinfo {author} {\bibfnamefont {Emmanuel}\
  \bibnamefont {Chang}}, \bibinfo {author} {\bibfnamefont {Zohreh}\
  \bibnamefont {Davoudi}}, \bibinfo {author} {\bibfnamefont {William}\
  \bibnamefont {Detmold}}, \bibinfo {author} {\bibfnamefont {Kostas}\
  \bibnamefont {Orginos}}, \bibinfo {author} {\bibfnamefont {Martin~J.}\
  \bibnamefont {Savage}}, \ and\ \bibinfo {author} {\bibfnamefont {Phiala~E.}\
  \bibnamefont {Shanahan}},\ }\bibfield  {title} {\enquote {\bibinfo {title}
  {{Double-$\beta$ Decay Matrix Elements from Lattice Quantum
  Chromodynamics}},}\ }\href {\doibase 10.1103/PhysRevD.96.054505} {\bibfield
  {journal} {\bibinfo  {journal} {Phys. Rev. D}\ }\textbf {\bibinfo {volume}
  {96}},\ \bibinfo {pages} {054505} (\bibinfo {year} {2017})},\ \Eprint
  {http://arxiv.org/abs/1702.02929} {arXiv:1702.02929 [hep-lat]} \BibitemShut
  {NoStop}%
\bibitem [{\citenamefont {Tews}\ \emph {et~al.}(2020)\citenamefont {Tews},
  \citenamefont {Davoudi}, \citenamefont {Ekström}, \citenamefont {Holt},\ and\
  \citenamefont {Lynn}}]{Tews:2020hgp}%
  \BibitemOpen
  \bibfield  {author} {\bibinfo {author} {\bibfnamefont {Ingo}\ \bibnamefont
  {Tews}}, \bibinfo {author} {\bibfnamefont {Zohreh}\ \bibnamefont {Davoudi}},
  \bibinfo {author} {\bibfnamefont {Andreas}\ \bibnamefont {Ekström}}, \bibinfo
  {author} {\bibfnamefont {Jason~D.}\ \bibnamefont {Holt}}, \ and\ \bibinfo
  {author} {\bibfnamefont {Joel~E.}\ \bibnamefont {Lynn}},\ }\bibfield  {title}
  {\enquote {\bibinfo {title} {{New Ideas in Constraining Nuclear Forces}},}\ \
  }(\bibinfo {year} {2020})\ \Eprint {http://arxiv.org/abs/2001.03334}
  {arXiv:2001.03334 [nucl-th]} \BibitemShut {NoStop}%
\bibitem [{\citenamefont {Kaplan}\ \emph
  {et~al.}(1998{\natexlab{a}})\citenamefont {Kaplan}, \citenamefont {Savage},\
  and\ \citenamefont {Wise}}]{Kaplan:1998tg}%
  \BibitemOpen
  \bibfield  {author} {\bibinfo {author} {\bibfnamefont {David~B.}\
  \bibnamefont {Kaplan}}, \bibinfo {author} {\bibfnamefont {Martin~J.}\
  \bibnamefont {Savage}}, \ and\ \bibinfo {author} {\bibfnamefont {Mark~B.}\
  \bibnamefont {Wise}},\ }\bibfield  {title} {\enquote {\bibinfo {title} {{A
  New expansion for nucleon-nucleon interactions}},}\ }\href {\doibase
  10.1016/S0370-2693(98)00210-X} {\bibfield  {journal} {\bibinfo  {journal}
  {Phys. Lett.}\ }\textbf {\bibinfo {volume} {B424}},\ \bibinfo {pages}
  {390--396} (\bibinfo {year} {1998}{\natexlab{a}})},\ \Eprint
  {http://arxiv.org/abs/nucl-th/9801034} {arXiv:nucl-th/9801034 [nucl-th]}
  \BibitemShut {NoStop}%
\bibitem [{\citenamefont {Kaplan}\ \emph
  {et~al.}(1998{\natexlab{b}})\citenamefont {Kaplan}, \citenamefont {Savage},\
  and\ \citenamefont {Wise}}]{Kaplan:1998we}%
  \BibitemOpen
  \bibfield  {author} {\bibinfo {author} {\bibfnamefont {David~B.}\
  \bibnamefont {Kaplan}}, \bibinfo {author} {\bibfnamefont {Martin~J.}\
  \bibnamefont {Savage}}, \ and\ \bibinfo {author} {\bibfnamefont {Mark~B.}\
  \bibnamefont {Wise}},\ }\bibfield  {title} {\enquote {\bibinfo {title} {{Two
  nucleon systems from effective field theory}},}\ }\href {\doibase
  10.1016/S0550-3213(98)00440-4} {\bibfield  {journal} {\bibinfo  {journal}
  {Nucl. Phys.}\ }\textbf {\bibinfo {volume} {B534}},\ \bibinfo {pages}
  {329--355} (\bibinfo {year} {1998}{\natexlab{b}})},\ \Eprint
  {http://arxiv.org/abs/nucl-th/9802075} {arXiv:nucl-th/9802075 [nucl-th]}
  \BibitemShut {NoStop}%
\bibitem [{\citenamefont {van Kolck}(1999)}]{vanKolck:1998bw}%
  \BibitemOpen
  \bibfield  {author} {\bibinfo {author} {\bibfnamefont {U.}~\bibnamefont {van
  Kolck}},\ }\bibfield  {title} {\enquote {\bibinfo {title} {{Effective field
  theory of short range forces}},}\ }\href {\doibase
  10.1016/S0375-9474(98)00612-5} {\bibfield  {journal} {\bibinfo  {journal}
  {Nucl. Phys. A}\ }\textbf {\bibinfo {volume} {645}},\ \bibinfo {pages}
  {273--302} (\bibinfo {year} {1999})},\ \Eprint
  {http://arxiv.org/abs/nucl-th/9808007} {arXiv:nucl-th/9808007} \BibitemShut
  {NoStop}%
\bibitem [{\citenamefont {Chen}\ \emph {et~al.}(1999)\citenamefont {Chen},
  \citenamefont {Rupak},\ and\ \citenamefont {Savage}}]{Chen:1999tn}%
  \BibitemOpen
  \bibfield  {author} {\bibinfo {author} {\bibfnamefont {Jiunn-Wei}\
  \bibnamefont {Chen}}, \bibinfo {author} {\bibfnamefont {Gautam}\ \bibnamefont
  {Rupak}}, \ and\ \bibinfo {author} {\bibfnamefont {Martin~J.}\ \bibnamefont
  {Savage}},\ }\bibfield  {title} {\enquote {\bibinfo {title} {{Nucleon-nucleon
  effective field theory without pions}},}\ }\href {\doibase
  10.1016/S0375-9474(99)00298-5} {\bibfield  {journal} {\bibinfo  {journal}
  {Nucl. Phys.}\ }\textbf {\bibinfo {volume} {A653}},\ \bibinfo {pages}
  {386--412} (\bibinfo {year} {1999})},\ \Eprint
  {http://arxiv.org/abs/nucl-th/9902056} {arXiv:nucl-th/9902056 [nucl-th]}
  \BibitemShut {NoStop}%
\bibitem [{\citenamefont {Hammer}\ \emph {et~al.}(2020)\citenamefont {Hammer},
  \citenamefont {König},\ and\ \citenamefont {van Kolck}}]{Hammer:2019poc}%
  \BibitemOpen
  \bibfield  {author} {\bibinfo {author} {\bibfnamefont {H.-W.}\ \bibnamefont
  {Hammer}}, \bibinfo {author} {\bibfnamefont {S.}~\bibnamefont {König}}, \
  and\ \bibinfo {author} {\bibfnamefont {U.}~\bibnamefont {van Kolck}},\
  }\bibfield  {title} {\enquote {\bibinfo {title} {{Nuclear effective field
  theory: status and perspectives}},}\ }\href {\doibase
  10.1103/RevModPhys.92.025004} {\bibfield  {journal} {\bibinfo  {journal}
  {Rev. Mod. Phys.}\ }\textbf {\bibinfo {volume} {92}},\ \bibinfo {pages}
  {025004} (\bibinfo {year} {2020})},\ \Eprint
  {http://arxiv.org/abs/1906.12122} {arXiv:1906.12122 [nucl-th]} \BibitemShut
  {NoStop}%
\bibitem [{\citenamefont {Fleming}\ \emph {et~al.}(2000)\citenamefont
  {Fleming}, \citenamefont {Mehen},\ and\ \citenamefont
  {Stewart}}]{Fleming:1999ee}%
  \BibitemOpen
  \bibfield  {author} {\bibinfo {author} {\bibfnamefont {Sean}\ \bibnamefont
  {Fleming}}, \bibinfo {author} {\bibfnamefont {Thomas}\ \bibnamefont {Mehen}},
  \ and\ \bibinfo {author} {\bibfnamefont {Iain~W.}\ \bibnamefont {Stewart}},\
  }\bibfield  {title} {\enquote {\bibinfo {title} {{NNLO corrections to
  nucleon-nucleon scattering and perturbative pions}},}\ }\href {\doibase
  10.1016/S0375-9474(00)00221-9} {\bibfield  {journal} {\bibinfo  {journal}
  {Nucl. Phys.}\ }\textbf {\bibinfo {volume} {A677}},\ \bibinfo {pages}
  {313--366} (\bibinfo {year} {2000})},\ \Eprint
  {http://arxiv.org/abs/nucl-th/9911001} {arXiv:nucl-th/9911001 [nucl-th]}
  \BibitemShut {NoStop}%
\bibitem [{\citenamefont {Beane}\ \emph {et~al.}(2002)\citenamefont {Beane},
  \citenamefont {Bedaque}, \citenamefont {Savage},\ and\ \citenamefont {van
  Kolck}}]{Beane:2001bc}%
  \BibitemOpen
  \bibfield  {author} {\bibinfo {author} {\bibfnamefont {S.R.}\ \bibnamefont
  {Beane}}, \bibinfo {author} {\bibfnamefont {Paulo~F.}\ \bibnamefont
  {Bedaque}}, \bibinfo {author} {\bibfnamefont {M.J.}\ \bibnamefont {Savage}},
  \ and\ \bibinfo {author} {\bibfnamefont {U.}~\bibnamefont {van Kolck}},\
  }\bibfield  {title} {\enquote {\bibinfo {title} {{Towards a perturbative
  theory of nuclear forces}},}\ }\href {\doibase 10.1016/S0375-9474(01)01324-0}
  {\bibfield  {journal} {\bibinfo  {journal} {Nucl. Phys. A}\ }\textbf
  {\bibinfo {volume} {700}},\ \bibinfo {pages} {377--402} (\bibinfo {year}
  {2002})},\ \Eprint {http://arxiv.org/abs/nucl-th/0104030}
  {arXiv:nucl-th/0104030} \BibitemShut {NoStop}%
\bibitem [{\citenamefont {Cirigliano}\ \emph
  {et~al.}(2018{\natexlab{a}})\citenamefont {Cirigliano}, \citenamefont
  {Dekens}, \citenamefont {Mereghetti},\ and\ \citenamefont
  {Walker-Loud}}]{Cirigliano:2017tvr}%
  \BibitemOpen
  \bibfield  {author} {\bibinfo {author} {\bibfnamefont {Vincenzo}\
  \bibnamefont {Cirigliano}}, \bibinfo {author} {\bibfnamefont {Wouter}\
  \bibnamefont {Dekens}}, \bibinfo {author} {\bibfnamefont {Emanuele}\
  \bibnamefont {Mereghetti}}, \ and\ \bibinfo {author} {\bibfnamefont
  {Andr\'e}\ \bibnamefont {Walker-Loud}},\ }\bibfield  {title} {\enquote
  {\bibinfo {title} {{Neutrinoless double-\ensuremath{\beta} decay in effective
  field theory: The light-Majorana neutrino-exchange mechanism}},}\ }\href
  {\doibase 10.1103/PhysRevC.97.065501} {\bibfield  {journal} {\bibinfo
  {journal} {Phys. Rev. C}\ }\textbf {\bibinfo {volume} {97}},\ \bibinfo
  {pages} {065501} (\bibinfo {year} {2018}{\natexlab{a}})},\ \bibinfo {note}
  {[Erratum: Phys.Rev.C 100, 019903 (2019)]},\ \Eprint
  {http://arxiv.org/abs/1710.01729} {arXiv:1710.01729 [hep-ph]} \BibitemShut
  {NoStop}%
\bibitem [{\citenamefont {Cirigliano}\ \emph
  {et~al.}(2018{\natexlab{b}})\citenamefont {Cirigliano}, \citenamefont
  {Dekens}, \citenamefont {De~Vries}, \citenamefont {Graesser}, \citenamefont
  {Mereghetti}, \citenamefont {Pastore},\ and\ \citenamefont
  {Van~Kolck}}]{Cirigliano:2018hja}%
  \BibitemOpen
  \bibfield  {author} {\bibinfo {author} {\bibfnamefont {Vincenzo}\
  \bibnamefont {Cirigliano}}, \bibinfo {author} {\bibfnamefont {Wouter}\
  \bibnamefont {Dekens}}, \bibinfo {author} {\bibfnamefont {Jordy}\
  \bibnamefont {De~Vries}}, \bibinfo {author} {\bibfnamefont {Michael~L.}\
  \bibnamefont {Graesser}}, \bibinfo {author} {\bibfnamefont {Emanuele}\
  \bibnamefont {Mereghetti}}, \bibinfo {author} {\bibfnamefont {Saori}\
  \bibnamefont {Pastore}}, \ and\ \bibinfo {author} {\bibfnamefont {Ubirajara}\
  \bibnamefont {Van~Kolck}},\ }\bibfield  {title} {\enquote {\bibinfo {title}
  {{New Leading Contribution to Neutrinoless Double-$\beta$ Decay}},}\ }\href
  {\doibase 10.1103/PhysRevLett.120.202001} {\bibfield  {journal} {\bibinfo
  {journal} {Phys. Rev. Lett.}\ }\textbf {\bibinfo {volume} {120}},\ \bibinfo
  {pages} {202001} (\bibinfo {year} {2018}{\natexlab{b}})},\ \Eprint
  {http://arxiv.org/abs/1802.10097} {arXiv:1802.10097 [hep-ph]} \BibitemShut
  {NoStop}%
\bibitem [{\citenamefont {Cirigliano}\ \emph
  {et~al.}(2019{\natexlab{a}})\citenamefont {Cirigliano}, \citenamefont
  {Dekens}, \citenamefont {De~Vries}, \citenamefont {Graesser}, \citenamefont
  {Mereghetti}, \citenamefont {Pastore}, \citenamefont {Piarulli},
  \citenamefont {Van~Kolck},\ and\ \citenamefont
  {Wiringa}}]{Cirigliano:2019vdj}%
  \BibitemOpen
  \bibfield  {author} {\bibinfo {author} {\bibfnamefont {V.}~\bibnamefont
  {Cirigliano}}, \bibinfo {author} {\bibfnamefont {W.}~\bibnamefont {Dekens}},
  \bibinfo {author} {\bibfnamefont {J.}~\bibnamefont {De~Vries}}, \bibinfo
  {author} {\bibfnamefont {M.L.}\ \bibnamefont {Graesser}}, \bibinfo {author}
  {\bibfnamefont {E.}~\bibnamefont {Mereghetti}}, \bibinfo {author}
  {\bibfnamefont {S.}~\bibnamefont {Pastore}}, \bibinfo {author} {\bibfnamefont
  {M.}~\bibnamefont {Piarulli}}, \bibinfo {author} {\bibfnamefont
  {U.}~\bibnamefont {Van~Kolck}}, \ and\ \bibinfo {author} {\bibfnamefont
  {R.B.}\ \bibnamefont {Wiringa}},\ }\bibfield  {title} {\enquote {\bibinfo
  {title} {{Renormalized approach to neutrinoless double- $\beta$ decay}},}\
  }\href {\doibase 10.1103/PhysRevC.100.055504} {\bibfield  {journal} {\bibinfo
   {journal} {Phys. Rev. C}\ }\textbf {\bibinfo {volume} {100}},\ \bibinfo
  {pages} {055504} (\bibinfo {year} {2019}{\natexlab{a}})},\ \Eprint
  {http://arxiv.org/abs/1907.11254} {arXiv:1907.11254 [nucl-th]} \BibitemShut
  {NoStop}%
\bibitem [{\citenamefont {Beane}\ and\ \citenamefont
  {Savage}(2001)}]{Beane:2000fi}%
  \BibitemOpen
  \bibfield  {author} {\bibinfo {author} {\bibfnamefont {Silas~R.}\
  \bibnamefont {Beane}}\ and\ \bibinfo {author} {\bibfnamefont {Martin~J.}\
  \bibnamefont {Savage}},\ }\bibfield  {title} {\enquote {\bibinfo {title}
  {{Rearranging pionless effective field theory}},}\ }\href {\doibase
  10.1016/S0375-9474(01)01088-0} {\bibfield  {journal} {\bibinfo  {journal}
  {Nucl. Phys. A}\ }\textbf {\bibinfo {volume} {694}},\ \bibinfo {pages}
  {511--524} (\bibinfo {year} {2001})},\ \Eprint
  {http://arxiv.org/abs/nucl-th/0011067} {arXiv:nucl-th/0011067} \BibitemShut
  {NoStop}%
\bibitem [{\citenamefont {Phillips}\ \emph {et~al.}(2000)\citenamefont
  {Phillips}, \citenamefont {Rupak},\ and\ \citenamefont
  {Savage}}]{Phillips:1999hh}%
  \BibitemOpen
  \bibfield  {author} {\bibinfo {author} {\bibfnamefont {Daniel~R.}\
  \bibnamefont {Phillips}}, \bibinfo {author} {\bibfnamefont {Gautam}\
  \bibnamefont {Rupak}}, \ and\ \bibinfo {author} {\bibfnamefont {Martin~J.}\
  \bibnamefont {Savage}},\ }\bibfield  {title} {\enquote {\bibinfo {title}
  {{Improving the convergence of N N effective field theory}},}\ }\href
  {\doibase 10.1016/S0370-2693(99)01496-3} {\bibfield  {journal} {\bibinfo
  {journal} {Phys. Lett. B}\ }\textbf {\bibinfo {volume} {473}},\ \bibinfo
  {pages} {209--218} (\bibinfo {year} {2000})},\ \Eprint
  {http://arxiv.org/abs/nucl-th/9908054} {arXiv:nucl-th/9908054} \BibitemShut
  {NoStop}%
\bibitem [{\citenamefont {Kong}\ and\ \citenamefont
  {Ravndal}(1999)}]{Kong:1999tw}%
  \BibitemOpen
  \bibfield  {author} {\bibinfo {author} {\bibfnamefont {Xinwei}\ \bibnamefont
  {Kong}}\ and\ \bibinfo {author} {\bibfnamefont {Finn}\ \bibnamefont
  {Ravndal}},\ }\bibfield  {title} {\enquote {\bibinfo {title} {{Proton proton
  fusion in leading order of effective field theory}},}\ }\href {\doibase
  10.1016/S0375-9474(99)00314-0} {\bibfield  {journal} {\bibinfo  {journal}
  {Nucl. Phys. A}\ }\textbf {\bibinfo {volume} {656}},\ \bibinfo {pages}
  {421--429} (\bibinfo {year} {1999})},\ \Eprint
  {http://arxiv.org/abs/nucl-th/9902064} {arXiv:nucl-th/9902064} \BibitemShut
  {NoStop}%
\bibitem [{\citenamefont {Kong}\ and\ \citenamefont
  {Ravndal}(2001)}]{Kong:2000px}%
  \BibitemOpen
  \bibfield  {author} {\bibinfo {author} {\bibfnamefont {Xinwei}\ \bibnamefont
  {Kong}}\ and\ \bibinfo {author} {\bibfnamefont {Finn}\ \bibnamefont
  {Ravndal}},\ }\bibfield  {title} {\enquote {\bibinfo {title} {{Proton proton
  fusion in effective field theory}},}\ }\href {\doibase
  10.1103/PhysRevC.64.044002} {\bibfield  {journal} {\bibinfo  {journal} {Phys.
  Rev. C}\ }\textbf {\bibinfo {volume} {64}},\ \bibinfo {pages} {044002}
  (\bibinfo {year} {2001})},\ \Eprint {http://arxiv.org/abs/nucl-th/0004038}
  {arXiv:nucl-th/0004038} \BibitemShut {NoStop}%
\bibitem [{\citenamefont {Butler}\ \emph {et~al.}(2001)\citenamefont {Butler},
  \citenamefont {Chen},\ and\ \citenamefont {Kong}}]{Butler:2000zp}%
  \BibitemOpen
  \bibfield  {author} {\bibinfo {author} {\bibfnamefont {Malcolm}\ \bibnamefont
  {Butler}}, \bibinfo {author} {\bibfnamefont {Jiunn-Wei}\ \bibnamefont
  {Chen}}, \ and\ \bibinfo {author} {\bibfnamefont {Xinwei}\ \bibnamefont
  {Kong}},\ }\bibfield  {title} {\enquote {\bibinfo {title} {{Neutrino deuteron
  scattering in effective field theory at next-to-next-to-leading order}},}\
  }\href {\doibase 10.1103/PhysRevC.63.035501} {\bibfield  {journal} {\bibinfo
  {journal} {Phys. Rev.}\ }\textbf {\bibinfo {volume} {C63}},\ \bibinfo {pages}
  {035501} (\bibinfo {year} {2001})},\ \Eprint
  {http://arxiv.org/abs/nucl-th/0008032} {arXiv:nucl-th/0008032 [nucl-th]}
  \BibitemShut {NoStop}%
\bibitem [{\citenamefont {Maiani}\ and\ \citenamefont
  {Testa}(1990)}]{Maiani:1990ca}%
  \BibitemOpen
  \bibfield  {author} {\bibinfo {author} {\bibfnamefont {L.}~\bibnamefont
  {Maiani}}\ and\ \bibinfo {author} {\bibfnamefont {M.}~\bibnamefont {Testa}},\
  }\bibfield  {title} {\enquote {\bibinfo {title} {{Final state interactions
  from Euclidean correlation functions}},}\ }\href {\doibase
  10.1016/0370-2693(90)90695-3} {\bibfield  {journal} {\bibinfo  {journal}
  {Phys. Lett. B}\ }\textbf {\bibinfo {volume} {245}},\ \bibinfo {pages}
  {585--590} (\bibinfo {year} {1990})}\BibitemShut {NoStop}%
\bibitem [{\citenamefont {{M. L\"uscher}}(1986)}]{Luscher:1986pf}%
  \BibitemOpen
  \bibfield  {author} {\bibinfo {author} {\bibnamefont {{M. L\"uscher}}},\
  }\bibfield  {title} {\enquote {\bibinfo {title} {{Volume Dependence of the
  Energy Spectrum in Massive Quantum Field Theories. 2. Scattering States}},}\
  }\href {\doibase 10.1007/bf01211097} {\bibfield  {journal} {\bibinfo
  {journal} {Commun.Math.Phys.}\ }\textbf {\bibinfo {volume} {105}},\ \bibinfo
  {pages} {153--188} (\bibinfo {year} {1986})}\BibitemShut {NoStop}%
\bibitem [{\citenamefont {{M. L\"uscher}}(1991)}]{Luscher:1990ux}%
  \BibitemOpen
  \bibfield  {author} {\bibinfo {author} {\bibnamefont {{M. L\"uscher}}},\
  }\bibfield  {title} {\enquote {\bibinfo {title} {{Two particle states on a
  torus and their relation to the scattering matrix}},}\ }\href {\doibase
  10.1016/0550-3213(91)90366-6} {\bibfield  {journal} {\bibinfo  {journal}
  {Nucl.Phys}\ }\textbf {\bibinfo {volume} {B354}},\ \bibinfo {pages}
  {531--578} (\bibinfo {year} {1991})}\BibitemShut {NoStop}%
\bibitem [{\citenamefont {Rummukainen}\ and\ \citenamefont
  {Gottlieb}(1995)}]{Rummukainen:1995vs}%
  \BibitemOpen
  \bibfield  {author} {\bibinfo {author} {\bibfnamefont {K.}~\bibnamefont
  {Rummukainen}}\ and\ \bibinfo {author} {\bibfnamefont {Steven~A.}\
  \bibnamefont {Gottlieb}},\ }\bibfield  {title} {\enquote {\bibinfo {title}
  {Resonance scattering phase shifts on a nonrest frame lattice},}\ }\href
  {\doibase 10.1016/0550-3213(95)00313-h} {\bibfield  {journal} {\bibinfo
  {journal} {Nucl. Phys.}\ }\textbf {\bibinfo {volume} {B450}},\ \bibinfo
  {pages} {397--436} (\bibinfo {year} {1995})},\ \Eprint
  {http://arxiv.org/abs/hep-lat/9503028} {arXiv:hep-lat/9503028} \BibitemShut
  {NoStop}%
\bibitem [{\citenamefont {Kim}\ \emph {et~al.}(2005)\citenamefont {Kim},
  \citenamefont {Sachrajda},\ and\ \citenamefont {Sharpe}}]{Kim:2005gf}%
  \BibitemOpen
  \bibfield  {author} {\bibinfo {author} {\bibfnamefont {C.h.}\ \bibnamefont
  {Kim}}, \bibinfo {author} {\bibfnamefont {C.T.}\ \bibnamefont {Sachrajda}}, \
  and\ \bibinfo {author} {\bibfnamefont {Stephen~R.}\ \bibnamefont {Sharpe}},\
  }\bibfield  {title} {\enquote {\bibinfo {title} {{Finite-volume effects for
  two-hadron states in moving frames}},}\ }\href {\doibase
  10.1016/j.nuclphysb.2005.08.029} {\bibfield  {journal} {\bibinfo  {journal}
  {Nucl. Phys. B}\ }\textbf {\bibinfo {volume} {727}},\ \bibinfo {pages}
  {218--243} (\bibinfo {year} {2005})},\ \Eprint
  {http://arxiv.org/abs/hep-lat/0507006} {arXiv:hep-lat/0507006} \BibitemShut
  {NoStop}%
\bibitem [{\citenamefont {He}\ \emph {et~al.}(2005)\citenamefont {He},
  \citenamefont {Feng},\ and\ \citenamefont {Liu}}]{He:2005ey}%
  \BibitemOpen
  \bibfield  {author} {\bibinfo {author} {\bibfnamefont {Song}\ \bibnamefont
  {He}}, \bibinfo {author} {\bibfnamefont {Xu}~\bibnamefont {Feng}}, \ and\
  \bibinfo {author} {\bibfnamefont {Chuan}\ \bibnamefont {Liu}},\ }\bibfield
  {title} {\enquote {\bibinfo {title} {Two particle states and the s-matrix
  elements in multi-channel scattering},}\ }\href {\doibase
  10.1088/1126-6708/2005/07/011} {\bibfield  {journal} {\bibinfo  {journal}
  {Jhep}\ }\textbf {\bibinfo {volume} {7}},\ \bibinfo {pages} {11} (\bibinfo
  {year} {2005})},\ \Eprint {http://arxiv.org/abs/hep-lat/0504019}
  {arXiv:hep-lat/0504019 [hep-lat]} \BibitemShut {NoStop}%
\bibitem [{\citenamefont {Davoudi}\ and\ \citenamefont
  {Savage}(2011)}]{Davoudi:2011md}%
  \BibitemOpen
  \bibfield  {author} {\bibinfo {author} {\bibfnamefont {Zohreh}\ \bibnamefont
  {Davoudi}}\ and\ \bibinfo {author} {\bibfnamefont {Martin~J.}\ \bibnamefont
  {Savage}},\ }\bibfield  {title} {\enquote {\bibinfo {title} {{Improving the
  Volume Dependence of Two-Body Binding Energies Calculated with Lattice
  QCD}},}\ }\href {\doibase 10.1103/PhysRevD.84.114502} {\bibfield  {journal}
  {\bibinfo  {journal} {Phys. Rev.}\ }\textbf {\bibinfo {volume} {D84}},\
  \bibinfo {pages} {114502} (\bibinfo {year} {2011})},\ \Eprint
  {http://arxiv.org/abs/1108.5371} {arXiv:1108.5371 [hep-lat]} \BibitemShut
  {NoStop}%
\bibitem [{\citenamefont {Leskovec}\ and\ \citenamefont
  {Prelovsek}(2012)}]{Leskovec:2012gb}%
  \BibitemOpen
  \bibfield  {author} {\bibinfo {author} {\bibfnamefont {Luka}\ \bibnamefont
  {Leskovec}}\ and\ \bibinfo {author} {\bibfnamefont {Sasa}\ \bibnamefont
  {Prelovsek}},\ }\bibfield  {title} {\enquote {\bibinfo {title} {{Scattering
  phase shifts for two particles of different mass and non-zero total momentum
  in lattice QCD}},}\ }\href {\doibase 10.1103/PhysRevD.85.114507} {\bibfield
  {journal} {\bibinfo  {journal} {Phys. Rev.}\ }\textbf {\bibinfo {volume}
  {D85}},\ \bibinfo {pages} {114507} (\bibinfo {year} {2012})},\ \Eprint
  {http://arxiv.org/abs/1202.2145} {arXiv:1202.2145 [hep-lat]} \BibitemShut
  {NoStop}%
\bibitem [{\citenamefont {Hansen}\ and\ \citenamefont
  {Sharpe}(2012)}]{Hansen:2012tf}%
  \BibitemOpen
  \bibfield  {author} {\bibinfo {author} {\bibfnamefont {Maxwell~T.}\
  \bibnamefont {Hansen}}\ and\ \bibinfo {author} {\bibfnamefont {Stephen~R.}\
  \bibnamefont {Sharpe}},\ }\bibfield  {title} {\enquote {\bibinfo {title}
  {Multiple-channel generalization of lellouch-luscher formula},}\ }\href
  {\doibase 10.1103/PhysRevD.86.016007} {\bibfield  {journal} {\bibinfo
  {journal} {Phys.Rev.}\ }\textbf {\bibinfo {volume} {D86}},\ \bibinfo {pages}
  {16007} (\bibinfo {year} {2012})},\ \Eprint {http://arxiv.org/abs/1204.0826}
  {arXiv:1204.0826 [hep-lat]} \BibitemShut {NoStop}%
\bibitem [{\citenamefont {Briceno}\ and\ \citenamefont
  {Davoudi}(2013{\natexlab{a}})}]{Briceno:2012yi}%
  \BibitemOpen
  \bibfield  {author} {\bibinfo {author} {\bibfnamefont {Raul~A.}\ \bibnamefont
  {Briceno}}\ and\ \bibinfo {author} {\bibfnamefont {Zohreh}\ \bibnamefont
  {Davoudi}},\ }\bibfield  {title} {\enquote {\bibinfo {title} {{Moving
  multichannel systems in a finite volume with application to proton-proton
  fusion}},}\ }\href {\doibase 10.1103/PhysRevD.88.094507} {\bibfield
  {journal} {\bibinfo  {journal} {Phys. Rev.}\ }\textbf {\bibinfo {volume}
  {D88}},\ \bibinfo {pages} {94507} (\bibinfo {year} {2013}{\natexlab{a}})},\
  \Eprint {http://arxiv.org/abs/1204.1110} {arXiv:1204.1110 [hep-lat]}
  \BibitemShut {NoStop}%
\bibitem [{\citenamefont {Gockeler}\ \emph {et~al.}(2012)\citenamefont
  {Gockeler}, \citenamefont {Horsley}, \citenamefont {Lage}, \citenamefont
  {Meissner}, \citenamefont {Rakow}, \citenamefont {Rusetsky}, \citenamefont
  {Schierholz},\ and\ \citenamefont {Zanotti}}]{Gockeler:2012yj}%
  \BibitemOpen
  \bibfield  {author} {\bibinfo {author} {\bibfnamefont {M.}~\bibnamefont
  {Gockeler}}, \bibinfo {author} {\bibfnamefont {R.}~\bibnamefont {Horsley}},
  \bibinfo {author} {\bibfnamefont {M.}~\bibnamefont {Lage}}, \bibinfo {author}
  {\bibfnamefont {U.~G.}\ \bibnamefont {Meissner}}, \bibinfo {author}
  {\bibfnamefont {P.~E.~L.}\ \bibnamefont {Rakow}}, \bibinfo {author}
  {\bibfnamefont {A.}~\bibnamefont {Rusetsky}}, \bibinfo {author}
  {\bibfnamefont {G.}~\bibnamefont {Schierholz}}, \ and\ \bibinfo {author}
  {\bibfnamefont {J.~M.}\ \bibnamefont {Zanotti}},\ }\bibfield  {title}
  {\enquote {\bibinfo {title} {{Scattering phases for meson and baryon
  resonances on general moving-frame lattices}},}\ }\href {\doibase
  10.1103/PhysRevD.86.094513} {\bibfield  {journal} {\bibinfo  {journal} {Phys.
  Rev.}\ }\textbf {\bibinfo {volume} {D86}},\ \bibinfo {pages} {94513}
  (\bibinfo {year} {2012})},\ \Eprint {http://arxiv.org/abs/1206.4141}
  {arXiv:1206.4141 [hep-lat]} \BibitemShut {NoStop}%
\bibitem [{\citenamefont {Briceno}\ \emph
  {et~al.}(2013{\natexlab{a}})\citenamefont {Briceno}, \citenamefont
  {Davoudi},\ and\ \citenamefont {Luu}}]{Briceno:2013lba}%
  \BibitemOpen
  \bibfield  {author} {\bibinfo {author} {\bibfnamefont {Raul~A.}\ \bibnamefont
  {Briceno}}, \bibinfo {author} {\bibfnamefont {Zohreh}\ \bibnamefont
  {Davoudi}}, \ and\ \bibinfo {author} {\bibfnamefont {Thomas~C.}\ \bibnamefont
  {Luu}},\ }\bibfield  {title} {\enquote {\bibinfo {title} {{Two-Nucleon
  Systems in a Finite Volume: (I) Quantization Conditions}},}\ }\href {\doibase
  10.1103/PhysRevD.88.034502} {\bibfield  {journal} {\bibinfo  {journal} {Phys.
  Rev.}\ }\textbf {\bibinfo {volume} {D88}},\ \bibinfo {pages} {34502}
  (\bibinfo {year} {2013}{\natexlab{a}})},\ \Eprint
  {http://arxiv.org/abs/1305.4903} {arXiv:1305.4903 [hep-lat]} \BibitemShut
  {NoStop}%
\bibitem [{\citenamefont {Feng}\ \emph {et~al.}(2004)\citenamefont {Feng},
  \citenamefont {Li},\ and\ \citenamefont {Liu}}]{Feng:2004ua}%
  \BibitemOpen
  \bibfield  {author} {\bibinfo {author} {\bibfnamefont {Xu}~\bibnamefont
  {Feng}}, \bibinfo {author} {\bibfnamefont {Xin}\ \bibnamefont {Li}}, \ and\
  \bibinfo {author} {\bibfnamefont {Chuan}\ \bibnamefont {Liu}},\ }\bibfield
  {title} {\enquote {\bibinfo {title} {{Two particle states in an asymmetric
  box and the elastic scattering phases}},}\ }\href {\doibase
  10.1103/PhysRevD.70.014505} {\bibfield  {journal} {\bibinfo  {journal} {Phys.
  Rev.}\ }\textbf {\bibinfo {volume} {D70}},\ \bibinfo {pages} {14505}
  (\bibinfo {year} {2004})},\ \Eprint {http://arxiv.org/abs/hep-lat/0404001}
  {arXiv:hep-lat/0404001 [hep-lat]} \BibitemShut {NoStop}%
\bibitem [{\citenamefont {Lee}\ and\ \citenamefont
  {Alexandru}(2017)}]{Lee:2017igf}%
  \BibitemOpen
  \bibfield  {author} {\bibinfo {author} {\bibfnamefont {Frank~X.}\
  \bibnamefont {Lee}}\ and\ \bibinfo {author} {\bibfnamefont {Andrei}\
  \bibnamefont {Alexandru}},\ }\bibfield  {title} {\enquote {\bibinfo {title}
  {{Scattering phase-shift formulas for mesons and baryons in elongated
  boxes}},}\ }\href {\doibase 10.1103/PhysRevD.96.054508} {\bibfield  {journal}
  {\bibinfo  {journal} {Phys. Rev.}\ }\textbf {\bibinfo {volume} {D96}},\
  \bibinfo {pages} {54508} (\bibinfo {year} {2017})},\ \Eprint
  {http://arxiv.org/abs/1706.00262} {arXiv:1706.00262 [hep-lat]} \BibitemShut
  {NoStop}%
\bibitem [{\citenamefont {Bedaque}(2004)}]{Bedaque:2004kc}%
  \BibitemOpen
  \bibfield  {author} {\bibinfo {author} {\bibfnamefont {Paulo~F.}\
  \bibnamefont {Bedaque}},\ }\bibfield  {title} {\enquote {\bibinfo {title}
  {{Aharonov-Bohm effect and nucleon nucleon phase shifts on the lattice}},}\
  }\href {\doibase 10.1016/j.physletb.2004.04.045} {\bibfield  {journal}
  {\bibinfo  {journal} {Phys. Lett.}\ }\textbf {\bibinfo {volume} {B593}},\
  \bibinfo {pages} {82--88} (\bibinfo {year} {2004})},\ \Eprint
  {http://arxiv.org/abs/nucl-th/0402051} {arXiv:nucl-th/0402051 [nucl-th]}
  \BibitemShut {NoStop}%
\bibitem [{\citenamefont {Luu}\ and\ \citenamefont
  {Savage}(2011)}]{Luu:2011ep}%
  \BibitemOpen
  \bibfield  {author} {\bibinfo {author} {\bibfnamefont {Thomas}\ \bibnamefont
  {Luu}}\ and\ \bibinfo {author} {\bibfnamefont {Martin~J.}\ \bibnamefont
  {Savage}},\ }\bibfield  {title} {\enquote {\bibinfo {title} {{Extracting
  Scattering Phase-Shifts in Higher Partial-Waves from Lattice QCD
  Calculations}},}\ }\href {\doibase 10.1103/PhysRevD.83.114508} {\bibfield
  {journal} {\bibinfo  {journal} {Phys. Rev. D}\ }\textbf {\bibinfo {volume}
  {83}},\ \bibinfo {pages} {114508} (\bibinfo {year} {2011})},\ \Eprint
  {http://arxiv.org/abs/1101.3347} {arXiv:1101.3347 [hep-lat]} \BibitemShut
  {NoStop}%
\bibitem [{\citenamefont {Briceno}\ \emph
  {et~al.}(2014{\natexlab{a}})\citenamefont {Briceno}, \citenamefont {Davoudi},
  \citenamefont {Luu},\ and\ \citenamefont {Savage}}]{Briceno:2013hya}%
  \BibitemOpen
  \bibfield  {author} {\bibinfo {author} {\bibfnamefont {Raul~A.}\ \bibnamefont
  {Briceno}}, \bibinfo {author} {\bibfnamefont {Zohreh}\ \bibnamefont
  {Davoudi}}, \bibinfo {author} {\bibfnamefont {Thomas~C.}\ \bibnamefont
  {Luu}}, \ and\ \bibinfo {author} {\bibfnamefont {Martin~J.}\ \bibnamefont
  {Savage}},\ }\bibfield  {title} {\enquote {\bibinfo {title} {{Two-Baryon
  Systems with Twisted Boundary Conditions}},}\ }\href {\doibase
  10.1103/PhysRevD.89.074509} {\bibfield  {journal} {\bibinfo  {journal} {Phys.
  Rev.}\ }\textbf {\bibinfo {volume} {D89}},\ \bibinfo {pages} {74509}
  (\bibinfo {year} {2014}{\natexlab{a}})},\ \Eprint
  {http://arxiv.org/abs/1311.7686} {arXiv:1311.7686 [hep-lat]} \BibitemShut
  {NoStop}%
\bibitem [{\citenamefont {Briceno}\ \emph
  {et~al.}(2013{\natexlab{b}})\citenamefont {Briceno}, \citenamefont {Davoudi},
  \citenamefont {Luu},\ and\ \citenamefont {Savage}}]{Briceno:2013bda}%
  \BibitemOpen
  \bibfield  {author} {\bibinfo {author} {\bibfnamefont {Raul~A.}\ \bibnamefont
  {Briceno}}, \bibinfo {author} {\bibfnamefont {Zohreh}\ \bibnamefont
  {Davoudi}}, \bibinfo {author} {\bibfnamefont {Thomas}\ \bibnamefont {Luu}}, \
  and\ \bibinfo {author} {\bibfnamefont {Martin~J.}\ \bibnamefont {Savage}},\
  }\bibfield  {title} {\enquote {\bibinfo {title} {{Two-nucleon systems in a
  finite volume. II. $^3S_1-^3D_1$ coupled channels and the deuteron}},}\
  }\href {\doibase 10.1103/PhysRevD.88.114507} {\bibfield  {journal} {\bibinfo
  {journal} {Phys. Rev.}\ }\textbf {\bibinfo {volume} {D88}},\ \bibinfo {pages}
  {114507} (\bibinfo {year} {2013}{\natexlab{b}})},\ \Eprint
  {http://arxiv.org/abs/1309.3556} {arXiv:1309.3556 [hep-lat]} \BibitemShut
  {NoStop}%
\bibitem [{\citenamefont {Briceno}(2014)}]{Briceno:2014oea}%
  \BibitemOpen
  \bibfield  {author} {\bibinfo {author} {\bibfnamefont {Raul~A.}\ \bibnamefont
  {Briceno}},\ }\bibfield  {title} {\enquote {\bibinfo {title} {{Two-particle
  multichannel systems in a finite volume with arbitrary spin}},}\ }\href
  {\doibase 10.1103/PhysRevD.89.074507} {\bibfield  {journal} {\bibinfo
  {journal} {Phys. Rev.}\ }\textbf {\bibinfo {volume} {D89}},\ \bibinfo {pages}
  {74507} (\bibinfo {year} {2014})},\ \Eprint {http://arxiv.org/abs/1401.3312}
  {arXiv:1401.3312 [hep-lat]} \BibitemShut {NoStop}%
\bibitem [{\citenamefont {Briceno}\ \emph {et~al.}(2018)\citenamefont
  {Briceno}, \citenamefont {Dudek},\ and\ \citenamefont
  {Young}}]{Briceno:2017max}%
  \BibitemOpen
  \bibfield  {author} {\bibinfo {author} {\bibfnamefont {Raul~A.}\ \bibnamefont
  {Briceno}}, \bibinfo {author} {\bibfnamefont {Jozef~J.}\ \bibnamefont
  {Dudek}}, \ and\ \bibinfo {author} {\bibfnamefont {Ross~D.}\ \bibnamefont
  {Young}},\ }\bibfield  {title} {\enquote {\bibinfo {title} {{Scattering
  processes and resonances from lattice QCD}},}\ }\href {\doibase
  10.1103/RevModPhys.90.025001} {\bibfield  {journal} {\bibinfo  {journal}
  {Rev. Mod. Phys.}\ }\textbf {\bibinfo {volume} {90}},\ \bibinfo {pages}
  {025001} (\bibinfo {year} {2018})},\ \Eprint
  {http://arxiv.org/abs/1706.06223} {arXiv:1706.06223 [hep-lat]} \BibitemShut
  {NoStop}%
\bibitem [{\citenamefont {Polejaeva}\ and\ \citenamefont
  {Rusetsky}(2012)}]{Polejaeva:2012ut}%
  \BibitemOpen
  \bibfield  {author} {\bibinfo {author} {\bibfnamefont {K.}~\bibnamefont
  {Polejaeva}}\ and\ \bibinfo {author} {\bibfnamefont {A.}~\bibnamefont
  {Rusetsky}},\ }\bibfield  {title} {\enquote {\bibinfo {title} {{Three
  particles in a finite volume}},}\ }\href {\doibase
  10.1140/epja/i2012-12067-8} {\bibfield  {journal} {\bibinfo  {journal} {Eur.
  Phys. J.}\ }\textbf {\bibinfo {volume} {A48}},\ \bibinfo {pages} {67}
  (\bibinfo {year} {2012})},\ \Eprint {http://arxiv.org/abs/1203.1241}
  {arXiv:1203.1241 [hep-lat]} \BibitemShut {NoStop}%
\bibitem [{\citenamefont {Briceno}\ and\ \citenamefont
  {Davoudi}(2013{\natexlab{b}})}]{Briceno:2012rv}%
  \BibitemOpen
  \bibfield  {author} {\bibinfo {author} {\bibfnamefont {Raul~A.}\ \bibnamefont
  {Briceno}}\ and\ \bibinfo {author} {\bibfnamefont {Zohreh}\ \bibnamefont
  {Davoudi}},\ }\bibfield  {title} {\enquote {\bibinfo {title} {{Three-particle
  scattering amplitudes from a finite volume formalism}},}\ }\href {\doibase
  10.1103/PhysRevD.87.094507} {\bibfield  {journal} {\bibinfo  {journal} {Phys.
  Rev.}\ }\textbf {\bibinfo {volume} {D87}},\ \bibinfo {pages} {94507}
  (\bibinfo {year} {2013}{\natexlab{b}})},\ \Eprint
  {http://arxiv.org/abs/1212.3398} {arXiv:1212.3398 [hep-lat]} \BibitemShut
  {NoStop}%
\bibitem [{\citenamefont {Beane}\ and\ \citenamefont
  {Savage}(2014)}]{Beane:2014qha}%
  \BibitemOpen
  \bibfield  {author} {\bibinfo {author} {\bibfnamefont {Silas~R.}\
  \bibnamefont {Beane}}\ and\ \bibinfo {author} {\bibfnamefont {Martin~J.}\
  \bibnamefont {Savage}},\ }\bibfield  {title} {\enquote {\bibinfo {title}
  {{Two-Particle Elastic Scattering in a Finite Volume Including QED}},}\
  }\href {\doibase 10.1103/PhysRevD.90.074511} {\bibfield  {journal} {\bibinfo
  {journal} {Phys. Rev. D}\ }\textbf {\bibinfo {volume} {90}},\ \bibinfo
  {pages} {074511} (\bibinfo {year} {2014})},\ \Eprint
  {http://arxiv.org/abs/1407.4846} {arXiv:1407.4846 [hep-lat]} \BibitemShut
  {NoStop}%
\bibitem [{\citenamefont {Hansen}\ and\ \citenamefont
  {Sharpe}(2014)}]{Hansen:2014eka}%
  \BibitemOpen
  \bibfield  {author} {\bibinfo {author} {\bibfnamefont {Maxwell~T.}\
  \bibnamefont {Hansen}}\ and\ \bibinfo {author} {\bibfnamefont {Stephen~R.}\
  \bibnamefont {Sharpe}},\ }\bibfield  {title} {\enquote {\bibinfo {title}
  {{Relativistic, model-independent, three-particle quantization condition}},}\
  }\href {\doibase 10.1103/PhysRevD.90.116003} {\bibfield  {journal} {\bibinfo
  {journal} {Phys. Rev.}\ }\textbf {\bibinfo {volume} {D90}},\ \bibinfo {pages}
  {116003} (\bibinfo {year} {2014})},\ \Eprint {http://arxiv.org/abs/1408.5933}
  {arXiv:1408.5933 [hep-lat]} \BibitemShut {NoStop}%
\bibitem [{\citenamefont {Hammer}\ \emph
  {et~al.}(2017{\natexlab{a}})\citenamefont {Hammer}, \citenamefont {Pang},\
  and\ \citenamefont {Rusetsky}}]{Hammer:2017uqm}%
  \BibitemOpen
  \bibfield  {author} {\bibinfo {author} {\bibfnamefont {Hans-Werner}\
  \bibnamefont {Hammer}}, \bibinfo {author} {\bibfnamefont {Jin-Yi}\
  \bibnamefont {Pang}}, \ and\ \bibinfo {author} {\bibfnamefont
  {A.}~\bibnamefont {Rusetsky}},\ }\bibfield  {title} {\enquote {\bibinfo
  {title} {{Three-particle quantization condition in a finite volume: 1. The
  role of the three-particle force}},}\ }\href {\doibase
  10.1007/jhep09(2017)109} {\bibfield  {journal} {\bibinfo  {journal} {Jhep}\
  }\textbf {\bibinfo {volume} {9}},\ \bibinfo {pages} {109} (\bibinfo {year}
  {2017}{\natexlab{a}})},\ \Eprint {http://arxiv.org/abs/1706.07700}
  {arXiv:1706.07700 [hep-lat]} \BibitemShut {NoStop}%
\bibitem [{\citenamefont {Hammer}\ \emph
  {et~al.}(2017{\natexlab{b}})\citenamefont {Hammer}, \citenamefont {Pang},\
  and\ \citenamefont {Rusetsky}}]{Hammer:2017kms}%
  \BibitemOpen
  \bibfield  {author} {\bibinfo {author} {\bibfnamefont {H.~W.}\ \bibnamefont
  {Hammer}}, \bibinfo {author} {\bibfnamefont {J.~Y.}\ \bibnamefont {Pang}}, \
  and\ \bibinfo {author} {\bibfnamefont {A.}~\bibnamefont {Rusetsky}},\
  }\bibfield  {title} {\enquote {\bibinfo {title} {{Three particle quantization
  condition in a finite volume: 2. general formalism and the analysis of
  data}},}\ }\href {\doibase 10.1007/jhep10(2017)115} {\bibfield  {journal}
  {\bibinfo  {journal} {Jhep}\ }\textbf {\bibinfo {volume} {10}},\ \bibinfo
  {pages} {115} (\bibinfo {year} {2017}{\natexlab{b}})},\ \Eprint
  {http://arxiv.org/abs/1707.02176} {arXiv:1707.02176 [hep-lat]} \BibitemShut
  {NoStop}%
\bibitem [{\citenamefont {Guo}\ and\ \citenamefont
  {Gasparian}(2017)}]{Guo:2017ism}%
  \BibitemOpen
  \bibfield  {author} {\bibinfo {author} {\bibfnamefont {Peng}\ \bibnamefont
  {Guo}}\ and\ \bibinfo {author} {\bibfnamefont {Vladimir}\ \bibnamefont
  {Gasparian}},\ }\bibfield  {title} {\enquote {\bibinfo {title} {An solvable
  three-body model in finite volume},}\ }\href {\doibase
  10.1016/j.physletb.2017.10.009} {\bibfield  {journal} {\bibinfo  {journal}
  {Phys. Lett.}\ }\textbf {\bibinfo {volume} {B774}},\ \bibinfo {pages}
  {441--445} (\bibinfo {year} {2017})},\ \Eprint
  {http://arxiv.org/abs/1701.00438} {arXiv:1701.00438 [hep-lat]} \BibitemShut
  {NoStop}%
\bibitem [{\citenamefont {Mai}\ and\ \citenamefont
  {Doring}(2017)}]{Mai:2017bge}%
  \BibitemOpen
  \bibfield  {author} {\bibinfo {author} {\bibfnamefont {M.}~\bibnamefont
  {Mai}}\ and\ \bibinfo {author} {\bibfnamefont {M.}~\bibnamefont {Doring}},\
  }\bibfield  {title} {\enquote {\bibinfo {title} {{Three-body Unitarity in the
  Finite Volume}},}\ }\href@noop {} {\  (\bibinfo {year} {2017})},\ \Eprint
  {http://arxiv.org/abs/1709.08222} {arXiv:1709.08222 [hep-lat]} \BibitemShut
  {NoStop}%
\bibitem [{\citenamefont {Briceno}\ \emph {et~al.}(2017)\citenamefont
  {Briceno}, \citenamefont {Hansen},\ and\ \citenamefont
  {Sharpe}}]{Briceno:2017tce}%
  \BibitemOpen
  \bibfield  {author} {\bibinfo {author} {\bibfnamefont {Raul~A.}\ \bibnamefont
  {Briceno}}, \bibinfo {author} {\bibfnamefont {Maxwell~T.}\ \bibnamefont
  {Hansen}}, \ and\ \bibinfo {author} {\bibfnamefont {Stephen~R.}\ \bibnamefont
  {Sharpe}},\ }\bibfield  {title} {\enquote {\bibinfo {title} {{Relating the
  finite-volume spectrum and the two-and-three-particle $S$ matrix for
  relativistic systems of identical scalar particles}},}\ }\href {\doibase
  10.1103/PhysRevD.95.074510} {\bibfield  {journal} {\bibinfo  {journal} {Phys.
  Rev.}\ }\textbf {\bibinfo {volume} {D95}},\ \bibinfo {pages} {74510}
  (\bibinfo {year} {2017})},\ \Eprint {http://arxiv.org/abs/1701.07465}
  {arXiv:1701.07465 [hep-lat]} \BibitemShut {NoStop}%
\bibitem [{\citenamefont {Döring}\ \emph {et~al.}(2018)\citenamefont
  {Döring}, \citenamefont {Hammer}, \citenamefont {Mai}, \citenamefont {Pang},
  \citenamefont {Rusetsky},\ and\ \citenamefont {Wu}}]{Doring:2018xxx}%
  \BibitemOpen
  \bibfield  {author} {\bibinfo {author} {\bibfnamefont {M.}~\bibnamefont
  {Döring}}, \bibinfo {author} {\bibfnamefont {H.~W.}\ \bibnamefont {Hammer}},
  \bibinfo {author} {\bibfnamefont {M.}~\bibnamefont {Mai}}, \bibinfo {author}
  {\bibfnamefont {J.~Y}\ \bibnamefont {Pang}}, \bibinfo {author} {\bibfnamefont
  {§~A.}\ \bibnamefont {Rusetsky}}, \ and\ \bibinfo {author} {\bibfnamefont
  {J.}~\bibnamefont {Wu}},\ }\bibfield  {title} {\enquote {\bibinfo {title}
  {{Three-body spectrum in a finite volume: the role of cubic symmetry}},}\
  }\href@noop {} {\  (\bibinfo {year} {2018})},\ \Eprint
  {http://arxiv.org/abs/1802.03362} {arXiv:1802.03362 [hep-lat]} \BibitemShut
  {NoStop}%
\bibitem [{\citenamefont {Briceno}\ \emph {et~al.}(2019)\citenamefont
  {Briceno}, \citenamefont {Hansen},\ and\ \citenamefont
  {Sharpe}}]{Briceno:2018aml}%
  \BibitemOpen
  \bibfield  {author} {\bibinfo {author} {\bibfnamefont {Raul~A.}\ \bibnamefont
  {Briceno}}, \bibinfo {author} {\bibfnamefont {Maxwell~T.}\ \bibnamefont
  {Hansen}}, \ and\ \bibinfo {author} {\bibfnamefont {Stephen~R.}\ \bibnamefont
  {Sharpe}},\ }\bibfield  {title} {\enquote {\bibinfo {title} {{Three-particle
  systems with resonant subprocesses in a finite volume}},}\ }\href {\doibase
  10.1103/PhysRevD.99.014516} {\bibfield  {journal} {\bibinfo  {journal} {Phys.
  Rev. D}\ }\textbf {\bibinfo {volume} {99}},\ \bibinfo {pages} {014516}
  (\bibinfo {year} {2019})},\ \Eprint {http://arxiv.org/abs/1810.01429}
  {arXiv:1810.01429 [hep-lat]} \BibitemShut {NoStop}%
\bibitem [{\citenamefont {Jackura}\ \emph {et~al.}(2019)\citenamefont
  {Jackura}, \citenamefont {Dawid}, \citenamefont {Fernández-Ramírez},
  \citenamefont {Mathieu}, \citenamefont {Mikhasenko}, \citenamefont {Pilloni},
  \citenamefont {Sharpe},\ and\ \citenamefont {Szczepaniak}}]{Jackura:2019bmu}%
  \BibitemOpen
  \bibfield  {author} {\bibinfo {author} {\bibfnamefont {A.~W.}\ \bibnamefont
  {Jackura}}, \bibinfo {author} {\bibfnamefont {S.~M.}\ \bibnamefont {Dawid}},
  \bibinfo {author} {\bibfnamefont {C.}~\bibnamefont {Fernández-Ramírez}},
  \bibinfo {author} {\bibfnamefont {V.}~\bibnamefont {Mathieu}}, \bibinfo
  {author} {\bibfnamefont {M.}~\bibnamefont {Mikhasenko}}, \bibinfo {author}
  {\bibfnamefont {A.}~\bibnamefont {Pilloni}}, \bibinfo {author} {\bibfnamefont
  {S.~R.}\ \bibnamefont {Sharpe}}, \ and\ \bibinfo {author} {\bibfnamefont
  {A.~P.}\ \bibnamefont {Szczepaniak}},\ }\bibfield  {title} {\enquote
  {\bibinfo {title} {{Equivalence of three-particle scattering formalisms}},}\
  }\href {\doibase 10.1103/PhysRevD.100.034508} {\bibfield  {journal} {\bibinfo
   {journal} {Phys. Rev.}\ }\textbf {\bibinfo {volume} {D100}},\ \bibinfo
  {pages} {34508} (\bibinfo {year} {2019})},\ \Eprint
  {http://arxiv.org/abs/1905.12007} {arXiv:1905.12007 [hep-ph]} \BibitemShut
  {NoStop}%
\bibitem [{\citenamefont {Hansen}\ \emph {et~al.}(2020)\citenamefont {Hansen},
  \citenamefont {Romero-López},\ and\ \citenamefont {Sharpe}}]{Hansen:2020zhy}%
  \BibitemOpen
  \bibfield  {author} {\bibinfo {author} {\bibfnamefont {Maxwell~T.}\
  \bibnamefont {Hansen}}, \bibinfo {author} {\bibfnamefont {Fernando}\
  \bibnamefont {Romero-López}}, \ and\ \bibinfo {author} {\bibfnamefont
  {Stephen~R.}\ \bibnamefont {Sharpe}},\ }\bibfield  {title} {\enquote
  {\bibinfo {title} {{Generalizing the relativistic quantization condition to
  include all three-pion isospin channels}},}\ }\href@noop {} {\  (\bibinfo
  {year} {2020})},\ \Eprint {http://arxiv.org/abs/2003.10974} {arXiv:2003.10974
  [hep-lat]} \BibitemShut {NoStop}%
\bibitem [{\citenamefont {Hansen}\ and\ \citenamefont
  {Sharpe}(2019)}]{Hansen:2019nir}%
  \BibitemOpen
  \bibfield  {author} {\bibinfo {author} {\bibfnamefont {Maxwell~T.}\
  \bibnamefont {Hansen}}\ and\ \bibinfo {author} {\bibfnamefont {Stephen~R.}\
  \bibnamefont {Sharpe}},\ }\bibfield  {title} {\enquote {\bibinfo {title}
  {{Lattice QCD and Three-particle Decays of Resonances}},}\ }\href {\doibase
  10.1146/annurev-nucl-101918-023723} {\bibfield  {journal} {\bibinfo
  {journal} {Ann. Rev. Nucl. Part. Sci.}\ }\textbf {\bibinfo {volume} {69}},\
  \bibinfo {pages} {65--107} (\bibinfo {year} {2019})},\ \Eprint
  {http://arxiv.org/abs/1901.00483} {arXiv:1901.00483 [hep-lat]} \BibitemShut
  {NoStop}%
\bibitem [{\citenamefont {Lellouch}\ and\ \citenamefont
  {Luscher}(2001)}]{Lellouch:2000pv}%
  \BibitemOpen
  \bibfield  {author} {\bibinfo {author} {\bibfnamefont {Laurent}\ \bibnamefont
  {Lellouch}}\ and\ \bibinfo {author} {\bibfnamefont {Martin}\ \bibnamefont
  {Luscher}},\ }\bibfield  {title} {\enquote {\bibinfo {title} {{Weak
  transition matrix elements from finite volume correlation functions}},}\
  }\href {\doibase 10.1007/s002200100410} {\bibfield  {journal} {\bibinfo
  {journal} {Commun. Math. Phys.}\ }\textbf {\bibinfo {volume} {219}},\
  \bibinfo {pages} {31--44} (\bibinfo {year} {2001})},\ \Eprint
  {http://arxiv.org/abs/hep-lat/0003023} {arXiv:hep-lat/0003023 [hep-lat]}
  \BibitemShut {NoStop}%
\bibitem [{\citenamefont {Detmold}\ and\ \citenamefont
  {Savage}(2004)}]{Detmold:2004qn}%
  \BibitemOpen
  \bibfield  {author} {\bibinfo {author} {\bibfnamefont {William}\ \bibnamefont
  {Detmold}}\ and\ \bibinfo {author} {\bibfnamefont {Martin~J.}\ \bibnamefont
  {Savage}},\ }\bibfield  {title} {\enquote {\bibinfo {title} {{Electroweak
  matrix elements in the two nucleon sector from lattice QCD}},}\ }\href
  {\doibase 10.1016/j.nuclphysa.2004.07.007} {\bibfield  {journal} {\bibinfo
  {journal} {Nucl. Phys.}\ }\textbf {\bibinfo {volume} {A743}},\ \bibinfo
  {pages} {170--193} (\bibinfo {year} {2004})},\ \Eprint
  {http://arxiv.org/abs/hep-lat/0403005} {arXiv:hep-lat/0403005 [hep-lat]}
  \BibitemShut {NoStop}%
\bibitem [{\citenamefont {Meyer}(2011)}]{Meyer:2011um}%
  \BibitemOpen
  \bibfield  {author} {\bibinfo {author} {\bibfnamefont {Harvey~B.}\
  \bibnamefont {Meyer}},\ }\bibfield  {title} {\enquote {\bibinfo {title}
  {Lattice qcd and the timelike pion form factor},}\ }\href {\doibase
  10.1103/PhysRevLett.107.072002} {\bibfield  {journal} {\bibinfo  {journal}
  {Phys. Rev. Lett.}\ }\textbf {\bibinfo {volume} {107}},\ \bibinfo {pages}
  {72002} (\bibinfo {year} {2011})},\ \Eprint {http://arxiv.org/abs/1105.1892}
  {arXiv:1105.1892 [hep-lat]} \BibitemShut {NoStop}%
\bibitem [{\citenamefont {Bernard}\ \emph {et~al.}(2012)\citenamefont
  {Bernard}, \citenamefont {Hoja}, \citenamefont {Meissner},\ and\
  \citenamefont {Rusetsky}}]{Bernard:2012bi}%
  \BibitemOpen
  \bibfield  {author} {\bibinfo {author} {\bibfnamefont {V.}~\bibnamefont
  {Bernard}}, \bibinfo {author} {\bibfnamefont {D.}~\bibnamefont {Hoja}},
  \bibinfo {author} {\bibfnamefont {U.~G.}\ \bibnamefont {Meissner}}, \ and\
  \bibinfo {author} {\bibfnamefont {A.}~\bibnamefont {Rusetsky}},\ }\bibfield
  {title} {\enquote {\bibinfo {title} {{Matrix elements of unstable states}},}\
  }\href {\doibase 10.1007/jhep09(2012)023} {\bibfield  {journal} {\bibinfo
  {journal} {Jhep}\ }\textbf {\bibinfo {volume} {9}},\ \bibinfo {pages} {23}
  (\bibinfo {year} {2012})},\ \Eprint {http://arxiv.org/abs/1205.4642}
  {arXiv:1205.4642 [hep-lat]} \BibitemShut {NoStop}%
\bibitem [{\citenamefont {Briceno}\ \emph
  {et~al.}(2014{\natexlab{b}})\citenamefont {Briceno}, \citenamefont {Hansen},\
  and\ \citenamefont {Walker-Loud}}]{Briceno:2014uqa}%
  \BibitemOpen
  \bibfield  {author} {\bibinfo {author} {\bibfnamefont {Raul~A.}\ \bibnamefont
  {Briceno}}, \bibinfo {author} {\bibfnamefont {Maxwell~T.}\ \bibnamefont
  {Hansen}}, \ and\ \bibinfo {author} {\bibfnamefont {Andre}\ \bibnamefont
  {Walker-Loud}},\ }\bibfield  {title} {\enquote {\bibinfo {title}
  {Multichannel one-to-two transition form factors in a finite volume},}\
  }\href@noop {} {\  (\bibinfo {year} {2014}{\natexlab{b}})},\ \Eprint
  {http://arxiv.org/abs/1406.5965} {arXiv:1406.5965 [hep-lat]} \BibitemShut
  {NoStop}%
\bibitem [{\citenamefont {Feng}\ \emph {et~al.}(2015)\citenamefont {Feng},
  \citenamefont {Aoki}, \citenamefont {Hashimoto},\ and\ \citenamefont
  {Kaneko}}]{Feng:2014gba}%
  \BibitemOpen
  \bibfield  {author} {\bibinfo {author} {\bibfnamefont {Xu}~\bibnamefont
  {Feng}}, \bibinfo {author} {\bibfnamefont {Sinya}\ \bibnamefont {Aoki}},
  \bibinfo {author} {\bibfnamefont {Shoji}\ \bibnamefont {Hashimoto}}, \ and\
  \bibinfo {author} {\bibfnamefont {Takashi}\ \bibnamefont {Kaneko}},\
  }\bibfield  {title} {\enquote {\bibinfo {title} {Timelike pion form factor in
  lattice qcd},}\ }\href {\doibase 10.1103/PhysRevD.91.054504} {\bibfield
  {journal} {\bibinfo  {journal} {Phys. Rev.}\ }\textbf {\bibinfo {volume}
  {D91}},\ \bibinfo {pages} {54504} (\bibinfo {year} {2015})},\ \Eprint
  {http://arxiv.org/abs/1412.6319} {arXiv:1412.6319 [hep-lat]} \BibitemShut
  {NoStop}%
\bibitem [{\citenamefont {Briceno}\ and\ \citenamefont
  {Hansen}(2015)}]{Briceno:2015csa}%
  \BibitemOpen
  \bibfield  {author} {\bibinfo {author} {\bibfnamefont {Raul~A.}\ \bibnamefont
  {Briceno}}\ and\ \bibinfo {author} {\bibfnamefont {Maxwell~T.}\ \bibnamefont
  {Hansen}},\ }\bibfield  {title} {\enquote {\bibinfo {title} {Multichannel 0
  $\to$ 2 and 1 $\to$ 2 transition amplitudes for arbitrary spin particles in a
  finite volume},}\ }\href {\doibase 10.1103/PhysRevD.92.074509} {\bibfield
  {journal} {\bibinfo  {journal} {Phys. Rev.}\ }\textbf {\bibinfo {volume}
  {D92}},\ \bibinfo {pages} {74509} (\bibinfo {year} {2015})},\ \Eprint
  {http://arxiv.org/abs/1502.04314} {arXiv:1502.04314 [hep-lat]} \BibitemShut
  {NoStop}%
\bibitem [{\citenamefont {Briceno}\ and\ \citenamefont
  {Hansen}(2016)}]{Briceno:2015tza}%
  \BibitemOpen
  \bibfield  {author} {\bibinfo {author} {\bibfnamefont {Raul~A.}\ \bibnamefont
  {Briceno}}\ and\ \bibinfo {author} {\bibfnamefont {Maxwell~T.}\ \bibnamefont
  {Hansen}},\ }\bibfield  {title} {\enquote {\bibinfo {title} {Relativistic,
  model-independent, multichannel $2\to 2$ transition amplitudes in a finite
  volume},}\ }\href {\doibase 10.1103/PhysRevD.94.013008} {\bibfield  {journal}
  {\bibinfo  {journal} {Phys. Rev.}\ }\textbf {\bibinfo {volume} {D94}},\
  \bibinfo {pages} {13008} (\bibinfo {year} {2016})},\ \Eprint
  {http://arxiv.org/abs/1509.08507} {arXiv:1509.08507 [hep-lat]} \BibitemShut
  {NoStop}%
\bibitem [{\citenamefont {Christ}(2010)}]{Christ:2010gi}%
  \BibitemOpen
  \bibfield  {author} {\bibinfo {author} {\bibfnamefont {Norman~H.}\
  \bibnamefont {Christ}} (\bibinfo {collaboration} {RBC Collaboration, UKQCD
  Collaboration}),\ }\bibfield  {title} {\enquote {\bibinfo {title}
  {Long-distance contributions to weak amplitudes},}\ }\href@noop {} {\
  (\bibinfo {year} {2010})},\ \Eprint {http://arxiv.org/abs/1012.6034}
  {arXiv:1012.6034 [hep-lat]} \BibitemShut {NoStop}%
\bibitem [{\citenamefont {Christ}\ \emph
  {et~al.}(2014{\natexlab{a}})\citenamefont {Christ}, \citenamefont
  {Martinelli},\ and\ \citenamefont {Sachrajda}}]{Christ:2014qaa}%
  \BibitemOpen
  \bibfield  {author} {\bibinfo {author} {\bibfnamefont {N.H.}\ \bibnamefont
  {Christ}}, \bibinfo {author} {\bibfnamefont {G.}~\bibnamefont {Martinelli}},
  \ and\ \bibinfo {author} {\bibfnamefont {C.T.}\ \bibnamefont {Sachrajda}},\
  }\bibfield  {title} {\enquote {\bibinfo {title} {{Finite-volume effects in
  the evaluation of the KL-KS mass difference}},}\ }\href@noop {} {\bibfield
  {journal} {\bibinfo  {journal} {PoS}\ }\textbf {\bibinfo {volume}
  {LATTICE2013}},\ \bibinfo {pages} {399} (\bibinfo {year}
  {2014}{\natexlab{a}})},\ \Eprint {http://arxiv.org/abs/1401.1362}
  {arXiv:1401.1362 [hep-lat]} \BibitemShut {NoStop}%
\bibitem [{\citenamefont {Christ}\ \emph {et~al.}(2015)\citenamefont {Christ},
  \citenamefont {Feng}, \citenamefont {Martinelli},\ and\ \citenamefont
  {Sachrajda}}]{Christ:2015pwa}%
  \BibitemOpen
  \bibfield  {author} {\bibinfo {author} {\bibfnamefont {Norman~H.}\
  \bibnamefont {Christ}}, \bibinfo {author} {\bibfnamefont {Xu}~\bibnamefont
  {Feng}}, \bibinfo {author} {\bibfnamefont {Guido}\ \bibnamefont
  {Martinelli}}, \ and\ \bibinfo {author} {\bibfnamefont {Christopher~T.}\
  \bibnamefont {Sachrajda}},\ }\bibfield  {title} {\enquote {\bibinfo {title}
  {{Effects of finite volume on the KL-KS mass difference}},}\ }\href {\doibase
  10.1103/PhysRevD.91.114510} {\bibfield  {journal} {\bibinfo  {journal} {Phys.
  Rev.}\ }\textbf {\bibinfo {volume} {D91}},\ \bibinfo {pages} {114510}
  (\bibinfo {year} {2015})},\ \Eprint {http://arxiv.org/abs/1504.01170}
  {arXiv:1504.01170 [hep-lat]} \BibitemShut {NoStop}%
\bibitem [{\citenamefont {Christ}\ \emph {et~al.}(2013)\citenamefont {Christ},
  \citenamefont {Izubuchi}, \citenamefont {Sachrajda}, \citenamefont {Soni},\
  and\ \citenamefont {Yu}}]{Christ:2012se}%
  \BibitemOpen
  \bibfield  {author} {\bibinfo {author} {\bibfnamefont {N.H.}\ \bibnamefont
  {Christ}}, \bibinfo {author} {\bibfnamefont {T.}~\bibnamefont {Izubuchi}},
  \bibinfo {author} {\bibfnamefont {C.T.}\ \bibnamefont {Sachrajda}}, \bibinfo
  {author} {\bibfnamefont {A.}~\bibnamefont {Soni}}, \ and\ \bibinfo {author}
  {\bibfnamefont {J.}~\bibnamefont {Yu}} (\bibinfo {collaboration} {RBC and
  UKQCD Collaborations}),\ }\bibfield  {title} {\enquote {\bibinfo {title}
  {Long distance contribution to the kl-ks mass difference},}\ }\href {\doibase
  10.1103/PhysRevD.88.014508} {\bibfield  {journal} {\bibinfo  {journal}
  {Phys.Rev.}\ }\textbf {\bibinfo {volume} {D88}},\ \bibinfo {pages} {014508}
  (\bibinfo {year} {2013})},\ \Eprint {http://arxiv.org/abs/1212.5931}
  {arXiv:1212.5931 [hep-lat]} \BibitemShut {NoStop}%
\bibitem [{\citenamefont {Bai}\ \emph {et~al.}(2014)\citenamefont {Bai},
  \citenamefont {Christ}, \citenamefont {Izubuchi}, \citenamefont {Sachrajda},
  \citenamefont {Soni} \emph {et~al.}}]{Bai:2014cva}%
  \BibitemOpen
  \bibfield  {author} {\bibinfo {author} {\bibfnamefont {Z.}~\bibnamefont
  {Bai}}, \bibinfo {author} {\bibfnamefont {N.H.}\ \bibnamefont {Christ}},
  \bibinfo {author} {\bibfnamefont {T.}~\bibnamefont {Izubuchi}}, \bibinfo
  {author} {\bibfnamefont {C.T.}\ \bibnamefont {Sachrajda}}, \bibinfo {author}
  {\bibfnamefont {A.}~\bibnamefont {Soni}},  \emph {et~al.},\ }\bibfield
  {title} {\enquote {\bibinfo {title} {Kl-ks mass difference from lattice
  qcd},}\ }\href {\doibase 10.1103/PhysRevLett.113.112003} {\bibfield
  {journal} {\bibinfo  {journal} {Phys.Rev.Lett.}\ }\textbf {\bibinfo {volume}
  {113}},\ \bibinfo {pages} {112003} (\bibinfo {year} {2014})},\ \Eprint
  {http://arxiv.org/abs/1406.0916} {arXiv:1406.0916 [hep-lat]} \BibitemShut
  {NoStop}%
\bibitem [{\citenamefont {Christ}\ \emph
  {et~al.}(2014{\natexlab{b}})\citenamefont {Christ}, \citenamefont {Izubuchi},
  \citenamefont {Sachrajda}, \citenamefont {Soni},\ and\ \citenamefont
  {Yu}}]{Christ:2014qwa}%
  \BibitemOpen
  \bibfield  {author} {\bibinfo {author} {\bibfnamefont {Norman}\ \bibnamefont
  {Christ}}, \bibinfo {author} {\bibfnamefont {Taku}\ \bibnamefont {Izubuchi}},
  \bibinfo {author} {\bibfnamefont {Christopher~T.}\ \bibnamefont {Sachrajda}},
  \bibinfo {author} {\bibfnamefont {Amarjit}\ \bibnamefont {Soni}}, \ and\
  \bibinfo {author} {\bibfnamefont {Jianglei}\ \bibnamefont {Yu}} (\bibinfo
  {collaboration} {RBC, UKQCD}),\ }\bibfield  {title} {\enquote {\bibinfo
  {title} {{Calculating the KL-KS mass difference and $\epsilon_K$ to
  sub-percent accuracy}},}\ }\href@noop {} {\bibfield  {journal} {\bibinfo
  {journal} {PoS}\ }\textbf {\bibinfo {volume} {LATTICE2013}},\ \bibinfo
  {pages} {397} (\bibinfo {year} {2014}{\natexlab{b}})},\ \Eprint
  {http://arxiv.org/abs/1402.2577} {arXiv:1402.2577 [hep-lat]} \BibitemShut
  {NoStop}%
\bibitem [{\citenamefont {Christ}\ \emph {et~al.}(2016)\citenamefont {Christ},
  \citenamefont {Feng}, \citenamefont {Juttner}, \citenamefont {Lawson},
  \citenamefont {Portelli},\ and\ \citenamefont {Sachrajda}}]{Christ:2016mmq}%
  \BibitemOpen
  \bibfield  {author} {\bibinfo {author} {\bibfnamefont {Norman~H.}\
  \bibnamefont {Christ}}, \bibinfo {author} {\bibfnamefont {Xu}~\bibnamefont
  {Feng}}, \bibinfo {author} {\bibfnamefont {Andreas}\ \bibnamefont {Juttner}},
  \bibinfo {author} {\bibfnamefont {Andrew}\ \bibnamefont {Lawson}}, \bibinfo
  {author} {\bibfnamefont {Antonin}\ \bibnamefont {Portelli}}, \ and\ \bibinfo
  {author} {\bibfnamefont {Christopher~T.}\ \bibnamefont {Sachrajda}},\
  }\bibfield  {title} {\enquote {\bibinfo {title} {{First exploratory
  calculation of the long-distance contributions to the rare kaon decays
  $K\to\pi\ell^+\ell^-$}},}\ }\href {\doibase 10.1103/PhysRevD.94.114516}
  {\bibfield  {journal} {\bibinfo  {journal} {Phys. Rev.}\ }\textbf {\bibinfo
  {volume} {D94}},\ \bibinfo {pages} {114516} (\bibinfo {year} {2016})},\
  \Eprint {http://arxiv.org/abs/1608.07585} {arXiv:1608.07585 [hep-lat]}
  \BibitemShut {NoStop}%
\bibitem [{\citenamefont {Bai}\ \emph {et~al.}(2017)\citenamefont {Bai},
  \citenamefont {Christ}, \citenamefont {Feng}, \citenamefont {Lawson},
  \citenamefont {Portelli},\ and\ \citenamefont {Sachrajda}}]{Bai:2017fkh}%
  \BibitemOpen
  \bibfield  {author} {\bibinfo {author} {\bibfnamefont {Ziyuan}\ \bibnamefont
  {Bai}}, \bibinfo {author} {\bibfnamefont {Norman~H.}\ \bibnamefont {Christ}},
  \bibinfo {author} {\bibfnamefont {Xu}~\bibnamefont {Feng}}, \bibinfo {author}
  {\bibfnamefont {Andrew}\ \bibnamefont {Lawson}}, \bibinfo {author}
  {\bibfnamefont {Antonin}\ \bibnamefont {Portelli}}, \ and\ \bibinfo {author}
  {\bibfnamefont {Christopher~T.}\ \bibnamefont {Sachrajda}},\ }\bibfield
  {title} {\enquote {\bibinfo {title} {{Exploratory Lattice QCD Study of the
  Rare Kaon Decay $K^+\to\pi^+\nu\bar{\nu}$}},}\ }\href {\doibase
  10.1103/PhysRevLett.118.252001} {\bibfield  {journal} {\bibinfo  {journal}
  {Phys. Rev. Lett.}\ }\textbf {\bibinfo {volume} {118}},\ \bibinfo {pages}
  {252001} (\bibinfo {year} {2017})},\ \Eprint
  {http://arxiv.org/abs/1701.02858} {arXiv:1701.02858 [hep-lat]} \BibitemShut
  {NoStop}%
\bibitem [{\citenamefont {Bai}\ \emph {et~al.}(2018)\citenamefont {Bai},
  \citenamefont {Christ}, \citenamefont {Feng}, \citenamefont {Lawson},
  \citenamefont {Portelli},\ and\ \citenamefont {Sachrajda}}]{Bai:2018hqu}%
  \BibitemOpen
  \bibfield  {author} {\bibinfo {author} {\bibfnamefont {Ziyuan}\ \bibnamefont
  {Bai}}, \bibinfo {author} {\bibfnamefont {Norman~H.}\ \bibnamefont {Christ}},
  \bibinfo {author} {\bibfnamefont {Xu}~\bibnamefont {Feng}}, \bibinfo {author}
  {\bibfnamefont {Andrew}\ \bibnamefont {Lawson}}, \bibinfo {author}
  {\bibfnamefont {Antonin}\ \bibnamefont {Portelli}}, \ and\ \bibinfo {author}
  {\bibfnamefont {Christopher~T.}\ \bibnamefont {Sachrajda}},\ }\bibfield
  {title} {\enquote {\bibinfo {title} {{$K^+\to\pi^+\nu\bar{\nu}$ decay
  amplitude from lattice QCD}},}\ }\href {\doibase 10.1103/PhysRevD.98.074509}
  {\bibfield  {journal} {\bibinfo  {journal} {Phys. Rev.}\ }\textbf {\bibinfo
  {volume} {D98}},\ \bibinfo {pages} {074509} (\bibinfo {year} {2018})},\
  \Eprint {http://arxiv.org/abs/1806.11520} {arXiv:1806.11520 [hep-lat]}
  \BibitemShut {NoStop}%
\bibitem [{\citenamefont {Christ}\ \emph {et~al.}(2019)\citenamefont {Christ},
  \citenamefont {Feng},\ and\ \citenamefont {Sachrajda}}]{Christ:2019dxu}%
  \BibitemOpen
  \bibfield  {author} {\bibinfo {author} {\bibfnamefont {Norman~H.}\
  \bibnamefont {Christ}}, \bibinfo {author} {\bibfnamefont {Xu}~\bibnamefont
  {Feng}}, \ and\ \bibinfo {author} {\bibfnamefont {Christopher~T.}\
  \bibnamefont {Sachrajda}},\ }\bibfield  {title} {\enquote {\bibinfo {title}
  {{Lattice QCD study of the rare kaon decay $K^+\to\pi^+\nu\bar{\nu}$ at a
  near-physical pion mass}},}\ }\href@noop {} {\  (\bibinfo {year} {2019})},\
  \Eprint {http://arxiv.org/abs/1910.10644} {arXiv:1910.10644 [hep-lat]}
  \BibitemShut {NoStop}%
\bibitem [{\citenamefont {Briceno}\ \emph
  {et~al.}(2020{\natexlab{a}})\citenamefont {Briceno}, \citenamefont {Davoudi},
  \citenamefont {Hansen}, \citenamefont {Schindler},\ and\ \citenamefont
  {Baroni}}]{Briceno:2019opb}%
  \BibitemOpen
  \bibfield  {author} {\bibinfo {author} {\bibfnamefont {Raul~A.}\ \bibnamefont
  {Briceno}}, \bibinfo {author} {\bibfnamefont {Zohreh}\ \bibnamefont
  {Davoudi}}, \bibinfo {author} {\bibfnamefont {Maxwell~T.}\ \bibnamefont
  {Hansen}}, \bibinfo {author} {\bibfnamefont {Matthias~R.}\ \bibnamefont
  {Schindler}}, \ and\ \bibinfo {author} {\bibfnamefont {Alessandro}\
  \bibnamefont {Baroni}},\ }\bibfield  {title} {\enquote {\bibinfo {title}
  {{Long-range electroweak amplitudes of single hadrons from Euclidean
  finite-volume correlation functions}},}\ }\href {\doibase
  10.1103/PhysRevD.101.014509} {\bibfield  {journal} {\bibinfo  {journal}
  {Phys. Rev.}\ }\textbf {\bibinfo {volume} {D101}},\ \bibinfo {pages} {14509}
  (\bibinfo {year} {2020}{\natexlab{a}})},\ \Eprint
  {http://arxiv.org/abs/1911.04036} {arXiv:1911.04036 [hep-lat]} \BibitemShut
  {NoStop}%
\bibitem [{\citenamefont {Feng}\ \emph {et~al.}(2019)\citenamefont {Feng},
  \citenamefont {Jin}, \citenamefont {Tuo},\ and\ \citenamefont
  {Xia}}]{Feng:2018pdq}%
  \BibitemOpen
  \bibfield  {author} {\bibinfo {author} {\bibfnamefont {Xu}~\bibnamefont
  {Feng}}, \bibinfo {author} {\bibfnamefont {Lu-Chang}\ \bibnamefont {Jin}},
  \bibinfo {author} {\bibfnamefont {Xin-Yu}\ \bibnamefont {Tuo}}, \ and\
  \bibinfo {author} {\bibfnamefont {Shi-Cheng}\ \bibnamefont {Xia}},\
  }\bibfield  {title} {\enquote {\bibinfo {title} {{Light-Neutrino Exchange and
  Long-Distance Contributions to $0\nu2\beta$ Decays: An Exploratory Study on
  $\pi\pi\to ee$}},}\ }\href {\doibase 10.1103/PhysRevLett.122.022001}
  {\bibfield  {journal} {\bibinfo  {journal} {Phys. Rev. Lett.}\ }\textbf
  {\bibinfo {volume} {122}},\ \bibinfo {pages} {22001} (\bibinfo {year}
  {2019})},\ \Eprint {http://arxiv.org/abs/1809.10511} {arXiv:1809.10511
  [hep-lat]} \BibitemShut {NoStop}%
\bibitem [{\citenamefont {Tuo}\ \emph {et~al.}(2019)\citenamefont {Tuo},
  \citenamefont {Feng},\ and\ \citenamefont {Jin}}]{Tuo:2019bue}%
  \BibitemOpen
  \bibfield  {author} {\bibinfo {author} {\bibfnamefont {Xin-Yu}\ \bibnamefont
  {Tuo}}, \bibinfo {author} {\bibfnamefont {Xu}~\bibnamefont {Feng}}, \ and\
  \bibinfo {author} {\bibfnamefont {Lu-Chang}\ \bibnamefont {Jin}},\ }\bibfield
   {title} {\enquote {\bibinfo {title} {{Long-distance Contributions to
  Neutrinoless Double Beta Decay $\pi^- \to\pi^+ e e$}},}\ }\href@noop {} {\
  (\bibinfo {year} {2019})},\ \Eprint {http://arxiv.org/abs/1909.13525}
  {arXiv:1909.13525 [hep-lat]} \BibitemShut {NoStop}%
\bibitem [{\citenamefont {Detmold}\ and\ \citenamefont
  {Murphy}(2020)}]{Detmold:2020jqv}%
  \BibitemOpen
  \bibfield  {author} {\bibinfo {author} {\bibfnamefont {W.}~\bibnamefont
  {Detmold}}\ and\ \bibinfo {author} {\bibfnamefont {D.J.}\ \bibnamefont
  {Murphy}} (\bibinfo {collaboration} {NPLQCD}),\ }\bibfield  {title} {\enquote
  {\bibinfo {title} {{Neutrinoless Double Beta Decay from Lattice QCD: The
  Long-Distance $\pi^{-} \rightarrow \pi^{+} e^{-} e^{-}$ Amplitude}},}\
  }\href@noop {} {\  (\bibinfo {year} {2020})},\ \Eprint
  {http://arxiv.org/abs/2004.07404} {arXiv:2004.07404 [hep-lat]} \BibitemShut
  {NoStop}%
\bibitem [{\citenamefont {Feng}\ \emph {et~al.}(2020)\citenamefont {Feng},
  \citenamefont {Jin}, \citenamefont {Wang},\ and\ \citenamefont
  {Zhang}}]{Feng:2020nqj}%
  \BibitemOpen
  \bibfield  {author} {\bibinfo {author} {\bibfnamefont {Xu}~\bibnamefont
  {Feng}}, \bibinfo {author} {\bibfnamefont {Lu-Chang}\ \bibnamefont {Jin}},
  \bibinfo {author} {\bibfnamefont {Zi-Yu}\ \bibnamefont {Wang}}, \ and\
  \bibinfo {author} {\bibfnamefont {Zheng}\ \bibnamefont {Zhang}},\ }\bibfield
  {title} {\enquote {\bibinfo {title} {{Finite-volume formalism in the $2
  \xrightarrow[]{H_I+H_I} 2$ transition: an application to the lattice QCD
  calculation of double beta decays}},}\ }\href@noop {} {\  (\bibinfo {year}
  {2020})},\ \Eprint {http://arxiv.org/abs/2005.01956} {arXiv:2005.01956
  [hep-lat]} \BibitemShut {NoStop}%
\bibitem [{\citenamefont {Davoudi}\ and\ \citenamefont
  {Kadam}(2020)}]{Davoudi:2020}%
  \BibitemOpen
  \bibfield  {author} {\bibinfo {author} {\bibfnamefont {Zohreh}\ \bibnamefont
  {Davoudi}}\ and\ \bibinfo {author} {\bibfnamefont {Saurabh}\ \bibnamefont
  {Kadam}},\ }\href@noop {} {\  (\bibinfo {year} {2020})},\ \bibinfo {note} {in
  preparation.}\BibitemShut {Stop}%
\bibitem [{\citenamefont {Kaplan}\ \emph {et~al.}(1996)\citenamefont {Kaplan},
  \citenamefont {Savage},\ and\ \citenamefont {Wise}}]{Kaplan:1996xu}%
  \BibitemOpen
  \bibfield  {author} {\bibinfo {author} {\bibfnamefont {David~B.}\
  \bibnamefont {Kaplan}}, \bibinfo {author} {\bibfnamefont {Martin~J.}\
  \bibnamefont {Savage}}, \ and\ \bibinfo {author} {\bibfnamefont {Mark~B.}\
  \bibnamefont {Wise}},\ }\bibfield  {title} {\enquote {\bibinfo {title}
  {{Nucleon - nucleon scattering from effective field theory}},}\ }\href
  {\doibase 10.1016/0550-3213(96)00357-4} {\bibfield  {journal} {\bibinfo
  {journal} {Nucl. Phys. B}\ }\textbf {\bibinfo {volume} {478}},\ \bibinfo
  {pages} {629--659} (\bibinfo {year} {1996})},\ \Eprint
  {http://arxiv.org/abs/nucl-th/9605002} {arXiv:nucl-th/9605002} \BibitemShut
  {NoStop}%
\bibitem [{\citenamefont {Chen}\ \emph {et~al.}(2003)\citenamefont {Chen},
  \citenamefont {Heeger},\ and\ \citenamefont {Robertson}}]{Chen:2002pv}%
  \BibitemOpen
  \bibfield  {author} {\bibinfo {author} {\bibfnamefont {Jiunn-Wei}\
  \bibnamefont {Chen}}, \bibinfo {author} {\bibfnamefont {Karsten~M.}\
  \bibnamefont {Heeger}}, \ and\ \bibinfo {author} {\bibfnamefont
  {R.G.~Hamish}\ \bibnamefont {Robertson}},\ }\bibfield  {title} {\enquote
  {\bibinfo {title} {{Constraining the leading weak axial two-body current by
  SNO and super-K}},}\ }\href {\doibase 10.1103/PhysRevC.67.025801} {\bibfield
  {journal} {\bibinfo  {journal} {Phys. Rev. C}\ }\textbf {\bibinfo {volume}
  {67}},\ \bibinfo {pages} {025801} (\bibinfo {year} {2003})},\ \Eprint
  {http://arxiv.org/abs/nucl-th/0210073} {arXiv:nucl-th/0210073} \BibitemShut
  {NoStop}%
\bibitem [{\citenamefont {Chen}\ \emph {et~al.}(2005)\citenamefont {Chen},
  \citenamefont {Inoue}, \citenamefont {Ji},\ and\ \citenamefont
  {Li}}]{Chen:2005ak}%
  \BibitemOpen
  \bibfield  {author} {\bibinfo {author} {\bibfnamefont {Jiunn-Wei}\
  \bibnamefont {Chen}}, \bibinfo {author} {\bibfnamefont {Takashi}\
  \bibnamefont {Inoue}}, \bibinfo {author} {\bibfnamefont {Xiang-dong}\
  \bibnamefont {Ji}}, \ and\ \bibinfo {author} {\bibfnamefont {Ying-chuan}\
  \bibnamefont {Li}},\ }\bibfield  {title} {\enquote {\bibinfo {title} {{Fixing
  two-nucleon weak-axial coupling L(1,A) from mu- d capture}},}\ }\href
  {\doibase 10.1103/PhysRevC.72.061001} {\bibfield  {journal} {\bibinfo
  {journal} {Phys. Rev. C}\ }\textbf {\bibinfo {volume} {72}},\ \bibinfo
  {pages} {061001} (\bibinfo {year} {2005})},\ \Eprint
  {http://arxiv.org/abs/nucl-th/0506001} {arXiv:nucl-th/0506001} \BibitemShut
  {NoStop}%
\bibitem [{\citenamefont {Acharya}\ and\ \citenamefont
  {Bacca}(2020)}]{Acharya:2019fij}%
  \BibitemOpen
  \bibfield  {author} {\bibinfo {author} {\bibfnamefont {Bijaya}\ \bibnamefont
  {Acharya}}\ and\ \bibinfo {author} {\bibfnamefont {Sonia}\ \bibnamefont
  {Bacca}},\ }\bibfield  {title} {\enquote {\bibinfo {title}
  {{Neutrino-deuteron scattering: Uncertainty quantification and new $L_{1,A}$
  constraints}},}\ }\href {\doibase 10.1103/PhysRevC.101.015505} {\bibfield
  {journal} {\bibinfo  {journal} {Phys. Rev. C}\ }\textbf {\bibinfo {volume}
  {101}},\ \bibinfo {pages} {015505} (\bibinfo {year} {2020})},\ \Eprint
  {http://arxiv.org/abs/1911.12659} {arXiv:1911.12659 [nucl-th]} \BibitemShut
  {NoStop}%
\bibitem [{\citenamefont {Adelberger}\ \emph {et~al.}(2011)\citenamefont
  {Adelberger} \emph {et~al.}}]{Adelberger:2010qa}%
  \BibitemOpen
  \bibfield  {author} {\bibinfo {author} {\bibfnamefont {E.G.}\ \bibnamefont
  {Adelberger}} \emph {et~al.},\ }\bibfield  {title} {\enquote {\bibinfo
  {title} {{Solar fusion cross sections II: the pp chain and CNO cycles}},}\
  }\href {\doibase 10.1103/RevModPhys.83.195} {\bibfield  {journal} {\bibinfo
  {journal} {Rev. Mod. Phys.}\ }\textbf {\bibinfo {volume} {83}},\ \bibinfo
  {pages} {195} (\bibinfo {year} {2011})},\ \Eprint
  {http://arxiv.org/abs/1004.2318} {arXiv:1004.2318 [nucl-ex]} \BibitemShut
  {NoStop}%
\bibitem [{\citenamefont {De-Leon}\ \emph {et~al.}(2019)\citenamefont
  {De-Leon}, \citenamefont {Platter},\ and\ \citenamefont
  {Gazit}}]{De-Leon:2016wyu}%
  \BibitemOpen
  \bibfield  {author} {\bibinfo {author} {\bibfnamefont {Hilla}\ \bibnamefont
  {De-Leon}}, \bibinfo {author} {\bibfnamefont {Lucas}\ \bibnamefont
  {Platter}}, \ and\ \bibinfo {author} {\bibfnamefont {Doron}\ \bibnamefont
  {Gazit}},\ }\bibfield  {title} {\enquote {\bibinfo {title} {{Tritium
  $\beta$-decay in pionless effective field theory}},}\ }\href {\doibase
  10.1103/PhysRevC.100.055502} {\bibfield  {journal} {\bibinfo  {journal}
  {Phys. Rev. C}\ }\textbf {\bibinfo {volume} {100}},\ \bibinfo {pages}
  {055502} (\bibinfo {year} {2019})},\ \Eprint
  {http://arxiv.org/abs/1611.10004} {arXiv:1611.10004 [nucl-th]} \BibitemShut
  {NoStop}%
\bibitem [{\citenamefont {Baroni}\ \emph {et~al.}(2016)\citenamefont {Baroni},
  \citenamefont {Girlanda}, \citenamefont {Kievsky}, \citenamefont {Marcucci},
  \citenamefont {Schiavilla},\ and\ \citenamefont {Viviani}}]{Baroni:2016xll}%
  \BibitemOpen
  \bibfield  {author} {\bibinfo {author} {\bibfnamefont {A.}~\bibnamefont
  {Baroni}}, \bibinfo {author} {\bibfnamefont {L.}~\bibnamefont {Girlanda}},
  \bibinfo {author} {\bibfnamefont {A.}~\bibnamefont {Kievsky}}, \bibinfo
  {author} {\bibfnamefont {L.E.}\ \bibnamefont {Marcucci}}, \bibinfo {author}
  {\bibfnamefont {R.}~\bibnamefont {Schiavilla}}, \ and\ \bibinfo {author}
  {\bibfnamefont {M.}~\bibnamefont {Viviani}},\ }\bibfield  {title} {\enquote
  {\bibinfo {title} {{Tritium $\beta$-decay in chiral effective field
  theory}},}\ }\href {\doibase 10.1103/PhysRevC.94.024003} {\bibfield
  {journal} {\bibinfo  {journal} {Phys. Rev. C}\ }\textbf {\bibinfo {volume}
  {94}},\ \bibinfo {pages} {024003} (\bibinfo {year} {2016})},\ \bibinfo {note}
  {[Erratum: Phys.Rev.C 95, 059902 (2017)]},\ \Eprint
  {http://arxiv.org/abs/1605.01620} {arXiv:1605.01620 [nucl-th]} \BibitemShut
  {NoStop}%
\bibitem [{\citenamefont {Savage}\ \emph {et~al.}(2017)\citenamefont {Savage},
  \citenamefont {Shanahan}, \citenamefont {Tiburzi}, \citenamefont {Wagman},
  \citenamefont {Winter}, \citenamefont {Beane}, \citenamefont {Chang},
  \citenamefont {Davoudi}, \citenamefont {Detmold},\ and\ \citenamefont
  {Orginos}}]{Savage:2016kon}%
  \BibitemOpen
  \bibfield  {author} {\bibinfo {author} {\bibfnamefont {Martin~J.}\
  \bibnamefont {Savage}}, \bibinfo {author} {\bibfnamefont {Phiala~E.}\
  \bibnamefont {Shanahan}}, \bibinfo {author} {\bibfnamefont {Brian~C.}\
  \bibnamefont {Tiburzi}}, \bibinfo {author} {\bibfnamefont {Michael~L.}\
  \bibnamefont {Wagman}}, \bibinfo {author} {\bibfnamefont {Frank}\
  \bibnamefont {Winter}}, \bibinfo {author} {\bibfnamefont {Silas~R.}\
  \bibnamefont {Beane}}, \bibinfo {author} {\bibfnamefont {Emmanuel}\
  \bibnamefont {Chang}}, \bibinfo {author} {\bibfnamefont {Zohreh}\
  \bibnamefont {Davoudi}}, \bibinfo {author} {\bibfnamefont {William}\
  \bibnamefont {Detmold}}, \ and\ \bibinfo {author} {\bibfnamefont {Kostas}\
  \bibnamefont {Orginos}},\ }\bibfield  {title} {\enquote {\bibinfo {title}
  {{Proton-Proton Fusion and Tritium $\beta$ Decay from Lattice Quantum
  Chromodynamics}},}\ }\href {\doibase 10.1103/PhysRevLett.119.062002}
  {\bibfield  {journal} {\bibinfo  {journal} {Phys. Rev. Lett.}\ }\textbf
  {\bibinfo {volume} {119}},\ \bibinfo {pages} {062002} (\bibinfo {year}
  {2017})},\ \Eprint {http://arxiv.org/abs/1610.04545} {arXiv:1610.04545
  [hep-lat]} \BibitemShut {NoStop}%
\bibitem [{\citenamefont {Beane}\ \emph {et~al.}(2011)\citenamefont {Beane},
  \citenamefont {Detmold}, \citenamefont {Orginos},\ and\ \citenamefont
  {Savage}}]{Beane:2010em}%
  \BibitemOpen
  \bibfield  {author} {\bibinfo {author} {\bibfnamefont {S.R.}\ \bibnamefont
  {Beane}}, \bibinfo {author} {\bibfnamefont {W.}~\bibnamefont {Detmold}},
  \bibinfo {author} {\bibfnamefont {K.}~\bibnamefont {Orginos}}, \ and\
  \bibinfo {author} {\bibfnamefont {M.J.}\ \bibnamefont {Savage}},\ }\bibfield
  {title} {\enquote {\bibinfo {title} {{Nuclear Physics from Lattice QCD}},}\
  }\href {\doibase 10.1016/j.ppnp.2010.08.002} {\bibfield  {journal} {\bibinfo
  {journal} {Prog. Part. Nucl. Phys.}\ }\textbf {\bibinfo {volume} {66}},\
  \bibinfo {pages} {1--40} (\bibinfo {year} {2011})},\ \Eprint
  {http://arxiv.org/abs/1004.2935} {arXiv:1004.2935 [hep-lat]} \BibitemShut
  {NoStop}%
\bibitem [{\citenamefont {Beane}\ \emph {et~al.}(2015)\citenamefont {Beane},
  \citenamefont {Detmold}, \citenamefont {Orginos},\ and\ \citenamefont
  {Savage}}]{Beane:2014oea}%
  \BibitemOpen
  \bibfield  {author} {\bibinfo {author} {\bibfnamefont {Silas~R.}\
  \bibnamefont {Beane}}, \bibinfo {author} {\bibfnamefont {William}\
  \bibnamefont {Detmold}}, \bibinfo {author} {\bibfnamefont {Kostas}\
  \bibnamefont {Orginos}}, \ and\ \bibinfo {author} {\bibfnamefont {Martin~J.}\
  \bibnamefont {Savage}},\ }\bibfield  {title} {\enquote {\bibinfo {title}
  {{Uncertainty Quantification in Lattice QCD Calculations for Nuclear
  Physics}},}\ }\href {\doibase 10.1088/0954-3899/42/3/034022} {\bibfield
  {journal} {\bibinfo  {journal} {J. Phys. G}\ }\textbf {\bibinfo {volume}
  {42}},\ \bibinfo {pages} {034022} (\bibinfo {year} {2015})},\ \Eprint
  {http://arxiv.org/abs/1410.2937} {arXiv:1410.2937 [nucl-th]} \BibitemShut
  {NoStop}%
\bibitem [{\citenamefont {Briceno}\ \emph {et~al.}(2015)\citenamefont
  {Briceno}, \citenamefont {Davoudi},\ and\ \citenamefont
  {Luu}}]{Briceno:2014tqa}%
  \BibitemOpen
  \bibfield  {author} {\bibinfo {author} {\bibfnamefont {Raul~A.}\ \bibnamefont
  {Briceno}}, \bibinfo {author} {\bibfnamefont {Zohreh}\ \bibnamefont
  {Davoudi}}, \ and\ \bibinfo {author} {\bibfnamefont {Thomas~C.}\ \bibnamefont
  {Luu}},\ }\bibfield  {title} {\enquote {\bibinfo {title} {{Nuclear Reactions
  from Lattice QCD}},}\ }\href {\doibase 10.1088/0954-3899/42/2/023101}
  {\bibfield  {journal} {\bibinfo  {journal} {J. Phys. G}\ }\textbf {\bibinfo
  {volume} {42}},\ \bibinfo {pages} {023101} (\bibinfo {year} {2015})},\
  \Eprint {http://arxiv.org/abs/1406.5673} {arXiv:1406.5673 [hep-lat]}
  \BibitemShut {NoStop}%
\bibitem [{\citenamefont {Detmold}\ \emph {et~al.}(2019)\citenamefont
  {Detmold}, \citenamefont {Edwards}, \citenamefont {Dudek}, \citenamefont
  {Engelhardt}, \citenamefont {Lin}, \citenamefont {Meinel}, \citenamefont
  {Orginos},\ and\ \citenamefont {Shanahan}}]{Detmold:2019ghl}%
  \BibitemOpen
  \bibfield  {author} {\bibinfo {author} {\bibfnamefont {William}\ \bibnamefont
  {Detmold}}, \bibinfo {author} {\bibfnamefont {Robert~G.}\ \bibnamefont
  {Edwards}}, \bibinfo {author} {\bibfnamefont {Jozef~J.}\ \bibnamefont
  {Dudek}}, \bibinfo {author} {\bibfnamefont {Michael}\ \bibnamefont
  {Engelhardt}}, \bibinfo {author} {\bibfnamefont {Huey-Wen}\ \bibnamefont
  {Lin}}, \bibinfo {author} {\bibfnamefont {Stefan}\ \bibnamefont {Meinel}},
  \bibinfo {author} {\bibfnamefont {Kostas}\ \bibnamefont {Orginos}}, \ and\
  \bibinfo {author} {\bibfnamefont {Phiala}\ \bibnamefont {Shanahan}} (\bibinfo
  {collaboration} {USQCD}),\ }\bibfield  {title} {\enquote {\bibinfo {title}
  {{Hadrons and Nuclei}},}\ }\href {\doibase 10.1140/epja/i2019-12902-4}
  {\bibfield  {journal} {\bibinfo  {journal} {Eur. Phys. J. A}\ }\textbf
  {\bibinfo {volume} {55}},\ \bibinfo {pages} {193} (\bibinfo {year} {2019})},\
  \Eprint {http://arxiv.org/abs/1904.09512} {arXiv:1904.09512 [hep-lat]}
  \BibitemShut {NoStop}%
\bibitem [{\citenamefont {Cirigliano}\ \emph
  {et~al.}(2019{\natexlab{b}})\citenamefont {Cirigliano}, \citenamefont
  {Davoudi}, \citenamefont {Bhattacharya}, \citenamefont {Izubuchi},
  \citenamefont {Shanahan}, \citenamefont {Syritsyn},\ and\ \citenamefont
  {Wagman}}]{Cirigliano:2019jig}%
  \BibitemOpen
  \bibfield  {author} {\bibinfo {author} {\bibfnamefont {Vincenzo}\
  \bibnamefont {Cirigliano}}, \bibinfo {author} {\bibfnamefont {Zohreh}\
  \bibnamefont {Davoudi}}, \bibinfo {author} {\bibfnamefont {Tanmoy}\
  \bibnamefont {Bhattacharya}}, \bibinfo {author} {\bibfnamefont {Taku}\
  \bibnamefont {Izubuchi}}, \bibinfo {author} {\bibfnamefont {Phiala~E.}\
  \bibnamefont {Shanahan}}, \bibinfo {author} {\bibfnamefont {Sergey}\
  \bibnamefont {Syritsyn}}, \ and\ \bibinfo {author} {\bibfnamefont
  {Michael~L.}\ \bibnamefont {Wagman}} (\bibinfo {collaboration} {USQCD}),\
  }\bibfield  {title} {\enquote {\bibinfo {title} {{The Role of Lattice QCD in
  Searches for Violations of Fundamental Symmetries and Signals for New
  Physics}},}\ }\href {\doibase 10.1140/epja/i2019-12889-8} {\bibfield
  {journal} {\bibinfo  {journal} {Eur. Phys. J. A}\ }\textbf {\bibinfo {volume}
  {55}},\ \bibinfo {pages} {197} (\bibinfo {year} {2019}{\natexlab{b}})},\
  \Eprint {http://arxiv.org/abs/1904.09704} {arXiv:1904.09704 [hep-lat]}
  \BibitemShut {NoStop}%
\bibitem [{\citenamefont {Kronfeld}\ \emph {et~al.}(2019)\citenamefont
  {Kronfeld}, \citenamefont {Richards}, \citenamefont {Detmold}, \citenamefont
  {Gupta}, \citenamefont {Lin}, \citenamefont {Liu}, \citenamefont {Meyer},
  \citenamefont {Sufian},\ and\ \citenamefont {Syritsyn}}]{Kronfeld:2019nfb}%
  \BibitemOpen
  \bibfield  {author} {\bibinfo {author} {\bibfnamefont {Andreas~S.}\
  \bibnamefont {Kronfeld}}, \bibinfo {author} {\bibfnamefont {David~G.}\
  \bibnamefont {Richards}}, \bibinfo {author} {\bibfnamefont {William}\
  \bibnamefont {Detmold}}, \bibinfo {author} {\bibfnamefont {Rajan}\
  \bibnamefont {Gupta}}, \bibinfo {author} {\bibfnamefont {Huey-Wen}\
  \bibnamefont {Lin}}, \bibinfo {author} {\bibfnamefont {Keh-Fei}\ \bibnamefont
  {Liu}}, \bibinfo {author} {\bibfnamefont {Aaron~S.}\ \bibnamefont {Meyer}},
  \bibinfo {author} {\bibfnamefont {Raza}\ \bibnamefont {Sufian}}, \ and\
  \bibinfo {author} {\bibfnamefont {Sergey}\ \bibnamefont {Syritsyn}} (\bibinfo
  {collaboration} {USQCD}),\ }\bibfield  {title} {\enquote {\bibinfo {title}
  {{Lattice QCD and Neutrino-Nucleus Scattering}},}\ }\href {\doibase
  10.1140/epja/i2019-12916-x} {\bibfield  {journal} {\bibinfo  {journal} {Eur.
  Phys. J. A}\ }\textbf {\bibinfo {volume} {55}},\ \bibinfo {pages} {196}
  (\bibinfo {year} {2019})},\ \Eprint {http://arxiv.org/abs/1904.09931}
  {arXiv:1904.09931 [hep-lat]} \BibitemShut {NoStop}%
\bibitem [{\citenamefont {Cirigliano}\ \emph {et~al.}(2020)\citenamefont
  {Cirigliano}, \citenamefont {Detmold}, \citenamefont {Nicholson},\ and\
  \citenamefont {Shanahan}}]{Cirigliano:2020yhp}%
  \BibitemOpen
  \bibfield  {author} {\bibinfo {author} {\bibfnamefont {Vincenzo}\
  \bibnamefont {Cirigliano}}, \bibinfo {author} {\bibfnamefont {William}\
  \bibnamefont {Detmold}}, \bibinfo {author} {\bibfnamefont {Amy}\ \bibnamefont
  {Nicholson}}, \ and\ \bibinfo {author} {\bibfnamefont {Phiala}\ \bibnamefont
  {Shanahan}},\ }\bibfield  {title} {\enquote {\bibinfo {title} {{Lattice QCD
  Inputs for Nuclear Double Beta Decay}},}\ }\href {\doibase
  10.1016/j.ppnp.2020.103771} {\  (\bibinfo {year} {2020}),\
  10.1016/j.ppnp.2020.103771},\ \Eprint {http://arxiv.org/abs/2003.08493}
  {arXiv:2003.08493 [nucl-th]} \BibitemShut {NoStop}%
\bibitem [{\citenamefont {Drischler}\ \emph {et~al.}(2019)\citenamefont
  {Drischler}, \citenamefont {Haxton}, \citenamefont {McElvain}, \citenamefont
  {Mereghetti}, \citenamefont {Nicholson}, \citenamefont {Vranas},\ and\
  \citenamefont {Walker-Loud}}]{Drischler:2019xuo}%
  \BibitemOpen
  \bibfield  {author} {\bibinfo {author} {\bibfnamefont {Christian}\
  \bibnamefont {Drischler}}, \bibinfo {author} {\bibfnamefont {Wick}\
  \bibnamefont {Haxton}}, \bibinfo {author} {\bibfnamefont {Kenneth}\
  \bibnamefont {McElvain}}, \bibinfo {author} {\bibfnamefont {Emanuele}\
  \bibnamefont {Mereghetti}}, \bibinfo {author} {\bibfnamefont {Amy}\
  \bibnamefont {Nicholson}}, \bibinfo {author} {\bibfnamefont {Pavlos}\
  \bibnamefont {Vranas}}, \ and\ \bibinfo {author} {\bibfnamefont {André}\
  \bibnamefont {Walker-Loud}},\ }\bibfield  {title} {\enquote {\bibinfo {title}
  {{Towards grounding nuclear physics in QCD}},}\ \ }(\bibinfo {year} {2019})\
  \Eprint {http://arxiv.org/abs/1910.07961} {arXiv:1910.07961 [nucl-th]}
  \BibitemShut {NoStop}%
\bibitem [{\citenamefont {Briceno}\ \emph
  {et~al.}(2020{\natexlab{b}})\citenamefont {Briceno}, \citenamefont
  {Guerrero}, \citenamefont {Hansen},\ and\ \citenamefont
  {Sturzu}}]{Briceno:2020rar}%
  \BibitemOpen
  \bibfield  {author} {\bibinfo {author} {\bibfnamefont {Raul~A.}\ \bibnamefont
  {Briceno}}, \bibinfo {author} {\bibfnamefont {Juan~V.}\ \bibnamefont
  {Guerrero}}, \bibinfo {author} {\bibfnamefont {Maxwell~T.}\ \bibnamefont
  {Hansen}}, \ and\ \bibinfo {author} {\bibfnamefont {Alexandru}\ \bibnamefont
  {Sturzu}},\ }\bibfield  {title} {\enquote {\bibinfo {title} {{The role of
  boundary conditions in quantum computations of scattering observables}},}\
  }\href@noop {} {\  (\bibinfo {year} {2020}{\natexlab{b}})},\ \Eprint
  {http://arxiv.org/abs/2007.01155} {arXiv:2007.01155 [hep-lat]} \BibitemShut
  {NoStop}%
\end{thebibliography}%

\end{document}